\documentclass[usenatbib]{mn2e} 
\input psfig.sty 
\usepackage{epsfig}
\usepackage{amsmath,amssymb} 
\usepackage{psfig}

\newcommand{\Msol}{{\,\rm M}_\odot} 
\newcommand{\kpc} {{\,\rm kpc}} 
\newcommand{\pc} {{\,\rm pc}} 
\newcommand{\kms}{{\,\rm {km\,s^{-1}} }} 

\title[The formation of disc galaxies] {The formation of disc galaxies in a $\Lambda$CDM universe}\author[Oscar Agertz
  et al.] {\parbox[t]{\textwidth}{Oscar Agertz$^1$\thanks{agertz@physik.uzh.ch}, Romain Teyssier$^{1,2}$ and Ben Moore$^1$}\vspace*{6pt}\\$^1$ Institute for Theoretical Physics, University of Z\"urich, CH-8057 Z\"urich, Switzerland\\$^{2}$ CEA Saclay, DSM/IRFU/SAp, Batiment 709, 91191 Gif-sur-Yvette Cedex, France}
\date{\today}
\begin{document}
\maketitle
\begin{abstract} 
We study the formation of disc galaxies in a fully cosmological framework
using adaptive mesh refinement simulations. We perform an extensive parameter study of
the main subgrid processes that control how gas is converted into stars
and the coupled effect of supernovae feedback. We argue that previous
attempts to form disc galaxies have been unsuccessful because of the universal
adoption of strong feedback combined with high star formation
efficiencies. Unless extreme amounts of energy are injected into the interstellar medium during supernovae events, 
these star formation parameters result in bulge-dominated S0/Sa galaxies as star formation is too efficient at $z\sim 3$. 
We show that a low efficiency of star formation
more closely models the sub-parsec physical processes, especially at high redshift. 
We highlight the successful formation of extended disc galaxies with scale
lengths $r_{\rm d}=4-5\kpc$, flat rotation curves and bulge-to-disc ratios
of B/D$\,\sim1/4$. Not only do we resolve the formation of a Milky
Way-like spiral galaxy, we also observe the secular evolution of the disc as it forms a pseudo-bulge. The disc properties agree well with observations and
are compatible with the photometric and baryonic Tully-Fisher relations,
the $\Sigma_{\rm SFR}-\Sigma_{\rm gas}$
(Kennicutt-Schmidt) relation and the observed angular momentum content of
spiral galaxies. We conclude that the underlying small-scale star formation
physics plays a greater role than previously considered in simulations of
galaxy formation.
\end{abstract}


\section{Introduction}
\label{sect:intro}
The prevailing picture of galaxy formation emerged more than 30 yr ago \citep[][]{WhiteRees78,FallEfstathiou80}. Within the framework of the broadly accepted $\Lambda$ Cold Dark Matter ($\Lambda$CDM) scenario \citep{Komatsu2009}, gravity assembles structures in a bottom-up fashion. Haloes of dark matter acquire angular momentum via tidal torques \citep{Peebles69,FallEfstathiou80} from interacting structures, and as gas cools and condenses into their central parts, star-forming galaxies form. A realistic angular momentum content can be accounted for if most of the angular momentum is retained in the assembly process. In this picture, the host halo is responsible for the final galaxy characteristics \citep[e.g.][]{MoMaoWhite98}. While several aspects of the theory of galaxy formation are still being developed, e.g. the underlying physics of the missing satellite problem \citep{Klypin1999,moore99} and the role of cold stream accretion \citep[][]{Keres05,Keres09,Dekel09}, the model has proven successful for understanding global properties of galaxy assembly.

Given the complexity and non-linearity of the involved processes, computer simulations have become the ideal tool for studying the formation of structure. The formation of a late-type spiral galaxy, such as our own Milky Way, has been studied numerically in fully $\Lambda$CDM cosmological context by many authors \citep[e.g.][]{Abadi03a,SommerLarsen03,Governato04,Robertson04,Okamoto05,Governato07,croft09,Scannapieco09,Piontek09b,Agertz09b}. To date, no attempt has yielded a realistic candidate. The dominant reason for this is the so called "angular momentum problem" which leads to small, centrally concentrated discs dominated by large bulges \citep{NavarroBenz91,NavarroWhite1994}. Merging substructures lose angular momentum to the outer halo via dynamical friction, forcing the associated baryons to end up in the central parts of the proto-galaxy as a spheroid rather than a disc. This poses a problem for the theoretical understanding of extended late-type galaxies. This might in part stem from numerical issues:  the commonly used Smoothed Particle Hydrodynamics (SPH) \citep{GingoldMonaghan77,Lucy77} technique is known to incorrectly treat boundaries, hence poorly treating multiphase fluids \citep[e.g.][]{Agertz07,Read2010}. This can lead to artificial angular momentum transfer at the interface between cold disc and a hot halo \citep{Okamoto05}.

Many proposed solutions exists to the angular momentum problem, all amounting to the same process: keep the gas from cooling and forming stars too efficiently in the merging dark matter satellites at high redshift. One natural source is the cosmological UV background, being responsible for reionization at $z\gtrsim6$ which heats the gas, preventing it to cool efficiently into star-forming dwarf galaxies \citep{ThoulWeinberg96,Quinn96,gnedin00,hoeft06}. However, the impact on objects larger than $v_{\rm circ}\sim 10\kms$ is unclear due to e.g. self-shielding and efficient collisional cooling \citep{Dijkstra04}. 

Gas in low-mass haloes can also be blown out by supernova driven winds \citep{DekelSilk86,Efstathiou00}, hence lowering the resulting star formation efficiency (SFE), enriching the intergalactic medium (IGM) in the process. \cite{MacLowFerrara99} demonstrated that while dwarf galaxies of mass $10^6-10^9\Msol$ efficiently can expel metals in supernovae-driven winds, virtually no mass is lost for systems of mass $\gtrsim10^7\Msol$ \citep[see also][]{dubois08}. The inefficiency in driving winds from dwarfs was also reported by \cite{Marcolini06} who attributed this to the extended dark matter halo and efficient metal cooling. In this scenario, mass-loss and IGM enrichment will occur due to tidal and ram-pressure stripping \citep[e.g.][]{mori00}. Phenomenological models of e.g. momentum driven winds have proven successful in reproducing the high-$z$ IGM \citep{oppenheimerdave06} but it is uncertain how it regulates star formation and in what manner the expelled gas is re-accreted at later times \citep{oppenheimer10}.

Various recipes of supernovae feedback have been developed for numerical simulations \citep[e.g.][]{NavarroWhite93,Kay02,Scannapieco06}, and the methods have proven successful in removing low angular momentum material from central parts of galaxies \citep[e.g.][]{SommerLarsen03,Okamoto05,Governato07}, yielding more extended galaxies in comparison to models without feedback. However, it is unclear to what extent this way of reducing star formation can account for disc-dominated spiral galaxies like the Milky Way. Recently \cite{Scannapieco09} demonstrated numerically, in a fully cosmological setting, how a set of 8 Milky Way sized haloes failed to form significant discs. While half of the sample were early type galaxies resulting from late time mergers, the other half of the sample had less than 20 per cent of their stellar mass in discs. This can be a result of the inability of the adopted feedback to remove or redistribute low angular momentum material, but is also a strong indication that something else might regulate star formation at high redshift. On the same topic, \cite{Sawala10} argues that modern simulations of dwarf galaxy formation \citep{Valcke08,Stinson09,Governato10} all yield much larger stellar masses than expected from observations as well as gas-to-star conversion efficiencies almost an order of magnitude too large. \cite{Dutton09} found that, for SNe feedback to yield realistic galaxies, it must be very efficient, converting 25 per cent of the SN energy into outflows. If too strong feedback is employed, the discs can be destroyed by internal processes as too much material is ejected into the halo, preventing efficient disc reformation from cold gas, and possibly violating the upper bounds of halo gas found in X-ray surveys [see \cite{Bregman07} and references within]. In light of these studies, it is unclear if supernovae feedback is the sole agent in regulating star formation. Note that SNe explosions can regulate star formation in galaxies without expelling gas, being a driver of galactic turbulence \citep{maclow:review04}. 

Fundamentally, star formation is regulated by the availability of H$_2$. The observed Kennicutt-Schmidt (from now on \emph{K-S}) relation \citep{kennicutt98}, that relates $\Sigma_{\rm SFR}$ to $\Sigma_{\rm gas}$, varies strongly among individual spiral galaxies and can not be fit with a single power law \citep{bigiel2008}. $\Sigma_{\rm SFR}$ behaves very differently for $\Sigma_{\rm gas}$ greater or smaller than $\approx9\Msol\,{\rm pc}^{-2}$, marking the transition from atomic to fully molecular star-forming gas \citep{Leroy08}, and is dependent on gas metallicity, dust content, turbulence, small scale clumpiness and local dissociating UV field \citep{mckeeostriker07}. The inclusion of these processes and its impact on global star formation in discs has recently been studied both numerically \citep{RobertsonKravtsov08,Gnedin09,Pelupessy09} as well as analytically \citep[e.g.][]{Krumholz09}. A natural outcome of this treatment is an order of magnitude lower amplitude of the \emph{K-S} relation at high redshifts ($z\sim3$) \citep{GnedinKravtsov2010}. This agrees well with the observation of damped Ly$\alpha$ systems \citep[DLA:][]{WolfeChen06} as well as Lyman Break Galaxies \citep{Rafelski09}. This indicates that star formation can be made inefficient at high redshift, leaving gas for late-time star formation in a disc like environment, but not necessarily by expelling gas in supernova-driven winds. In addition, \cite{Murray10} argues that the disruption time-scale of giant molecular clouds (GMCs) due to jets, H{\,\small II} gas pressure, and radiation pressure also serves to regulate the SFE in galaxies. The disruption occurs well before the most massive stars exit the main sequence, meaning that supernovae in principle have little effect on GMC lifetimes.

In this paper we investigate to what extent supernovae feedback and the underlying small scale star-forming physics can affect the formation and evolution of realistic spiral galaxies in a fully cosmological setting. The former effect is studied via well tested numerical implementations of SNII, SNIa feedback coupled to metal enrichment, as well as stellar mass-loss. The latter influence is achieved by considering different normalizations of the Schmidt-law star formation efficiency. We conduct a comprehensive analysis of the resulting $z=0$ discs and compare them to observational relations. 

The paper is organized as follows. In Section\,\ref{sect:num}, we describe the numerical method used in this work, including the adopted feedback and star formation prescriptions. In Section\,\ref{sect:IC}, we present the cosmological initial conditions and discuss the free parameters of this work. Section\,\ref{sect:discs} outlines the disc analysis and summarizes the final properties of the simulation suite. In Section\,\ref{sect:effect} and Section\,\ref{sect:FB}, we present a detailed analysis of the impact of small-scale SFE and supernova feedback respectively. In Section\,\ref{sect:observations} we compare our simulations to modern observations. Finally, Section\,\ref{sect:discussion} summarizes and discusses our conclusions.
 
\section{Numerical framework}
\label{sect:num}
We use the Adaptive Mesh Refinement (AMR) code {\small RAMSES} \citep{teyssier02} to simulate the formation of a massive disc galaxy in a cosmological context including dark matter, gas and stars. The gas dynamics is calculated using a second-order unsplit Godunov method, while collisionless particles (including stars) are evolved using the particle-mesh technique. The equation of state of the gas is that of a perfect mono-atomic gas with an adiabatic index $\gamma=5/3$. Self-gravity of the gas is calculated by solving the Poisson equation using the multi-grid method \citep{brandt77} on the coarse grid and by the conjugate gradient method on finer ones. The modelling includes realistic recipes for star formation \citep{Rasera06}, supernova feedback and enrichment \citep{dubois08}. Details on these implementations are given below. Metals are advected as a passive scalar and are incorporated self-consistently in the cooling and heating routine. The code adopts the cooling function of \cite{sutherlanddopita93} for cooling at temperatures $10^4-10^{8.5}\,$K. We extend cooling down to 300 K using rates form \cite{rosenbregman95}. Gas metallicity is also accounted for in the cooling routines. A UV background is considered using the prescription of \cite{haardtmadau96}. In order to model a subgrid gaseous equation of state, hence avoiding artificial gas fragmentation, the gas is given a polytropic equation of state
\begin{equation} 
T=T_0\left(\frac{\rho}{\rho_0}\right)^{\gamma_0-1},
\end{equation}
for densities large than $\rho_0$. Throughout this paper we adopt $T_0=1000\,\rm{K}$ and $\gamma_0=2.0$. In this work, the polytrope density is set equal to the star formation threshold $n_0$. Following \cite{Agertz09b}, we adopt an initial metallicity of $Z=10^{-4}Z_\odot$ in the high-resolution region. This also serves as a flag for allowed regions of refinement. The refinement strategy is based on a quasi-Lagrangian approach, so that the number of particles per cell remains roughly constant, avoiding discreteness effects \citep[e.g.][]{Romeo08}. 

\subsection{Star formation} 
\label{sect:starform}
To model the conversion of gas into stars we adopt a Schmidt-law \citep{Schmidt1959} of the form
\begin{equation}
\label{eq:SF}
\dot{\rho}_{g}=-\epsilon_{\rm ff}\frac{\rho_{\rm g}}{t_{\rm ff}}\,\,{\rm for} \,\,\rho>\rho_0,
\end{equation}
where $\rho_{\rm g}$ is the gas density, $t_{\rm ff}=\sqrt{3\pi/32G\rho}$ is the local free-fall time, $\epsilon_{\rm ff}$ is the star formation efficiency per free-fall time and $\rho_0$ is the threshold for star formation. As soon as a cell is eligible for star formation, particles are spawned using a Poisson process where the stellar mass, $m_*$, is chosen to be a multiple of $\rho_0\Delta x^3$. Each formed star particles is treated as one stellar population with an associated initial mass function (IMF). This is a relevant approximation as the star particle masses are orders of magnitudes larger than the average stellar masses. We also ensure that no more than $90$ per cent of the gas in a cell is depleted by the star formation process.

The $\rho_0$ and $\epsilon_{\rm ff}$ parameters in Eq.\,\ref{eq:SF} are, in addition to being unconstrained physical parameters, resolution and hence scale dependent. There are in a sense two regimes of star formation in global simulations of disc galaxies as follows.

\begin{enumerate}
\item {\emph{The ISM is resolved:}} At parsec-scale resolution, star formation occurs in their natural sites i.e. massive clouds such as GMCs. Modern estimates of star formation efficiencies by \cite{krumholztan07} point towards values of $\epsilon_{\rm ff}=1-2$ per cent at densities of $n\sim 10^2-10^5\,{\rm cm}^{-3}$. To only allow for star formation to take place in the actual physical star formation sites, hence tracing the formation of ${\rm H}_2$, allows for more accurate predictions of e.g. the Kennicutt-Schmidt star formation relation with less of a requirement to tune numerical parameters \citep[see e.g.][]{Gnedin09}. This treatment leads to $\rho_{\rm g}\rightarrow\rho_{{\rm H}_2}$ in Eq.\,\ref{eq:SF}, which is equivalent to $\epsilon_{\rm ff}$ being dependent on the environment, due in part to the local H$_2$ fraction. On galactic scales, this means that the scaleheight of all ISM components are resolved using at least 10 resolution elements \citep{romeo94}. If this is not satisfied, the true disc stability will not be modelled accurately. This treatment is the goal of most simulations, but is due to the computational load beyond the capabilities of modern simulations attempting to study the assembly and evolution of large spiral galaxies to $z=0$. Isolated simulations of large spiral galaxies in a non-cosmological setting have successfully reached this resolution \citep{Agertz09,tasker09}, albeit with simplified physics. As the star formation sites become resolved, new physics becomes important e.g. radiative feedback in order to accurately treat the lifetimes of GMC structures \citep{Murray10}. 
\\
\item {\emph{The ISM is under-resolved:}} To radially resolve a Milky Way like galactic disc, i.e. sampling the scale radius with at least 10 resolution elements, a force and hydro resolution of a few $100\,\pc$ is necessary. At this resolution the scale height is captured with more or less one resolution element. The true disc stability can be affected as both the density and velocity structure (gas and stellar dispersion) are influenced numerically. This still allows for the disc to have the correct global properties such as gas and stellar mass compositions, thin and thick disc, and even to develop realistic spiral structure. In this case a statistical star formation recipe based on the local gas density and free-fall time is well motivated both theoretically and observationally.
\end{enumerate}

As we will describe in Section\,\ref{sect:free}, we are targeting the latter regime of subgrid star formation and will investigate some of the numerical caveats related to it. At resolutions of several 100 pc, the $\epsilon_{\rm ff}$ parameter absorbs the small scale physics regulating star formation, allowing for a qualitative influence on galaxy formation. We note that there are alternative formulation of gas to star conversion laws in the literature (see e.g. \cite{Leroy08} for a comprehensive summary). However, as they are all designed to fit an observed relation, and all include a normalization constant similar to $\epsilon_{\rm ff}$, we believe Eq.\,\ref{eq:SF} is a representative choice for this study.

\subsection{Supernovae and stellar feedback}
\label{sect:feedback}
The standard recipe for supernova feedback in {\small RAMSES} involves only Type II supernovae events (SNII). We have also implemented additional treatment of Type Ia events (SNIa) as well as mass-loss via stellar winds. Including all of these effects is important as a single stellar population can return up to $30-40$ per cent of its mass to the ISM during its lifetime. The implementations are as follows.

\subsubsection{Type II}
\label{sect:SNII}
Type II SN events are relevant for stellar masses of $8-40\,\Msol$ which represents $\sim10$ per cent of the mass of a stellar population, regardless of IMF. We assume that 10 Myr after a star particle is formed, 10 per cent of the star particle's mass is injected into the nearest gas cell together with a total energy of $E_{\rm SNII}=10^{51}(m_{\rm ejecta}/10\,\Msol)\,{\rm erg}$ in thermal energy. At low resolution, this energy would quickly radiate away \citep{Katz92} in the dense gas, without allowing for an adiabatic expansion of the supernova blast-wave \citep{McKeeOstriker1977}. To remedy this, we turn off cooling in cells containing young stars to allow for the blast-wave to grow and be resolved by few cells, hence converting thermal energy into $P{\rm d}V$ work \citep[see e.g.][]{Gerritsen1997PhDT}. In detail, for every star formation event, the inverse of the birth time, $1/t_{\rm bt}$, is stored in the computational grid, overwriting any previous value. This passive scalar field is advected with the hydro flow (the conserved quantity is $\rho_{\rm gas}/t_{\rm bt}$). Gas cooling at a simulation time $t_{\rm sim}$ is only allowed if $t_{\rm sim}-t_{\rm bt}>\Delta t_{\rm off}$, where $\Delta t_{\rm off}$ is the cooling shut-off time-scale. Calculations of relevant time-scales and numerical tests using the SPH formalism was carried out by \cite{Stinson06}. Relevant time-scales are on the order of tens of Myr and we adopt $\Delta t_{\rm off}=50\,{\rm Myr}$. 

\subsubsection{Type Ia}
We treat SNIa and stellar mass-loss using the prescription outlined in \cite{Raiteri1996}. The assumed IMF is the parametrization by \cite{Kroupa1993}, which for a star particle of mass $m_*$ reads
\begin{equation*} 
\label{eq:IMF}
\Phi(M) = m_*A\left\{
\begin{array}{rl} 
2^{0.9}M^{-1.3} & \text{for } 0.08\leq M < 0.5\,\Msol \\
M^{-2.2} & \text{for } 0.5\leq M < 1\,\Msol \\
M^{-2.7} & \text{for } M \geq 1\,\Msol,
\end{array} \right.
\end{equation*}
where $M$ is here the stellar mass in units of $\Msol$ and the normalization constant $A\approx 0.3029$. The adopted lower and upper limits are $0.08\Msol$ and $100\Msol$ respectively. At each simulation time-step, we calculate the mass fraction of each star particle ending its H and He burning phase, i.e. leaving the main sequence, using the fit
\begin{equation}
\label{eq:agemass}
\log t_*=a_0(Z)+a_1(Z)\log M+a_2(Z)(\log M)^2,
\end{equation}
where $t_*$ is the lifetime of the star and $Z$ the metallicity. The adopted coefficients and references to the original data can be found in \cite{Raiteri1996}. Progenitors of SNIa are carbon plus oxygen white dwarfs that accrete mass from binary companions. Stellar evolution theory predict the binary masses to be in the range of $\sim3-16\Msol$. The number of SNIa events within a star particle, at a given simulation time with an associated timestep $\Delta t$, is 
\begin{equation}
N_{\rm SNIa}=\int_{m_t}^{m_{t+\Delta t}}\hat{\Phi}(M_2){\rm d}M_2,
\end{equation}
where $m_t$ and $m_{t+\Delta t}$ is the mass interval of stars ending their life during the computational timestep. $\hat{\Phi}(M_2)$ is the IMF of the secondary star, i.e.
\begin{equation}
\hat{\Phi}(M_2)=m_*A'\int_{M_{\rm inf}}^{M_{\rm sup}}\left(\frac{M_2}{M_{\rm B}}\right)^2 M_{\rm B}^{-2.7} {\rm d}M_{\rm B},
\end{equation}
where $M_{\rm B}$ is the mass of the binary, $M_{\rm inf}={\rm max}(2M_2,3\Msol)$ and $M_{\rm sup}=M_2+8\Msol$. The constant $A'=0.16A$, and is a calibrated value for SNIa events in our Galaxy \citep{VandenberghMcclure94}. Each explosion is assumed to release $10^{51}$ erg (released as thermal energy in the nearest gas cell) and $0.76\Msol$ of metal enriched material ($0.13\Msol$ of ${}^{16}$O and $0.63\Msol$ of ${}^{56}$Fe) \citep{Thielemann86}.

\subsubsection{Stellar mass-loss}
\label{sect:massloss}
For each time-step $\Delta t$, and star particle, we calculate the average stellar mass, $\langle M\rangle$, exiting the main sequence using Eq.\,\ref{eq:agemass}. The mass-loss during the $\Delta t$ time-span is calculated using the best fit initial-final mass relation of \cite{Kalirai2008}:
\begin{equation}
M_{\rm wind}=0.891-0.394/\langle M\rangle.
\end{equation}
At each time interval $\Delta t$, the total mass-loss in winds is
\begin{equation}
M_{\rm tot,wind}= f_m m_*M_{\rm wind},
\end{equation}
where 
\begin{equation}
f_m=\int_{m_t}^{m_{t+\Delta t}}\Phi(M){\rm d}M.
\end{equation}
The lost stellar mass enters the gaseous mass in the nearest cell and the gas metallicity is updated consistently with the star particle's metallicity. 

\section{Initial conditions and simulation suite} 
\label{sect:IC}
The initial conditions used in this work are a subset the Silver River simulation suite (Potter et al. in preparation), aimed to study the pure dark matter assembly history of a Milky Ways size halo in much greater detail than here. We adopt a \emph{WMAP5} \citep{Komatsu2009} compatible cosmology, i.e. a $\Lambda$CDM Universe with $\Omega_{\Lambda}=0.73$, $\Omega_{\rm m}=0.27$, $\Omega_{\rm b}=0.045$, $\sigma_8=0.8$ and $H_0=70\,{\rm km\,s}^{-1}\,{\rm Mpc}^{-1}$. 

The upcoming work by Potter et al. will present the details concerning the initial condition generation. Briefly, a pure dark matter simulation was performed using a simulation cube of size $L_{\rm box}=179\,{\rm Mpc}$. At $z=0$, a halo of mass $M_{\rm 200,c}\approx 9.7\times10^{11}\,\Msol$ was selected for re-simulation at high resolution, and traced back to the initial conditions at $z=133$. $M_{\rm 200,c}$ is the virial mass of the halo, defined as the mass enclosed in a sphere with mean density 200 times the critical value. The corresponding virial radius is $r_{\rm 200,c}=205\,\kpc$. By using a definition based on 200 times the background density we obtain $M_{\rm 200,bg}=1.25\times10^{12}\,\Msol$ and $r_{\rm 200,bg}=340\kpc$. When baryons are included in the simulations, the final \emph{total} halo mass remains roughly the same.

The halo has a quiet merger history, i.e. it undergoes no major merger after $z=1$, which favours the formation of a late-type galaxy. A nested hierarchy of initial conditions for the dark matter and baryons was generated using the {\small GRAFIC++}\footnote{{\tt http://grafic.sourceforge.net/}} code, where we allow for the high resolution particles to extended to 3 virial radii from the centre of the halo at $z=0$. This avoids mixing of different mass dark matter particles in the inner parts of the domain. In this work, we focus on two sets of resolutions from the Silver River suite, referred to as SR5 and SR6. The simulations are identical apart form the number of dark matter particles, and hence particle mass, as well as maximal AMR refinement. In SR6 the dark matter particle mass is $m_{\rm DM}=2.5\times 10^6\,\Msol$ and in SR5  $m_{\rm DM}=3.2\times 10^5\,\Msol$. The mesh is refined if a cell contains more than eight dark matter particles, and similar criterion is employed for the baryonic component. At the maximum level of refinement, the simulations reach a physical resolution of $\Delta x=170\,\pc$ and $\Delta x=340\,\pc$ in SR5 and SR6, respectively.

\subsection{The free parameters}
\label{sect:free}
\begin{table*}
\caption{Summary of the numerical parameters. The simulations use a maximum physical cell resolution of $\Delta x= 340\pc$ (SR6) or $\Delta x= 170\pc$ (SR5), and the high-resolution region is occupied with dark matter particles of mass $m_{\rm DM}=2.5\times 10^6\Msol$ (SR6) or $m_{\rm DM}=3.25\times 10^5\Msol$ (SR5). All simulations use delayed cooling in regions of young stars, unless specified. When SNII feedback is used $E_{\rm SNII}=10^{51}\,{\rm erg}$, unless other values are indicated.}
\label{table:simsummary1}
\center
\begin{minipage}{140mm}
\center
\begin{tabular}{cccc}
\hline
Run & $\epsilon_{\rm ff}$ & Feedback & Star formation threshold, $n_0$ ($\rm{cm^{-3}}$) \\ 
\hline
SR6-n01e1 & $1$ & SNII & 0.1 \\ 
SR6-n01e2 & $2$ & SNII & 0.1 \\ 
SR6-n01e5 & $5$ & SNII & 0.1 \\ 
\\
SR6-n01e1ML & $1$ & SNII, SNIa, mass-loss & 0.1 \\ 
SR6-n01e2ML & $2$ & SNII, SNIa, mass-loss & 0.1 \\ 
SR6-n01e5ML & $5$ & SNII, SNIa, mass-loss & 0.1 \\ 
\\
SR6-n1e1 & $1$ & SNII & 1 \\ 
SR6-n1e2 & $2$ & SNII & 1 \\ 
SR6-n1e5 &  $5$ & SNII & 1\\ 
\\
SR6-n1e1ML & $1$ & SNII, SNIa, mass-loss & 1 \\ 
SR6-n1e2ML & $2$ & SNII, SNIa, mass-loss & 1 \\ 
\\
SR6-n01e1NFB & $1$ & No feedback, $E_{\rm SNII}=0$  & 0.1 \\ 
SR6-n01e5NFB & $5$ & No feedback, $E_{\rm SNII}=0$ & 0.1 \\ 
SR6-n01e5NFBmet & $1$ & No feedback but metal enrichment & 0.1 \\ 
SR6-n01e5SN2 & $5$ & SNII, $E_{\rm SNII}=2\times 10^{51}\,{\rm erg}$ & 0.1 \\ 
SR6-n01e5SN5 & $5$ & SNII, $E_{\rm SNII}=5\times 10^{51}\,{\rm erg}$ & 0.1 \\ 
\\
SR5-n1e1ML & $1$ & SNII, SNIa, mass-loss & 1 \\ 
\hline
\end{tabular}
\end{minipage}
\end{table*}

\begin{table*}
\caption{Summary of disc characteristics at $z=0$. The mass of the components are obtained by fitting the stellar surface density (see text), and are in units of $10^{10} \Msol$. Note that we consider \emph{all} gas phases for the gas mass and all stars for the stellar masses. (1) Fitted scalelength of stellar disc. Large uncertainties exist for $r_{\rm d}>10\,\kpc$ as the stellar discs are small and feature almost flat stellar surface density profiles. (2) $f_{\rm gas}=M_{\rm disc,g}/(M_{\rm disc,tot}+M_{\rm bulge})$. (3) Total measured specific angular momentum of the baryons in the disc and bulge in units of ${\rm km\,s}^{-1}\kpc$.}
\label{table:simsummary2}
\begin{minipage}{180mm}
\center
\begin{tabular}{ccccccccccc}
\hline 
Run & $M_{\rm disc,s}$  & $M_{\rm disc,g}$ & $M_{\rm bulge,s}$ & $r_{\rm d}$ (kpc) (1) & $f_{\rm gas}$ (2) & B/D & B/T & $j_{\rm bar}$ (3) \\ 
\hline 
SR6-n01e1 	& 8.6 	& 1.6 	 	& 2.0 & 3.8 & 0.13	&	0.23  & 0.19 & 1920 \\ 
SR6-n01e2 	& 7.4	& 1.3		& 4.6 & 7.6 & 0.10	&	0.62 & 0.38 & 1655 \\ 
SR6-n01e5 	& 5.6	& 0.72		& 7.0 & $\sim15.0$ & 0.05	&	1.25 & 0.56 &1305 \\ 
\\
SR6-n01e1ML 	& 8.0	& 2.3	 	& 2.2 & 5.0 & 0.18	&	0.27	& 0.21 &1960 \\  
SR6-n01e2ML 	& 8.1 	& 1.6		& 3.8 & 5.0 & 0.12	&	0.47 & 0.32 &1718 \\
SR6-n01e5ML 	& 5.5 	& 0.93		& 7.2 & $\sim15.0$ & 0.07	& 	1.30   & 0.57 &	1464 \\
\\
SR6-n1e1 	& 6.6 	& 3.3 		& 2.9 & 2.7 & 0.26 	& 	0.44 & 0.31 & 1594 \\  
SR6-n1e2 	& 6.4 	& 2.4		& 4.3 & 2.5 & 0.18  	&	0.67 & 0.40 &1804 \\ 
SR6-n1e5 	& 6.0 	& 2.1		& 5.2 & 2.7 & 0.16	&	0.87 & 0.46 & 1643 \\ 
\\
SR6-n1e1ML 	& 6.5 	& 3.6 		& 2.7 & 2.7 & 0.28	&	0.42	& 0.29 &1618 \\ 
SR6-n1e2ML 	& 6.3	& 2.9	 	& 4.3 & 2.7 & 0.21	&	0.68	& 0.41 & 1281 \\ 
\\
SR6-n01e1NFB & 7.8	& 1.6	 	& 3.1 & 4.0 & 0.13	&	0.40	& 0.28 & 1938 \\
SR6-n01e5NFB & 5.8       & 0.62 		& 6.5 & $\sim15.0$ & 0.05	& 	1.12  & 0.53 & 1394 \\
SR6-n01e5NFBmet 	& 5.6 	& 0.56	& 6.6 & 10.0 & 0.05  & 	1.18	& 0.54 & 1430 \\
SR6-n01e5SN2 	& 5.6 	& 0.55	& 6.5 & 4.5 & 0.05	& 	1.16	& 0.54 & 1266 \\
SR6-n01e5SN5 	& 6.6	& 0.81	& 2.3 & 2.8 & 0.09	& 	0.35	& 0.26 & 1387 \\
\\
SR5-n1e1ML 	& 9.0 		& 2.0  	& 2.2 & 2.8 & 0.18	&	0.24	& 0.20 &  \\ 
\hline

\end{tabular}
\end{minipage}
\end{table*}
\begin{figure*}
\center
\begin{tabular}{cccccc}
\psfig{file=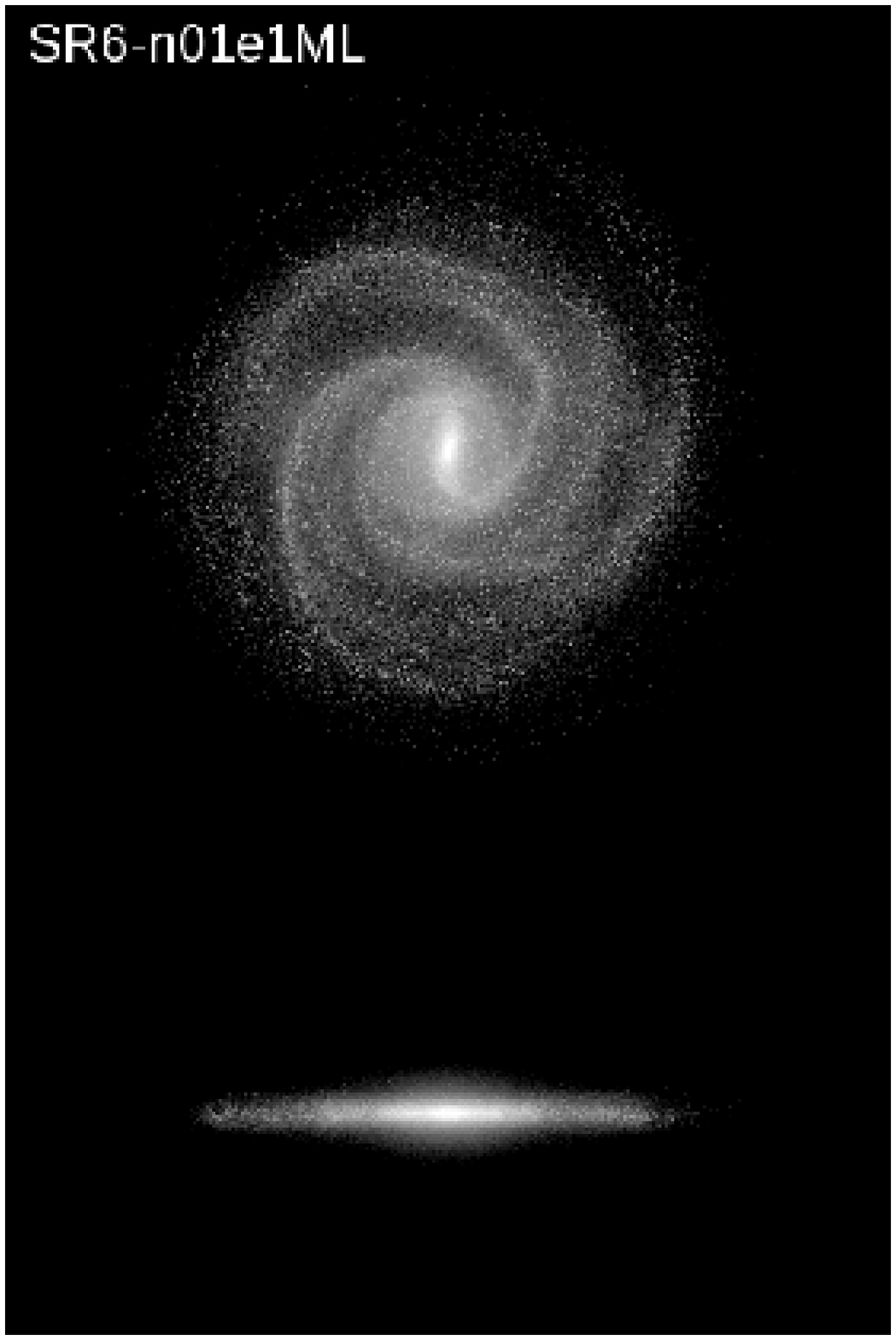,width=82pt} 
\psfig{file=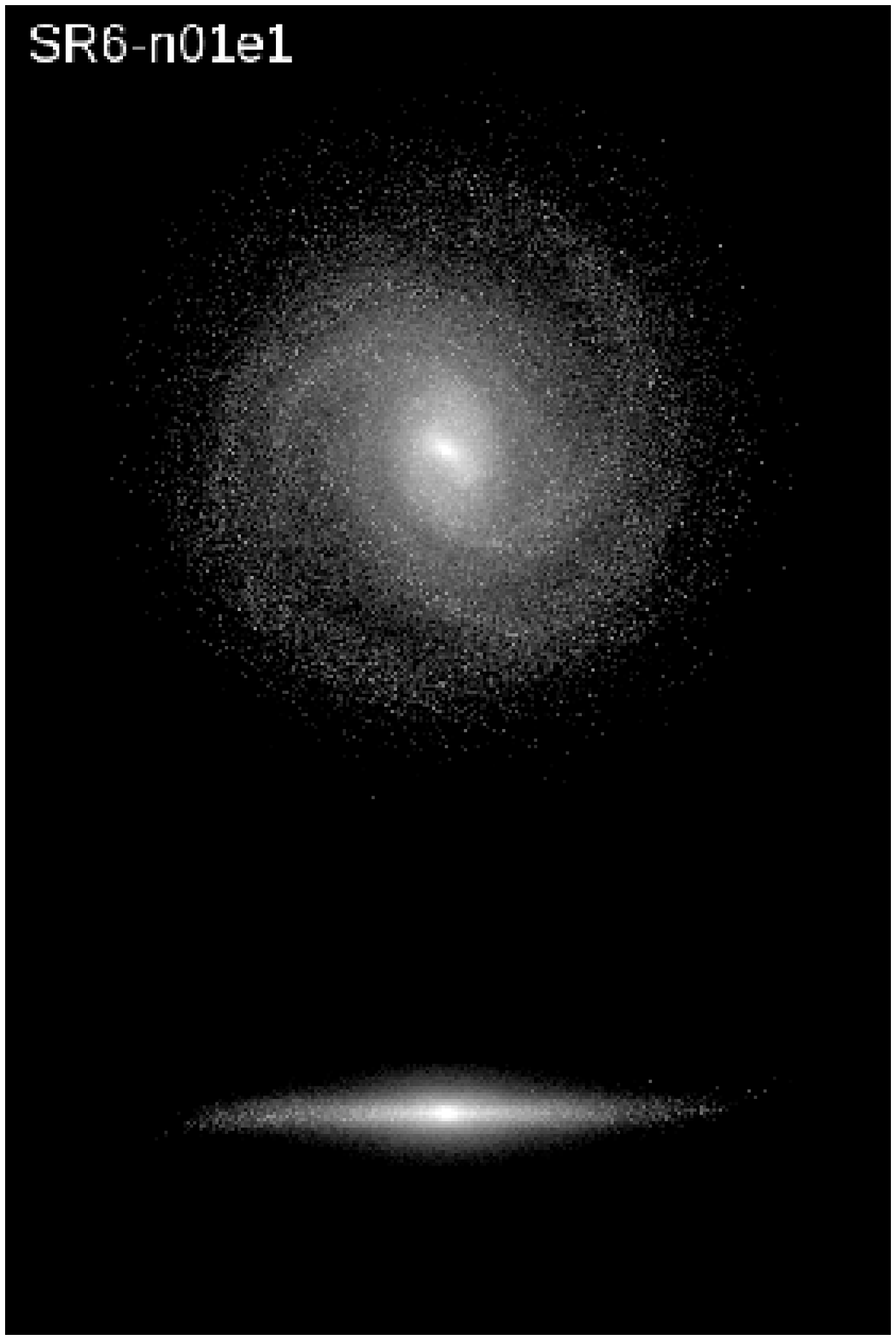,width=82pt} 
\psfig{file=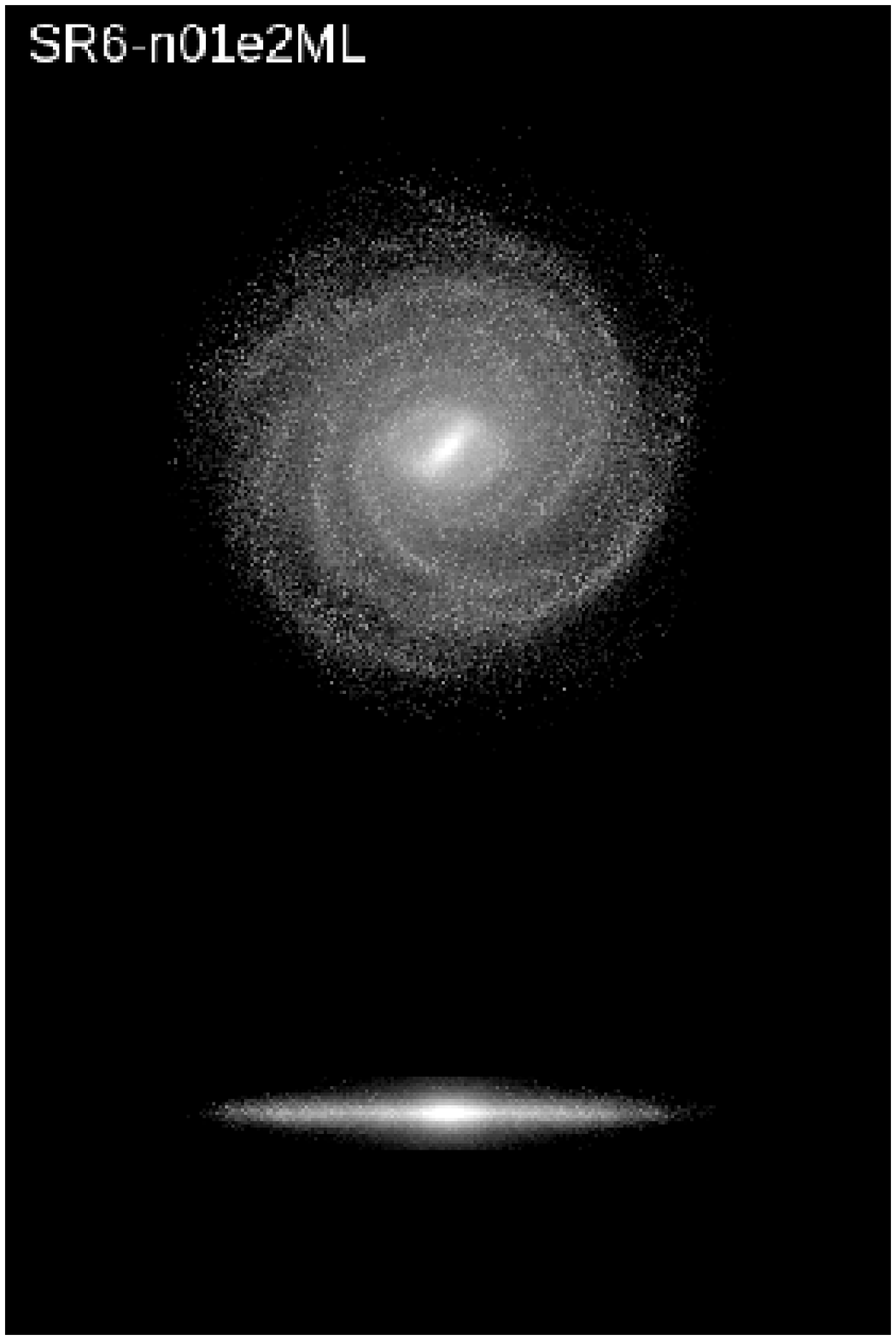,width=82pt} 
\psfig{file=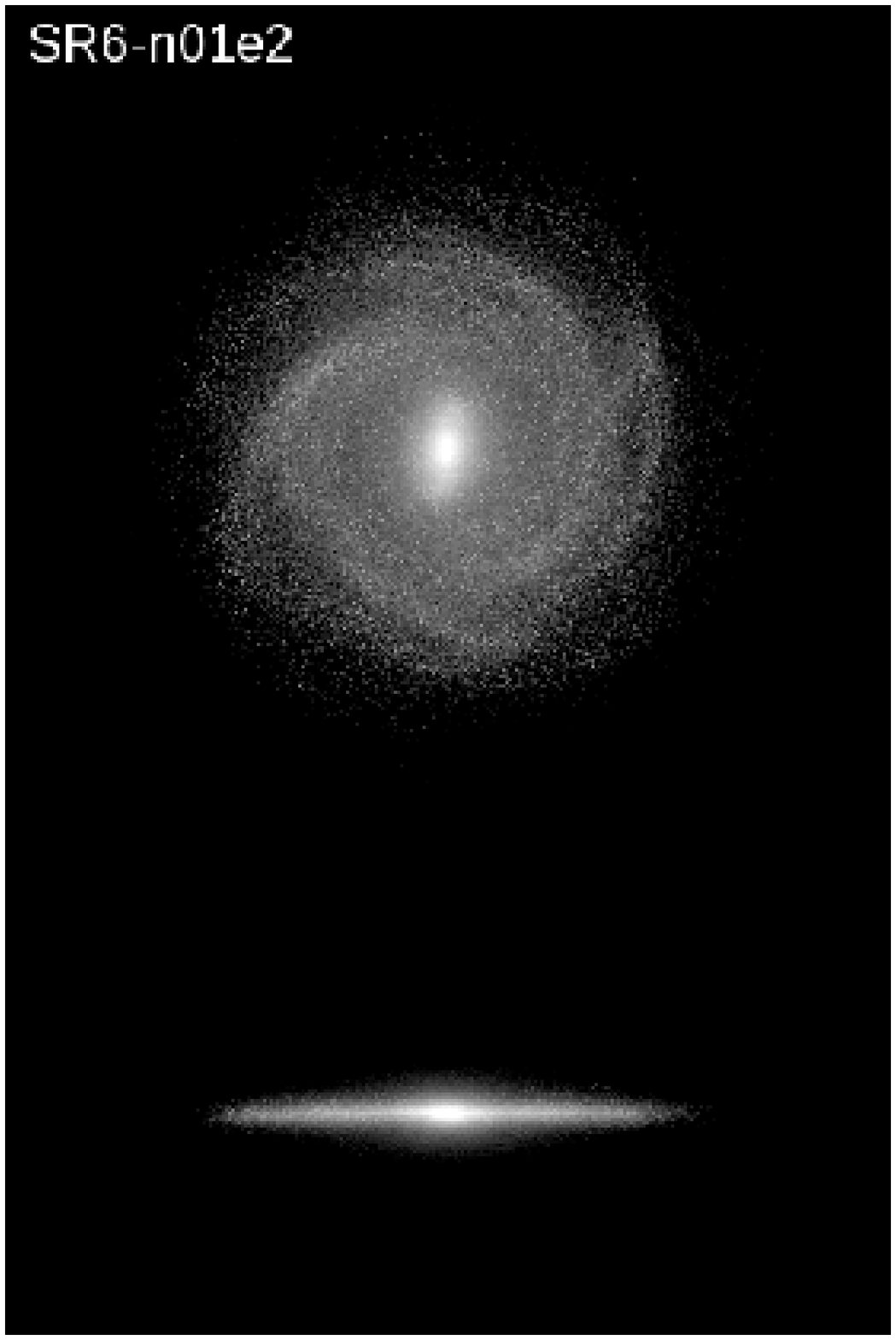,width=82pt} 
\psfig{file=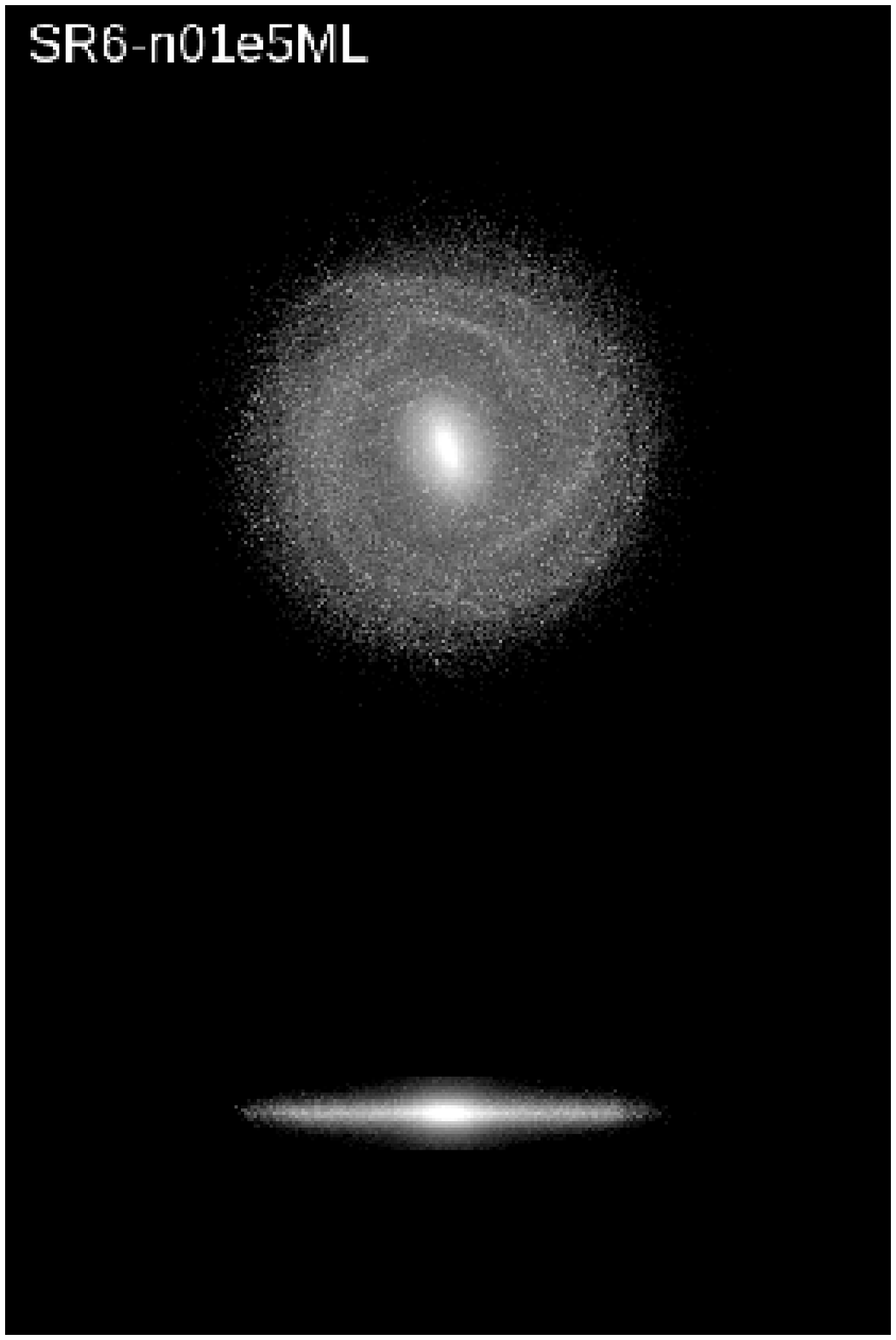,width=82pt}
\psfig{file=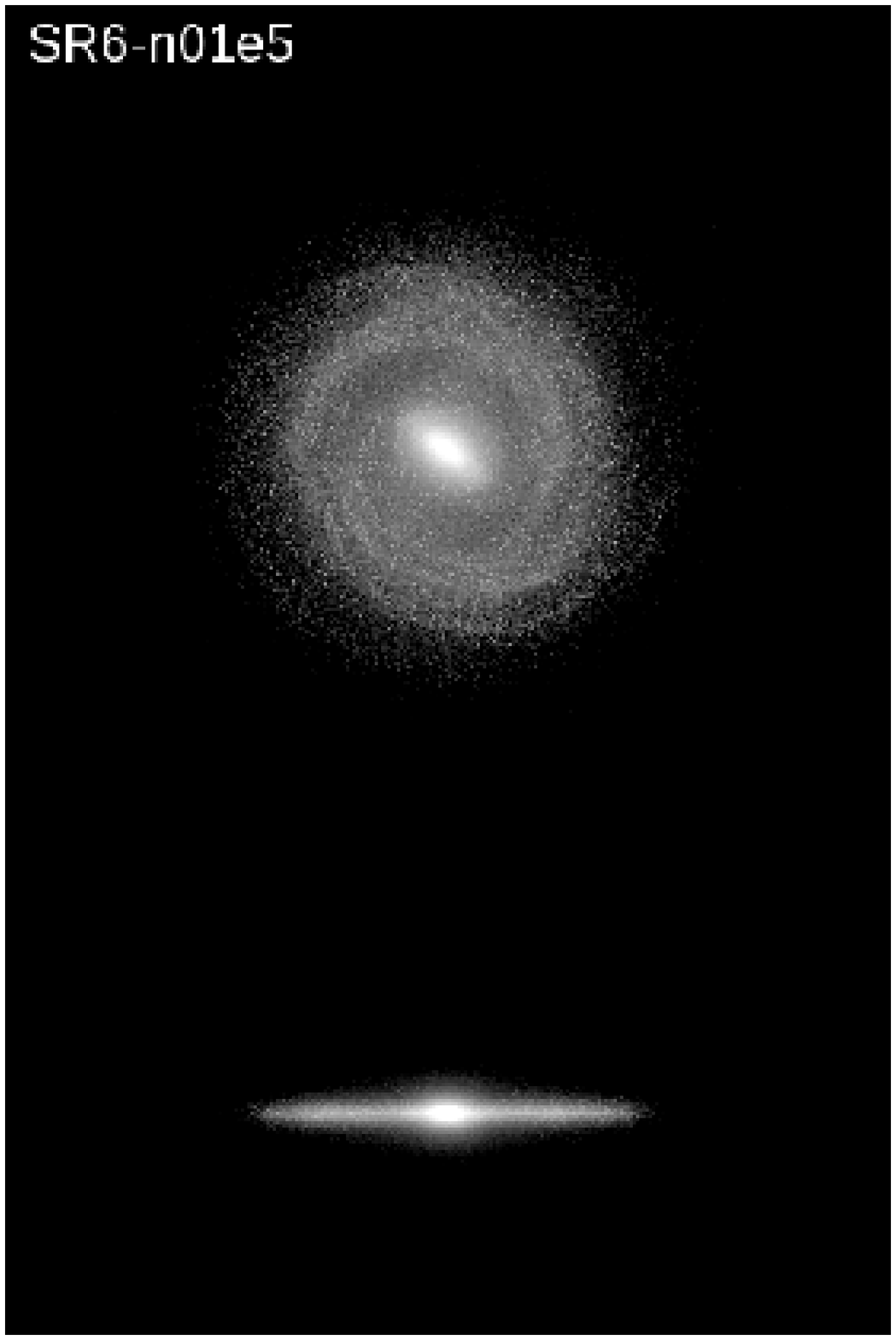,width=82pt} \\
\psfig{file=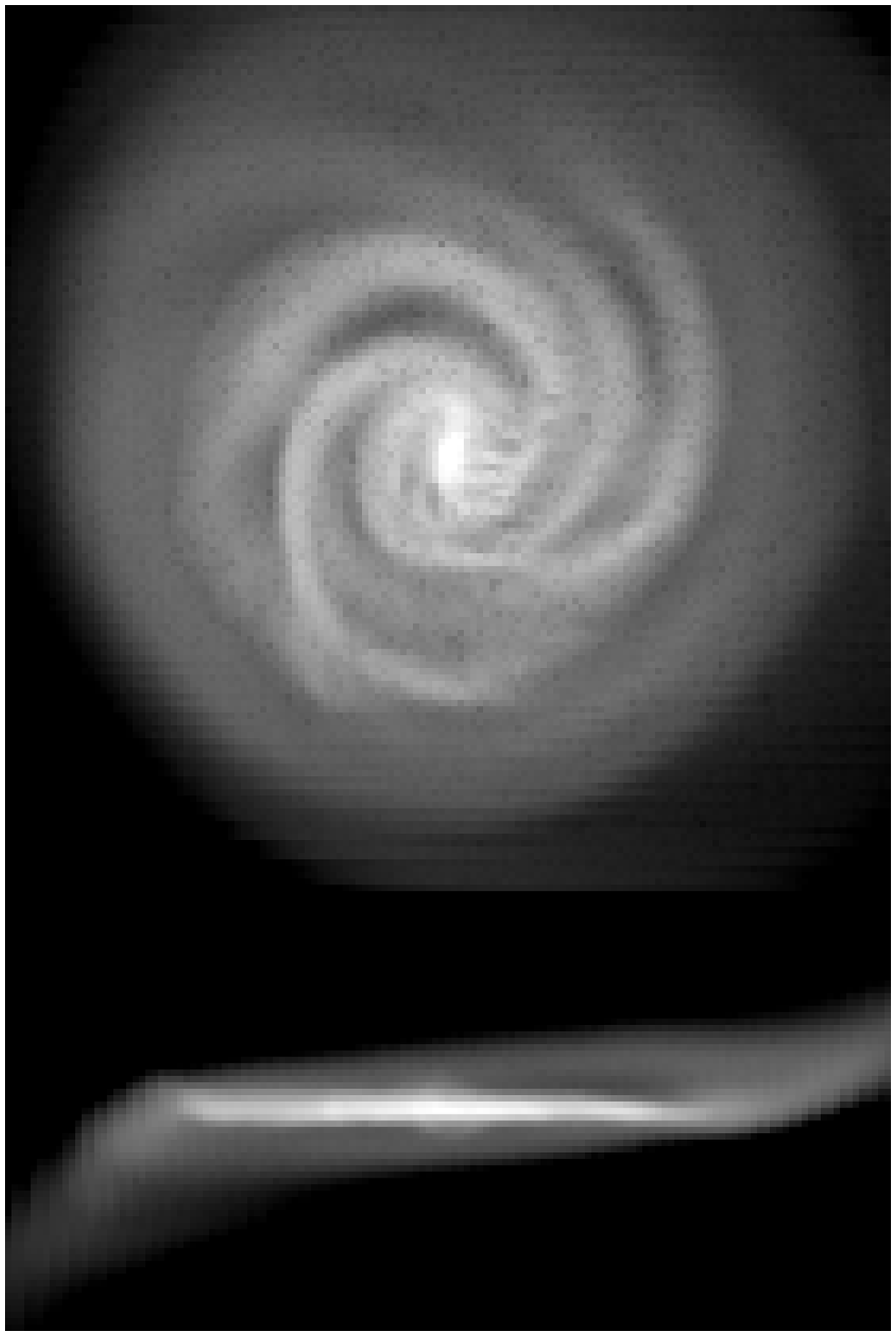,width=82pt} 
\psfig{file=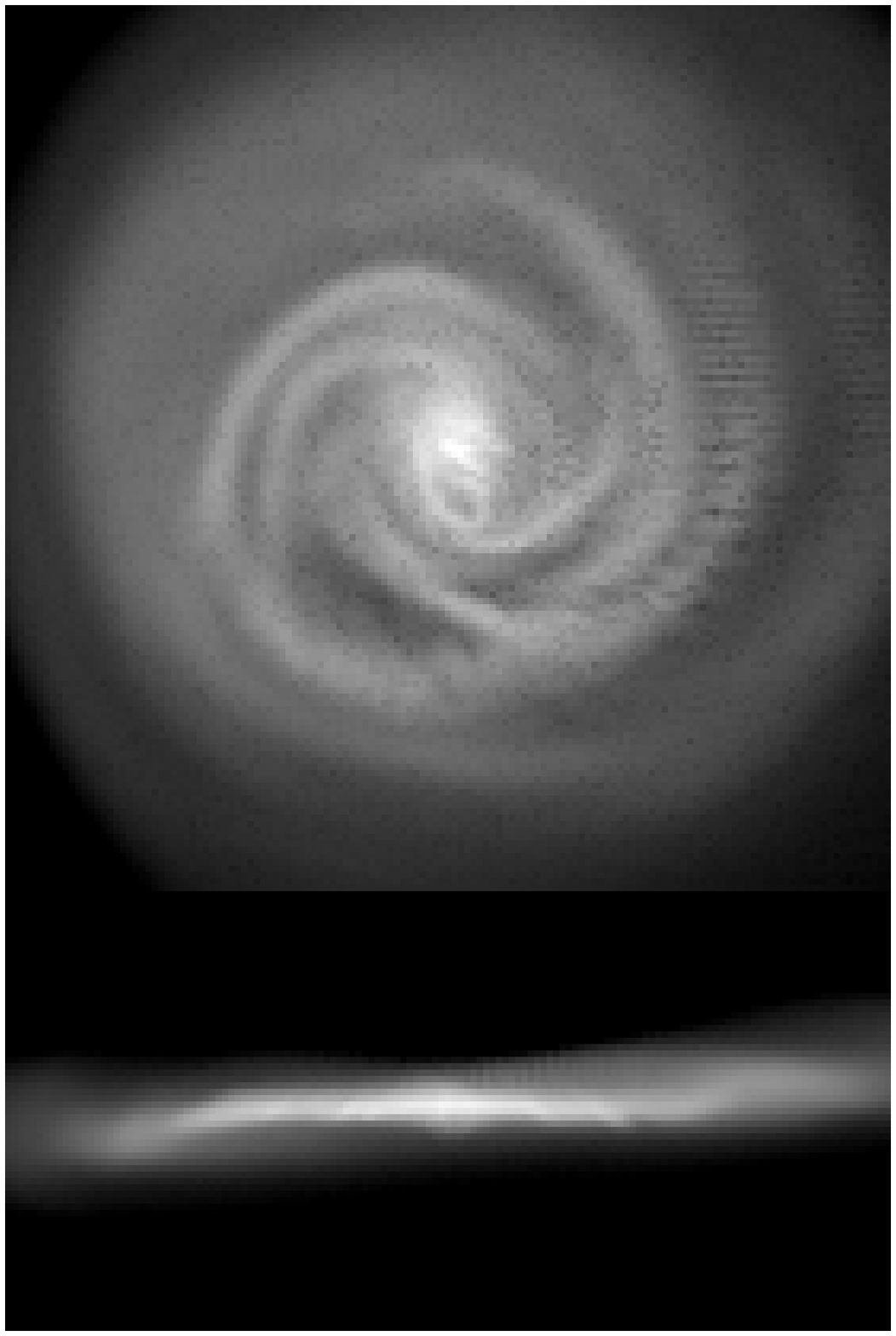,width=82pt}
\psfig{file=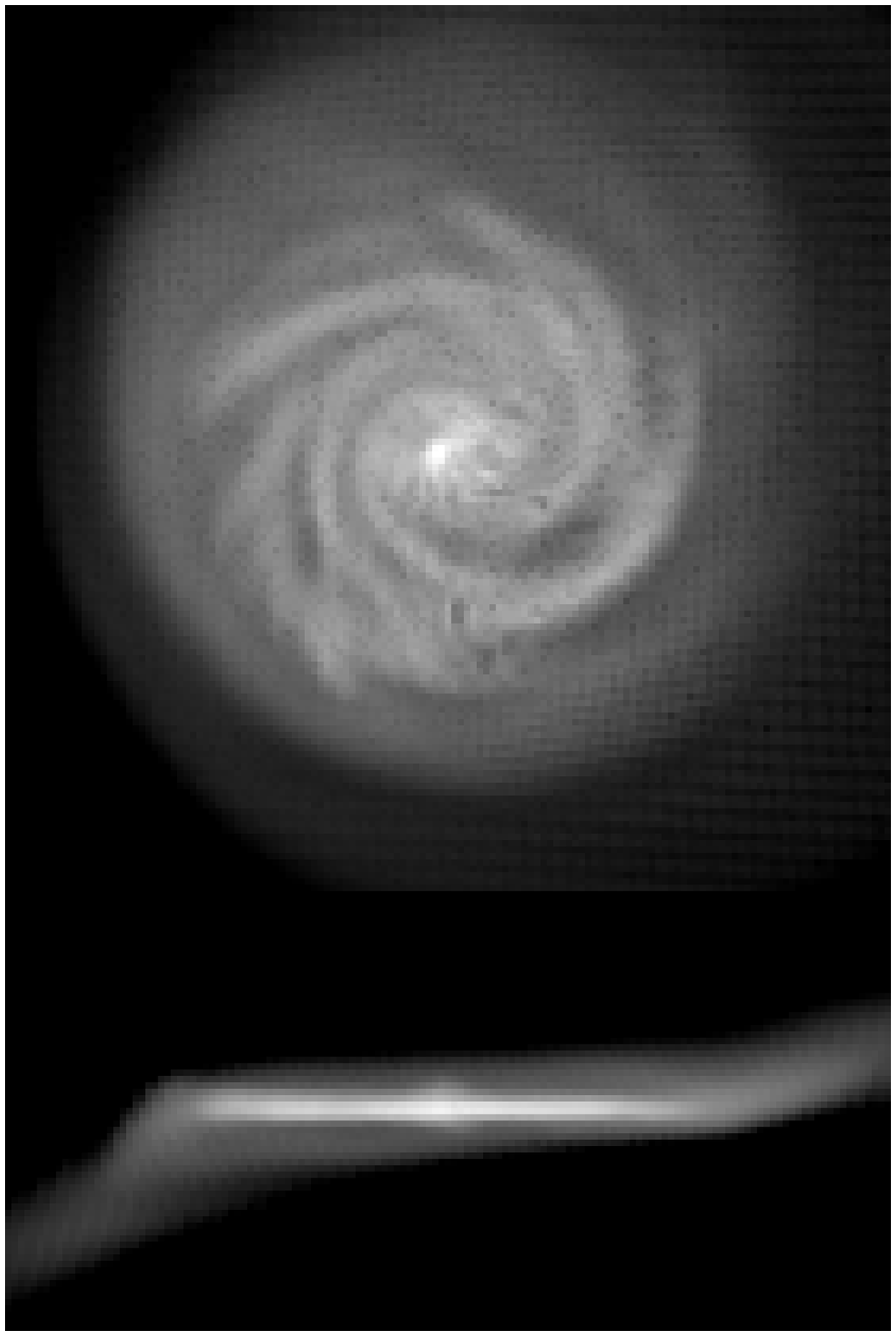,width=82pt} 
\psfig{file=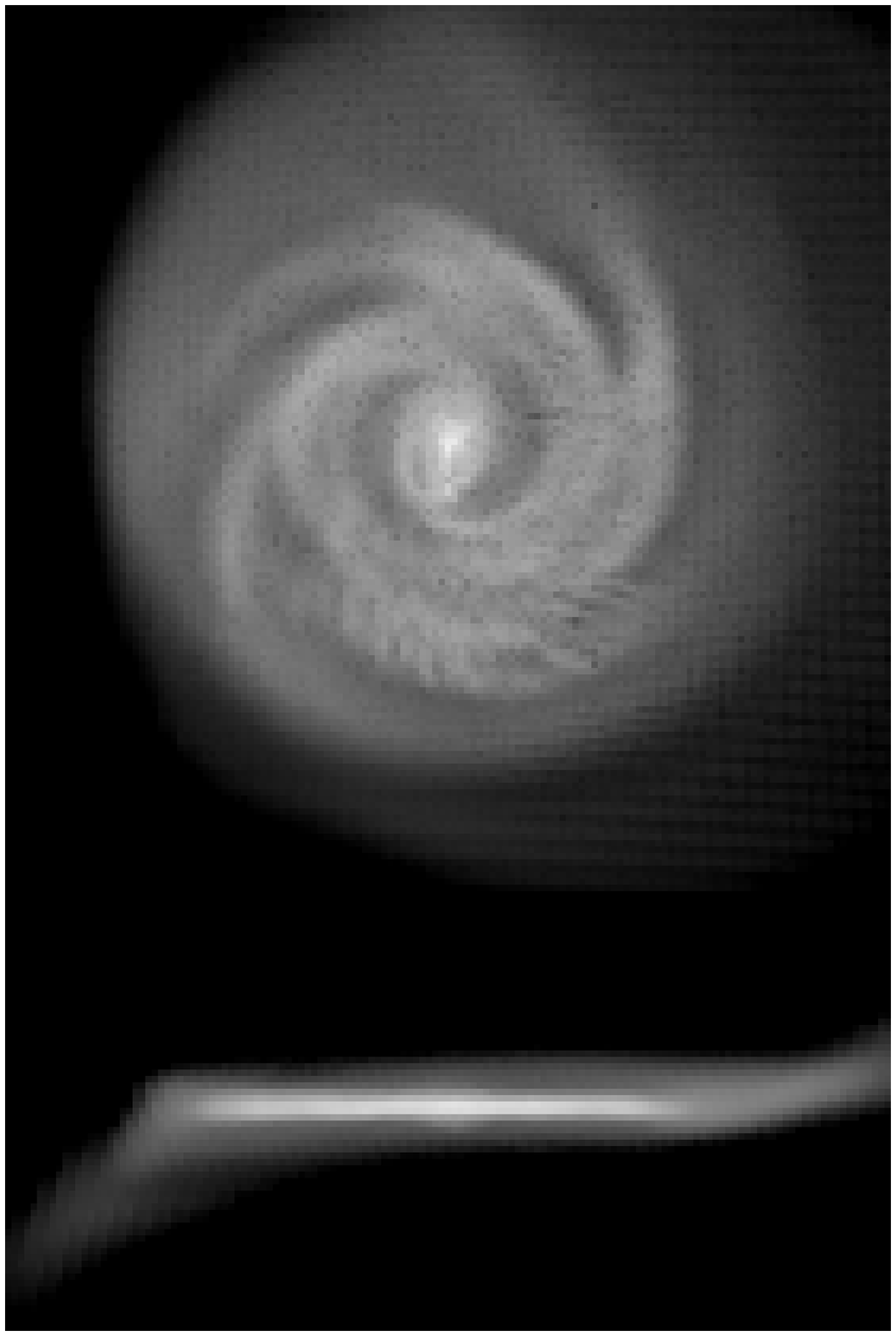,width=82pt} 
\psfig{file=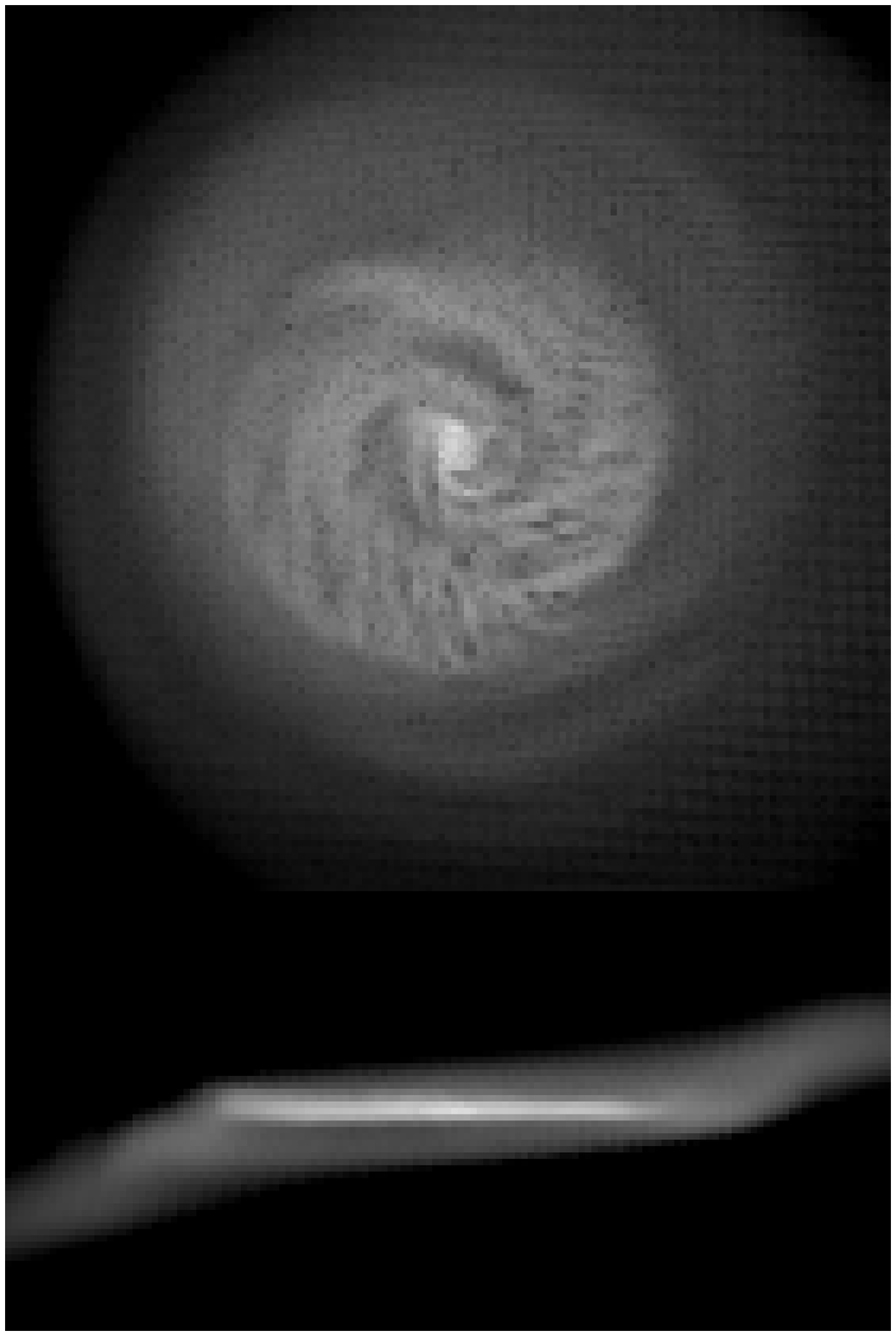,width=82pt} 
\psfig{file=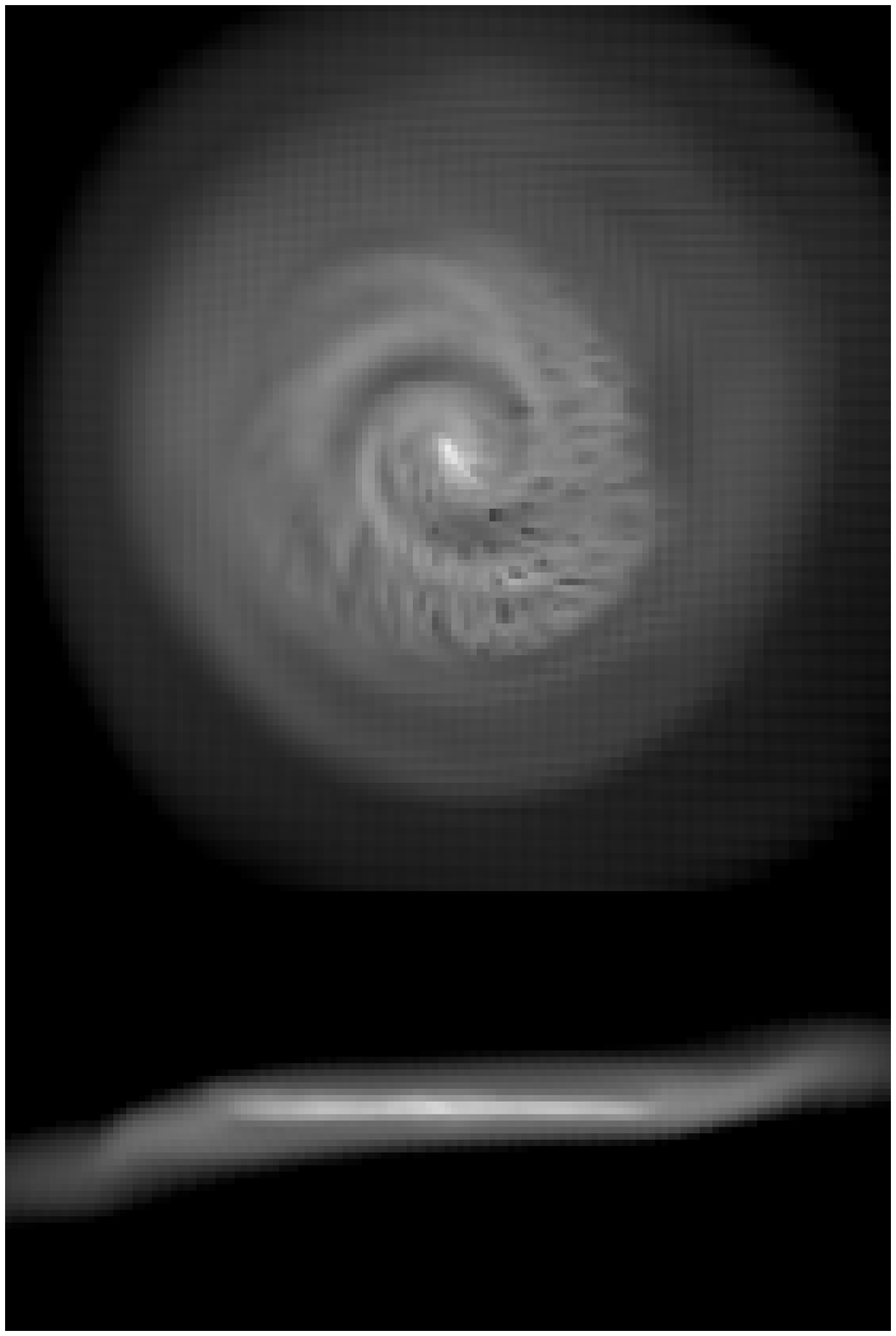,width=82pt}
\end{tabular}
\caption[]{Projected face-on and edge-on surface density maps of the stars (top) and gas (bottom) of the $z=0$ discs, where each panel is 60 kpc across. As the SFE is lowered and mass-loss is employed, spiral structure becomes more pronounced due to a less massive bulge. The Hubble type of the disc changes from an early type (S0 or Sa) disc, to a late-type spiral galaxy (Sb or Sbc) as we decrease $\epsilon_{\rm ff}$.}
\label{fig:maps}
\end{figure*}

\begin{figure*}
\center
\begin{tabular}{cccc}
\psfig{file=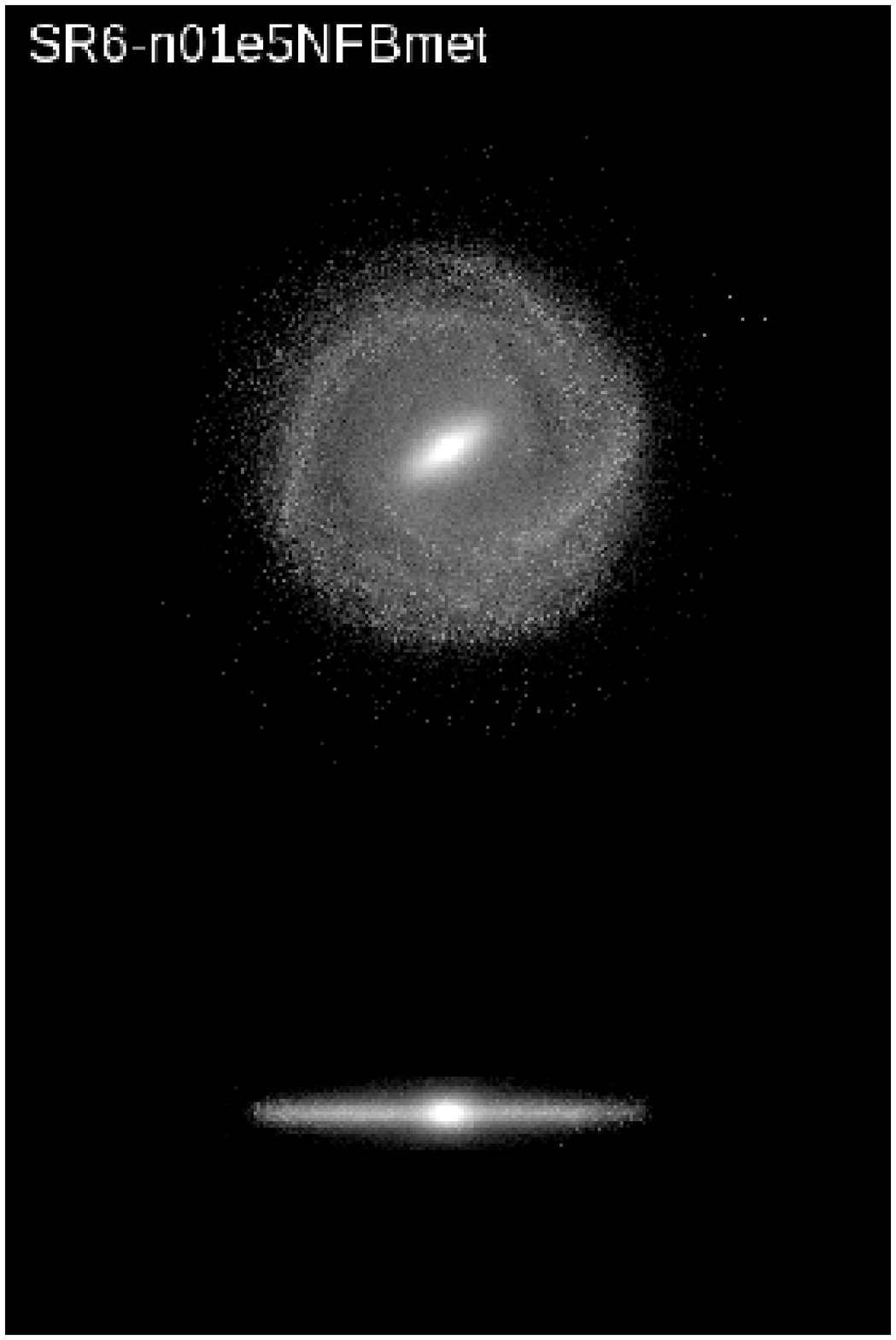,width=82pt} 
\psfig{file=n01e5star.ps,width=82pt} 
\psfig{file=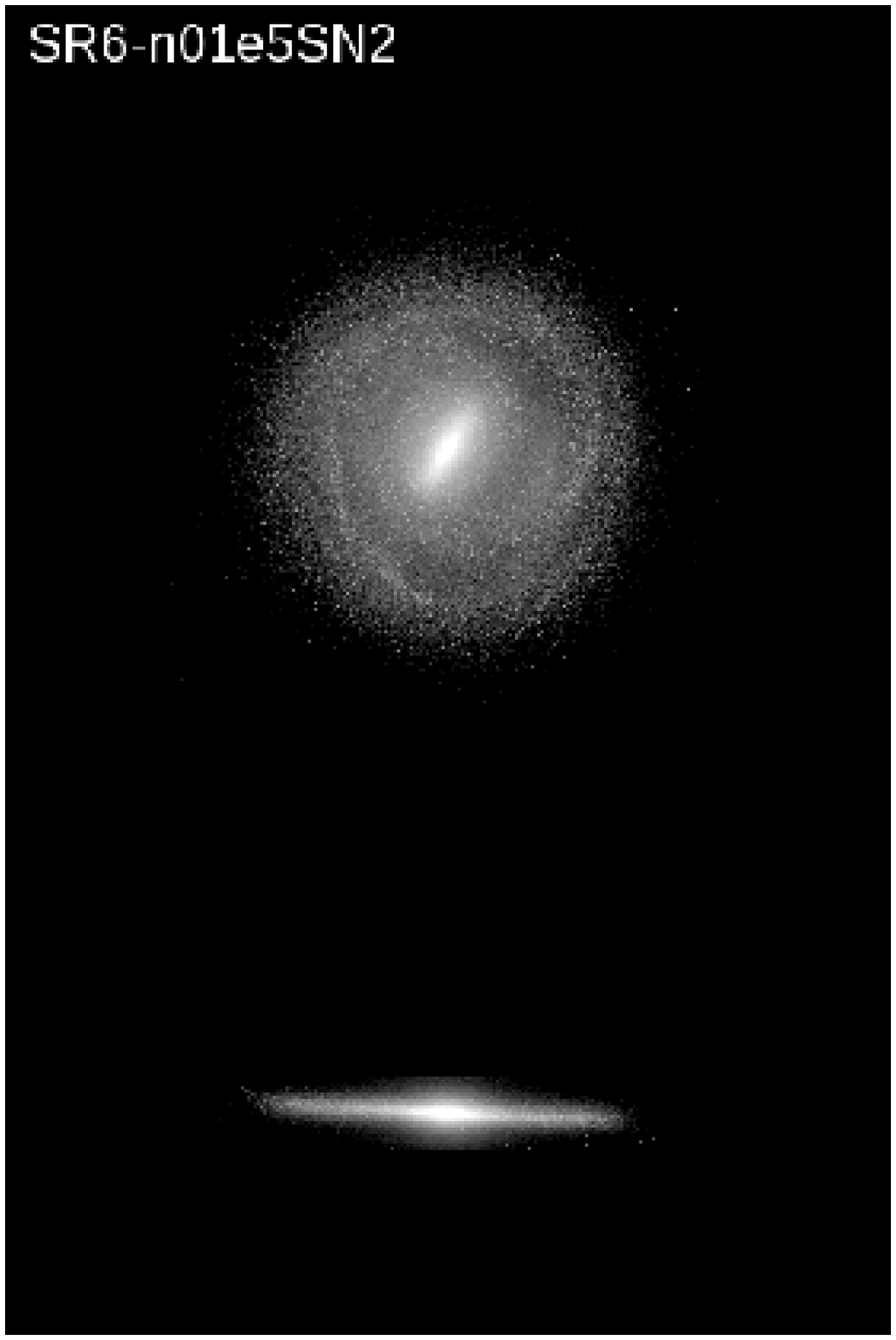,width=82pt} 
\psfig{file=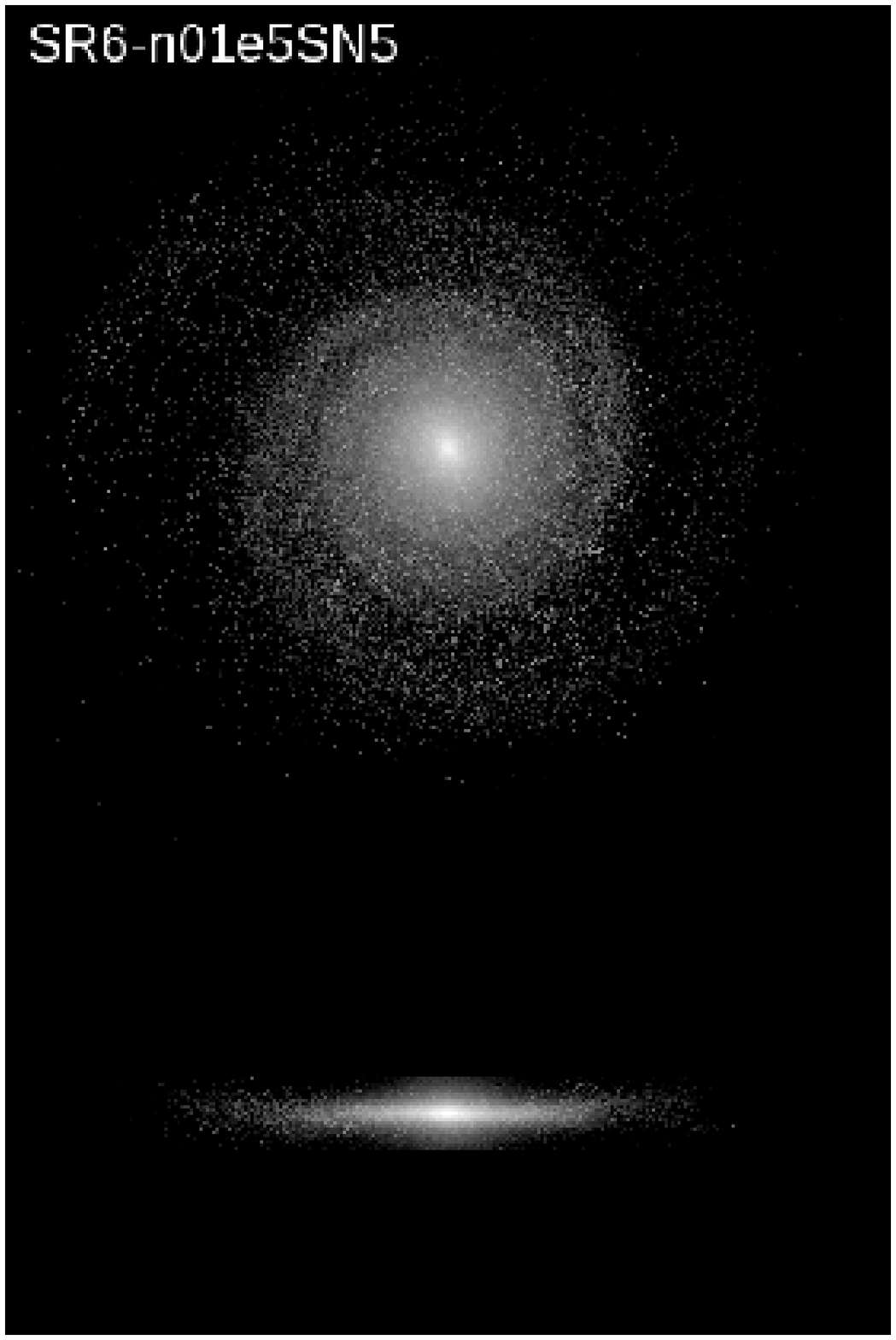,width=82pt} \\
\psfig{file=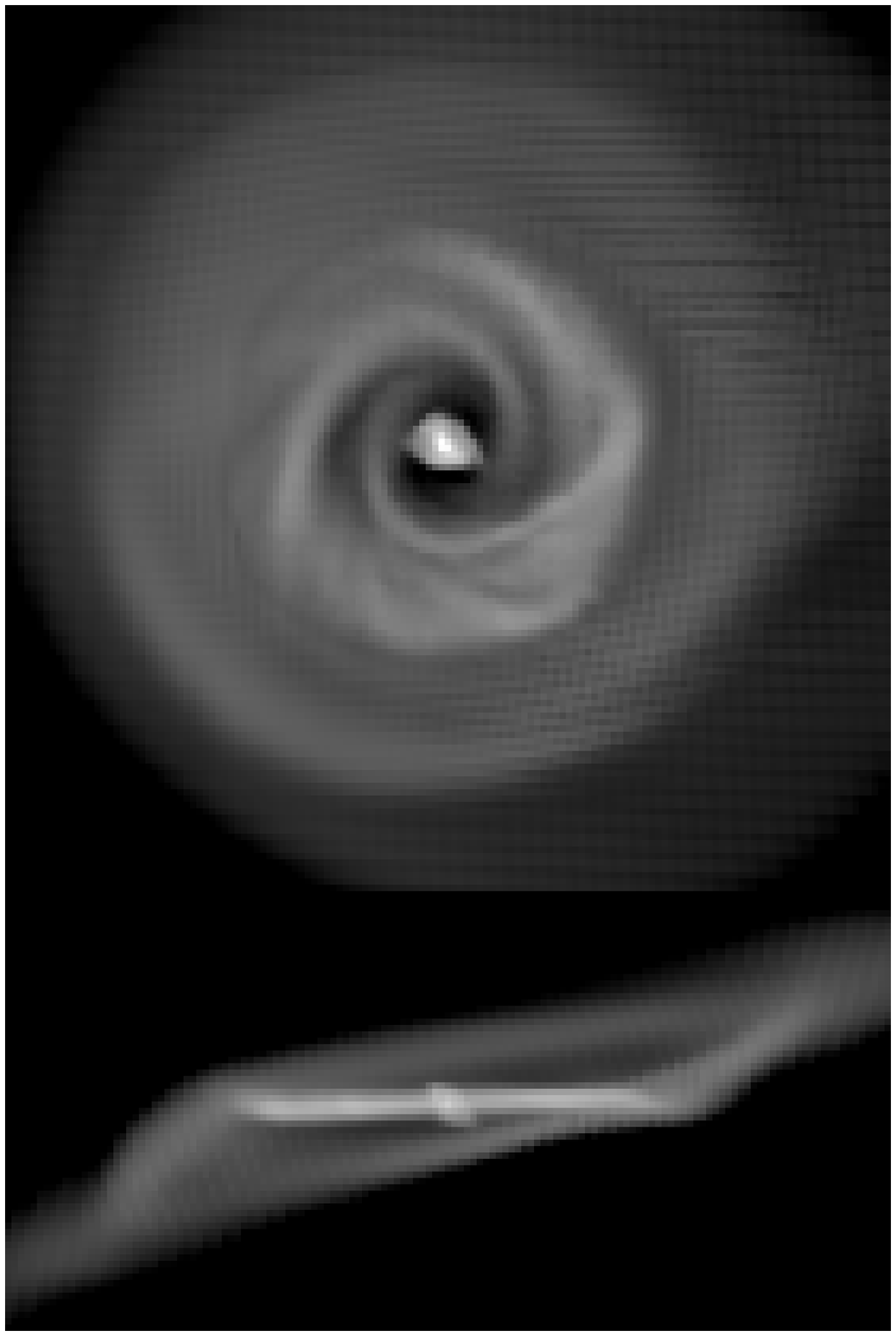,width=82pt}
\psfig{file=n01e5gas.ps,width=82pt}
\psfig{file=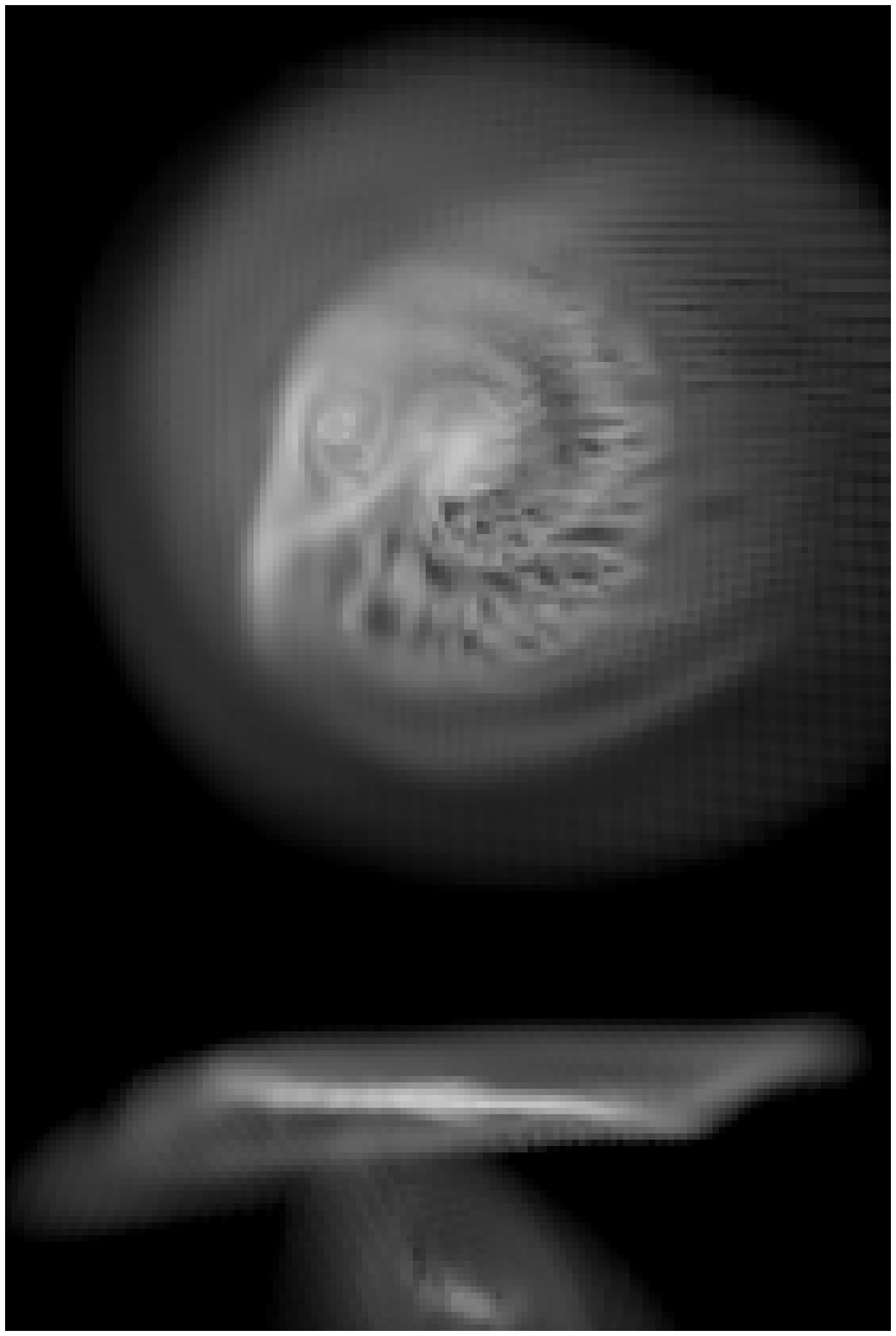,width=82pt}
\psfig{file=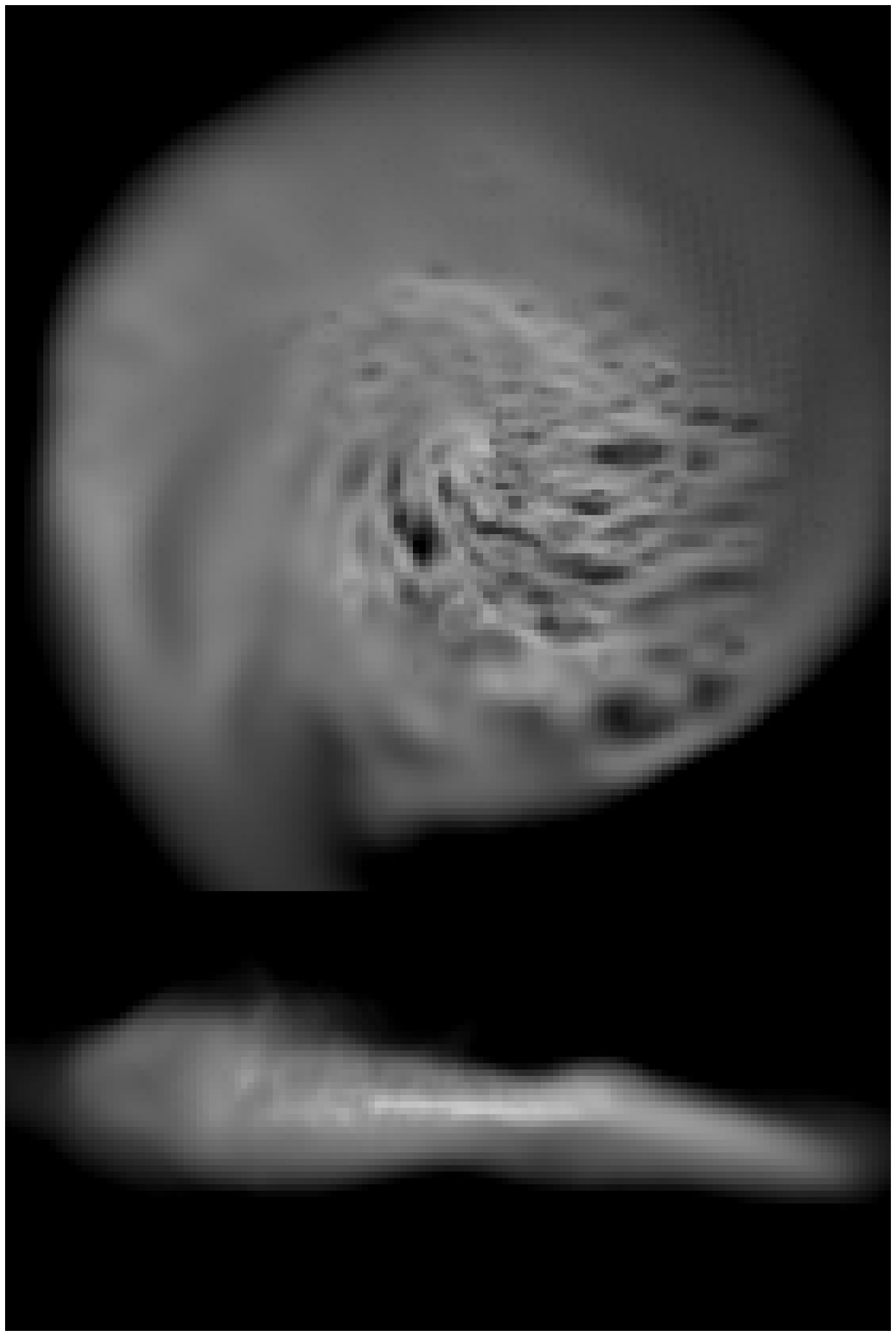,width=82pt}
\end{tabular}
\caption[]{Projected face-on and edge-on surface density maps of the stars (top) and gas (bottom) of the $z=0$ discs where each panel is 60 kpc across. From left to right, the feedback strength is $E_{\rm SNII}=0,1,2$ and $5\times10^{51}\,{\rm erg}$. As the injected amount of energy is increased, the gas component becomes more distorted.}
\label{fig:mapsFB}
\end{figure*}

The goal of this work is to study how the characteristics of disc galaxies change when standard numerical parameters governing star formation are modified. Following the discussion in Section\,\ref{sect:intro}, we consider star formation regulation in two different ways: small scale ($\sim100\pc$) physics such as H$_2$ abundance, GMC turbulence, metallicity, radiative effects etc., or via energy injection from supernovae explosions leading to gas expulsion in galactic winds. The first mechanism is modelled by varying the Schmidt-law (Eq.\,\ref{eq:SF}) SFE, $\epsilon_{\rm ff}$, which acts on a cell-by-cell basis. The latter is studied by increasing the injected SNII energy, $E_{\rm SNII}$. In addition we study the impact of the star formation threshold $n_0$, but in lesser detail. 

The traditional way of treating star formation in simulations of galaxy formation \citep[e.g.][]{Governato07,Piontek09b} is to tune the SFE parameter using an isolated disc model to match the observed \emph{K-S} relation, most commonly the fitting formula given by \cite{kennicutt98} of $z=0$ galaxies. In addition, the recipe for energy injection via supernovae and its efficiency is tuned simultaneously. These parameters are then used in fully cosmological simulation of galaxy formation. This scheme assumes that supernova explosions are the main sources of star formation regulation at high redshift. As argued in Section\,\ref{sect:intro}, the numerically assumed constant efficiency is strongly redshift dependent and a $z=0$ tuning is likely to over-predict star formation in more metal poor environment at higher redshift. We treat $\epsilon_{\rm ff}$ as a free, but constant, parameter and adopt $\epsilon_{\rm ff}=1,2$ or 5 per cent in the fully cosmological context. These values are in agreement with GMC estimates from \cite{krumholztan07}. As we will demonstrate below, lower values than what traditionally is adopted is preferred in order to form late-type galaxies. Note that $\epsilon_{\rm ff}\approx 2c_*$, where $c_*$ is the efficiency parameter used in e.g. \cite{Governato07} and \cite{Scannapieco09} (defined via $t_{\rm dyn}=1/\sqrt{4\pi G\rho}$). Values of $c_*=0.05-0.1$ are commonly employed i.e. a few times, up to an order of magnitude larger than what we consider here.

The standard SNII feedback described in Section\,\ref{sect:SNII} is the baseline feedback in all of our simulations. In a subset of simulations we add the additional stellar mass-loss and SNIa treatment. The high-efficiency simulation ($\epsilon_{\rm ff}=5\,$ per cent) is used as a template for the impact on the injected feedback SNII energy, which we set to $E_{\rm SNII}=1,2$ and $5\times10^{51}\,{\rm erg}$. Using energies that are several times larger than the canonical $10^{51}\,{\rm erg}$ might be perceived as unrealistic, but we believe it is illustrative to study the extreme cases of this type of feedback. In addition, the amount of SNII energy dissipated in cooling, after the shut-off time has passed, is complicated to measure. As a control set we also run the simulations without feedback, both with and without metal enrichment. 

The philosophy of the star formation threshold is as follows. In reality stars form in molecular clouds of average densities of $n>10^2\,{\rm cm}^{-3}$. Imposing a threshold of this magnitude would require a resolution on the order of parsecs to resolve the formation of the star-forming clouds, something that is beyond the scope of fully cosmological hydro+$N$-body simulations today \citep[but see][]{Gnedin09}. We adopt $n_0=0.1$ and $1\,{\rm cm}^{-3}$ for each setting of $\epsilon_{\rm ff}$, but the appropriate choice is fundamentally tied to resolution and can lead to spurious results. The ISM has been shown to be represented by a lognormal density probability distribution function (PDF) \citep[e.g.][]{Kravtsov2003,wada07}, or even a superposition of several lognormally distributed ISM phases \citep{RobertsonKravtsov08}. The amount of gas eligible for star formation is represented by the high density part of the PDF which in turn is a function of total disc gas mass and turbulence. A density threshold should be picked to allow for the high-density star-forming part of the PDF to be well resolved or at least contains, given an adopted numerical resolution, a converged amount of star-forming mass. If not, then chosen threshold will affect the numerical efficiency for global star formation. We will demonstrate this effect below.

In summary, the varied constants of interest is here the star formation threshold ($n_0$), the star formation efficiency per free-fall time ($\epsilon_{\rm ff}$) and the form of supernova feedback and injected energy ($E_{\rm SNII}$). Our main focus is the impact of these parameters at the SR6 level of resolution, and we present a brief resolution study in Appendix\,\ref{appendix:resolution}. We summarize our complete test suite in Table \ref{table:simsummary1}.

\section{The discs}
\begin{figure}
\center
\psfig{file=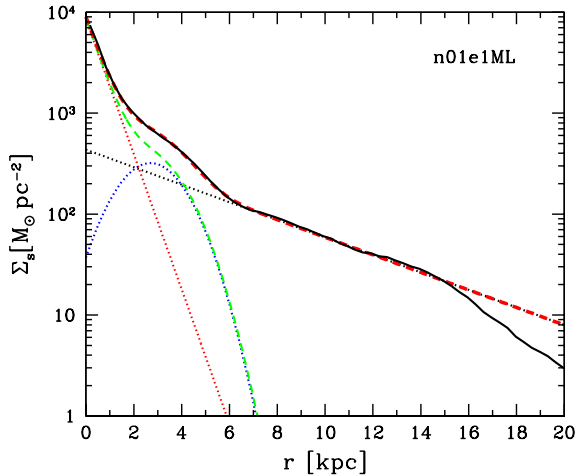,width=230pt} 
\caption[]{Example of a multi-component fit (red dashed line) to the stellar surface density (black solid line) of the $z=0$ disc in n01e1ML. The fit is composed of a bulge (red dotted line), a bar (blue dotted line) and a disc (black dotted line) component. The green dashed line shows the bulge and bar surface density contribution.}
\label{fig:fit}
\end{figure}

In this work, we focus primarily on the disc properties in the SR6 simulations at $z=0$. Details of the satellite galaxies and halo properties will be considered in a future work. Figs\,\ref{fig:maps} and \ref{fig:mapsFB} show projected face-on and edge-on stellar and gas density maps at $z=0$ for the galaxies in the star formation efficiency and feedback test suite, respectively. The discs show a wide range of spiral galaxy morphologies, and will return to this point in Section\,\ref{sect:hubble}. We decompose the resulting stellar discs into a bulge, bar and disc component and fit these simultaneously to the stellar surface density profile. The latter is calculated using all stars out to a height of $|z|=2.5\,\kpc$. For the bulge and disc component we assume exponential profiles, i.e.
\begin{equation}
\label{eq:discbulge}
\Sigma(r)=\Sigma_0\exp(-r/r_{\rm d}),
\end{equation}
where we fit for $\Sigma_0$ and the scale radius $r_{\rm d}$. The bar component is modelled using a simple Gaussian, 
\begin{equation}
\label{eq:bar}
\Sigma_{\rm bar}(r)=\frac{A_0}{\sigma}\exp\left(-\frac{(r-r_0)^2}{2\sigma^2}\right),
\end{equation} 
where we fit for the width $\sigma$, the central point $r_0$ and amplitude $A_0$. We consider this a conservative estimate of the bar mass as a Gaussian contribution falls off towards the centre of the disc, leaving more mass to be accounted for by the bulge. An example of the fitting procedure can be seen in Fig.\,\ref{fig:fit}. The necessity of a separate bar component is here clearly illustrated. In the more bulge-dominated cases the bar amplitude is decreased considerably, owing to the weaker disc self-gravity. 

The bulge mass, $M_{\rm bulge}=2\pi\Sigma_{\rm bulge}r_{\rm bulge}^2$, is obtained by integrating Eq.\,\ref{eq:discbulge}. A similar relation holds for the disc, where we also include the stellar disc mass past the break radius in the quoted disc stellar mass, $M_{\rm disc,s}$, but we only use the data within the break when fitting (see Fig.\,\ref{fig:fit}). The bar mass is simply the integrated mass from Eq.\,\ref{eq:bar}, and we consider the bar as a part of the disc component and include it in the quoted $M_{\rm disc,s}$. Doing this or not modifies the disc mass only slightly, especially in bulge-dominated galaxies. The gas is treated as a single component, and we simply consider the mass within $r=15\kpc$ and $|z|=2.5\,\kpc$ as the gaseous disc mass, $M_{\rm disc,g}$. We consider only the stars when calculating the bulge-to-disc (B/D) and bulge-to-total (B/T) ratios. All measured and derived quantities are summarized in Table \ref{table:simsummary2}. We note that this method of defining galactic components in simulations, as well as others e.g. via angular momentum \citep{Okamoto05,Scannapieco09}, carry uncertainties.

Our simulated discs span a large range of characteristics: stellar disc masses are in the range $M_{\rm disc,s}=5-9\times10^{10}\,\Msol$, bulge masses of $M_{\rm bulge,s}=2-7\times10^{10}\,\Msol$, B/D $\sim0.23-1.2$ and gas fractions $f_{\rm g}=0.05-0.28$. The scale radii of the discs, $r_{\rm d}$, vary from typically $4-5\,\kpc$ to $>10\,\kpc$ in the bulge-dominated systems. As we demonstrate below, extended disc galaxies of Sb, or even Sbc type, form only when star formation is numerically resolved in the whole disc ($n_0=0.1\,{\rm cm}^{-3}$), and a low efficiency of $\epsilon_{\rm ff}\sim 1$ per cent (or \emph{very} strong feedback) is adopted. At larger efficiencies we observe how the B/D ratios increase, the discs are less extended, the rotational velocities peak at very large values and the spiral patterns become more tightly wound and less pronounced. This indicates a shift towards early type discs like Sa or even S0. 

\label{sect:discs}
\begin{figure*}
\center
\begin{tabular}{ccc}
\psfig{file=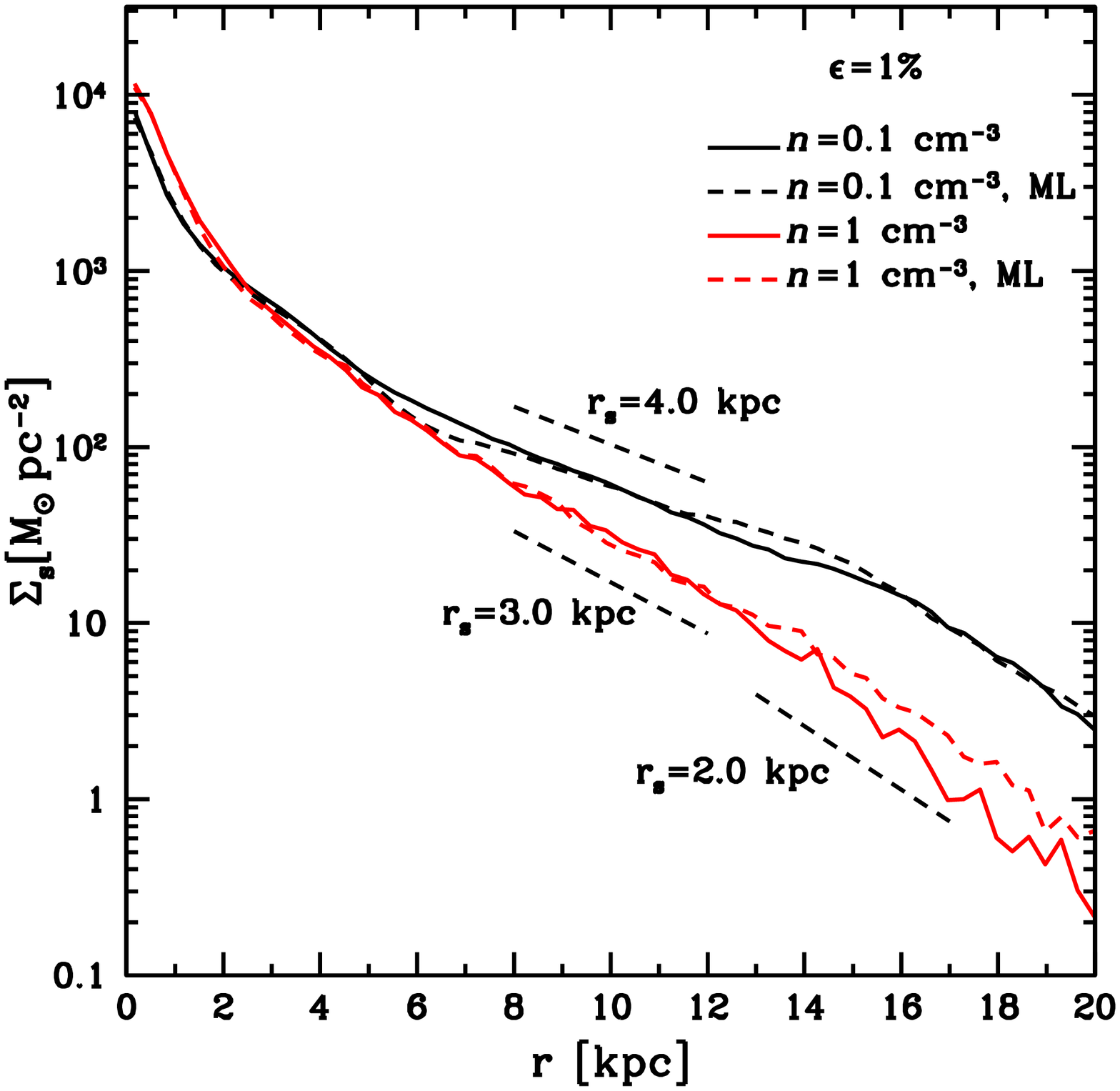,width=156pt} &
\psfig{file=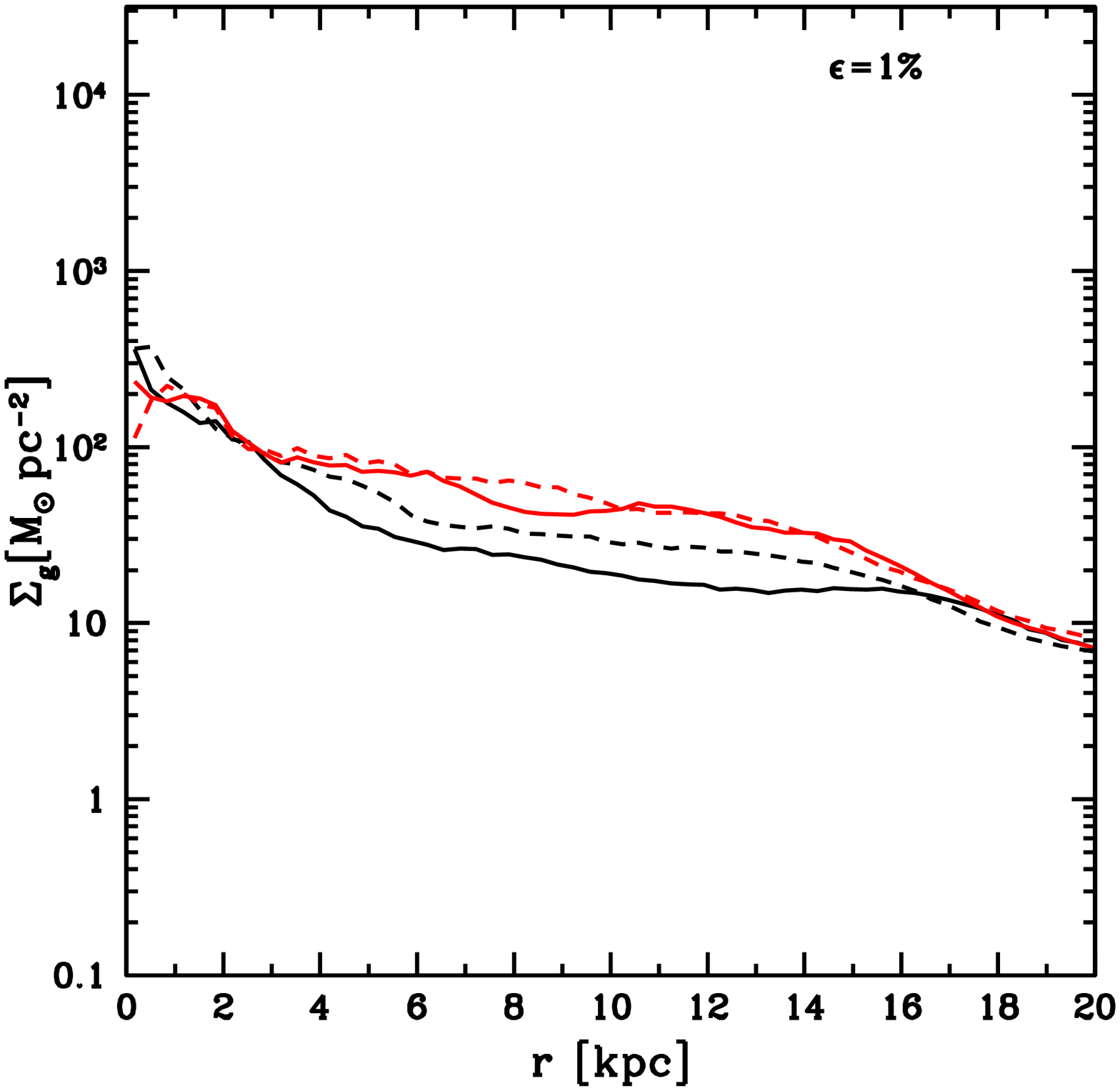,width=156pt} &
\psfig{file=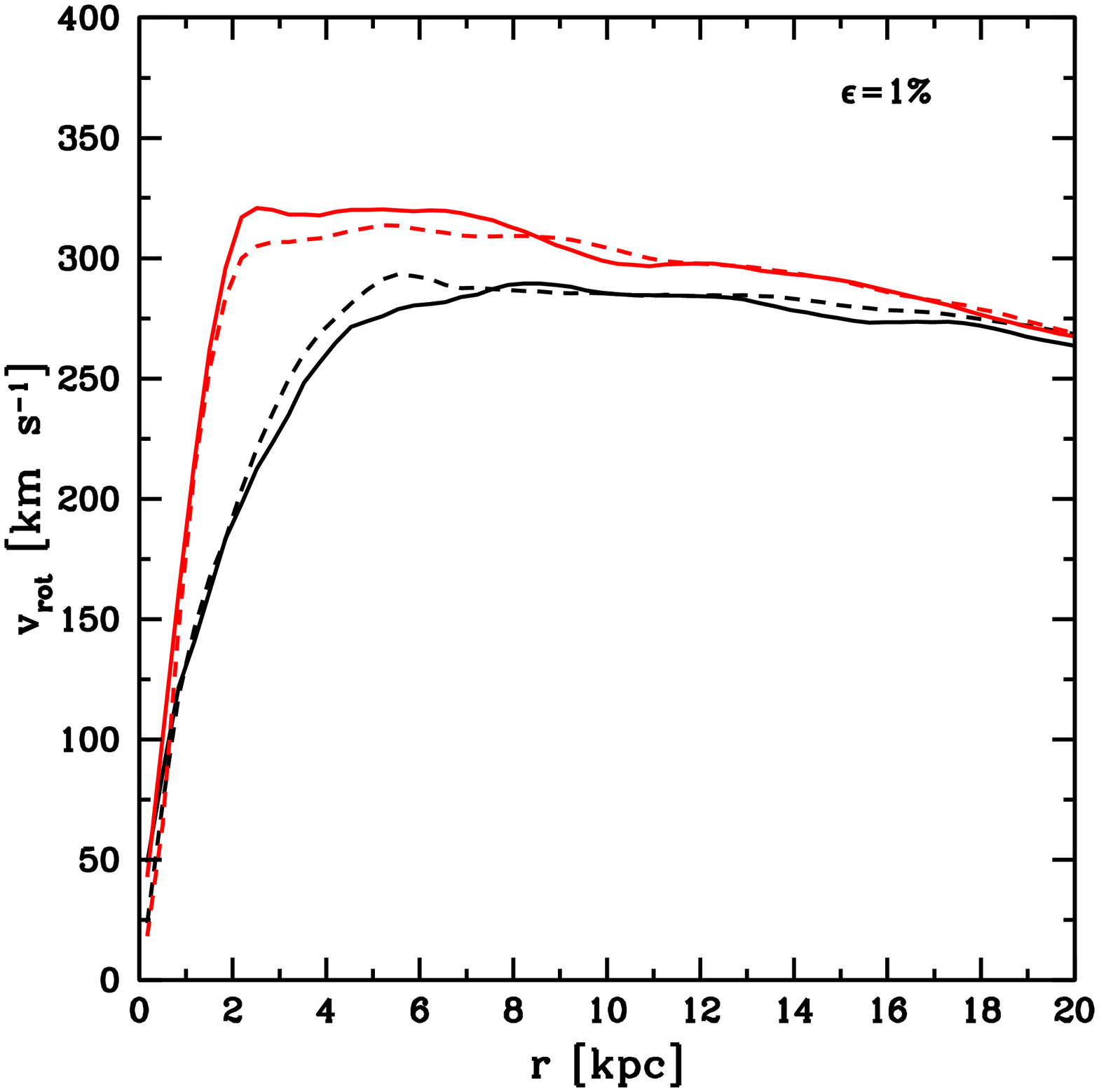,width=156pt} \\
\end{tabular}
\caption[]{The effect of the Schmidt-law star formation density threshold. The panels show the stellar surface density (left), gas surface density (middle) and rotational velocity measured from the gas (right). We consider all material within a height of $|z|<2.5\,\kpc$ for all components. The star formation efficiency is $\epsilon_{\rm ff}=1$ per cent in all simulations. The different colours are described in the first panel, and a dashed line indicates that we use the extended feedback model (see text).}
\label{fig:sigma}
\end{figure*}
\begin{figure*}
\center
\begin{tabular}{ccc}
\psfig{file=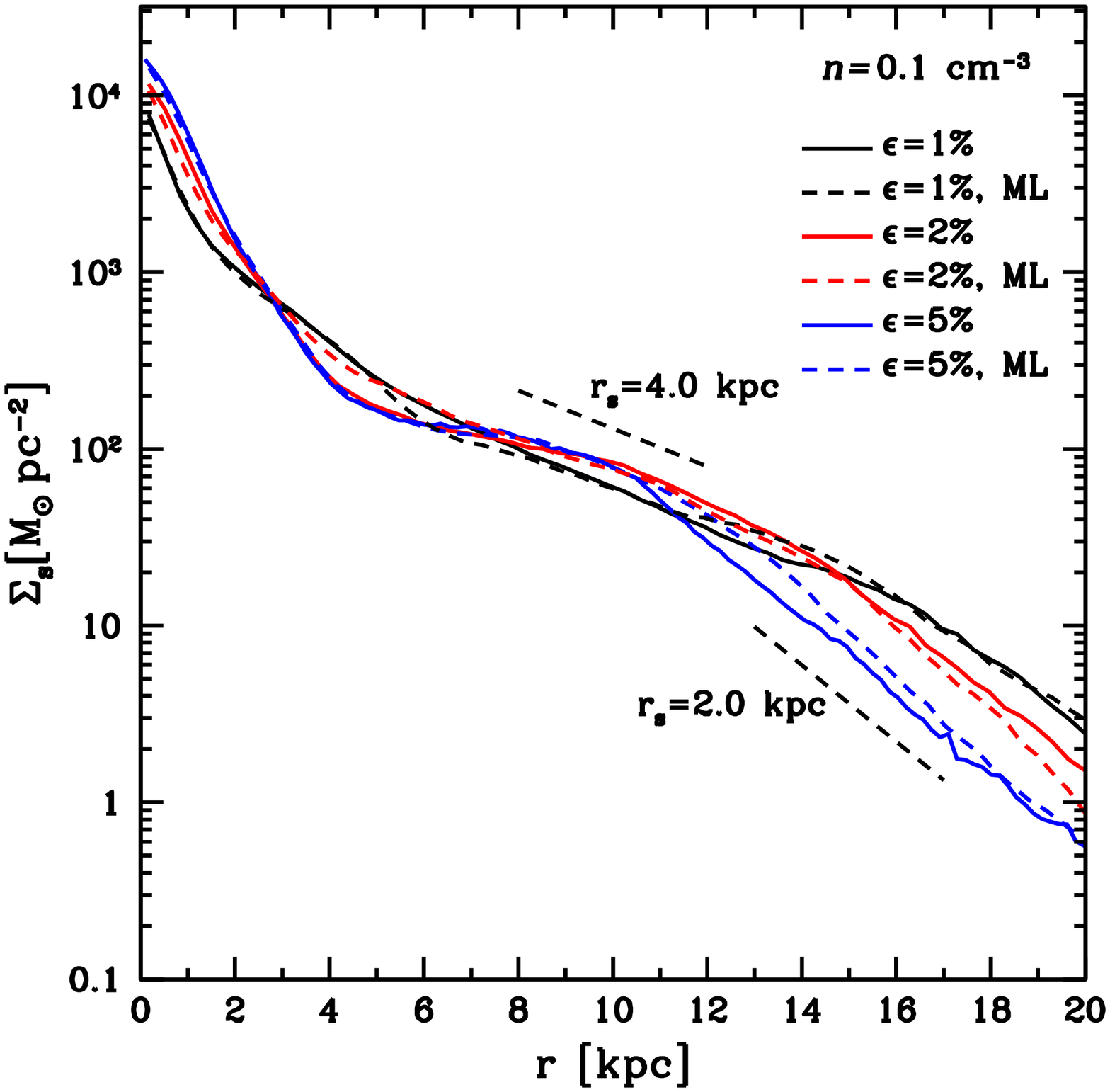,width=156pt} &
\psfig{file=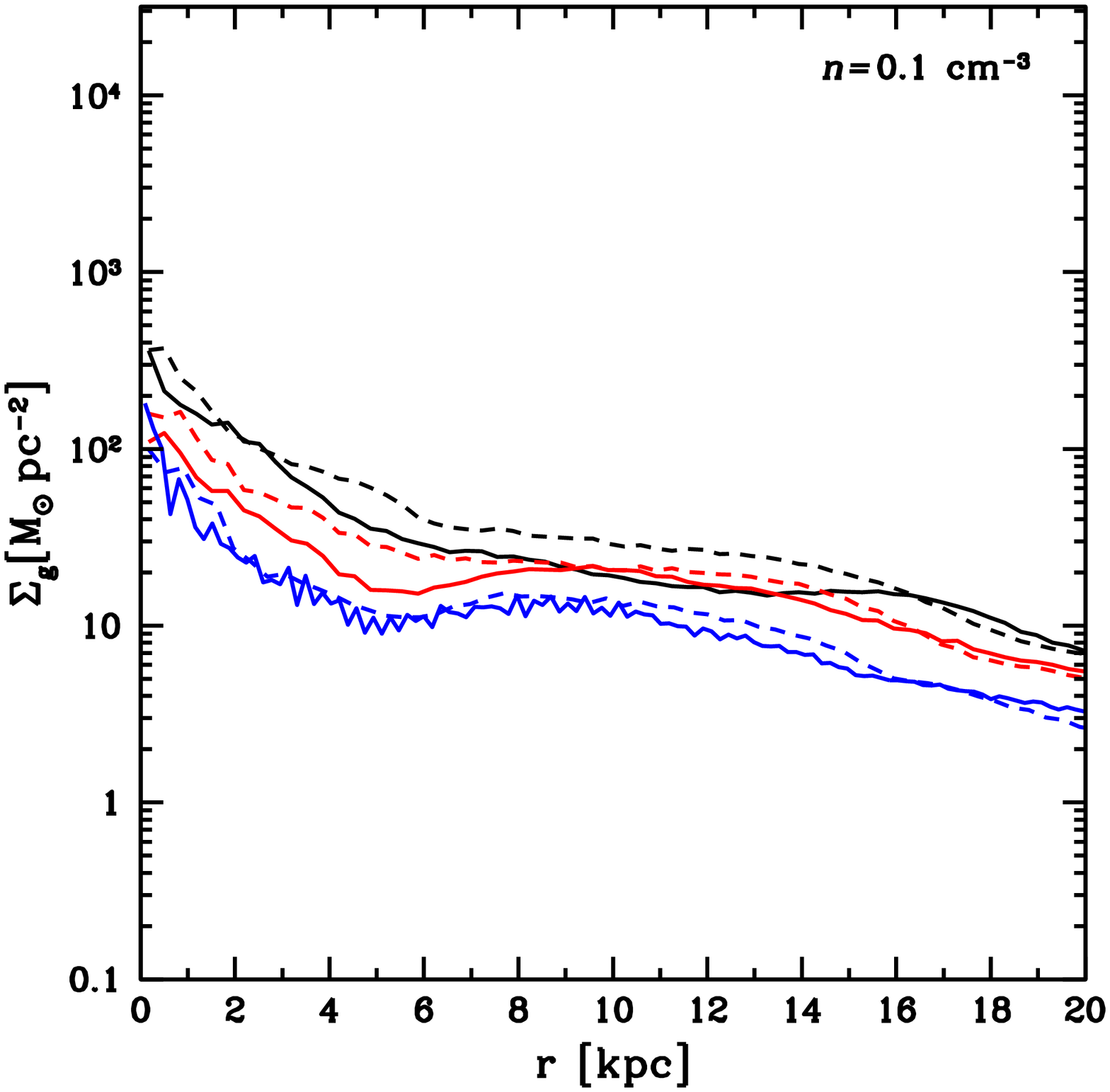,width=156pt} &
\psfig{file=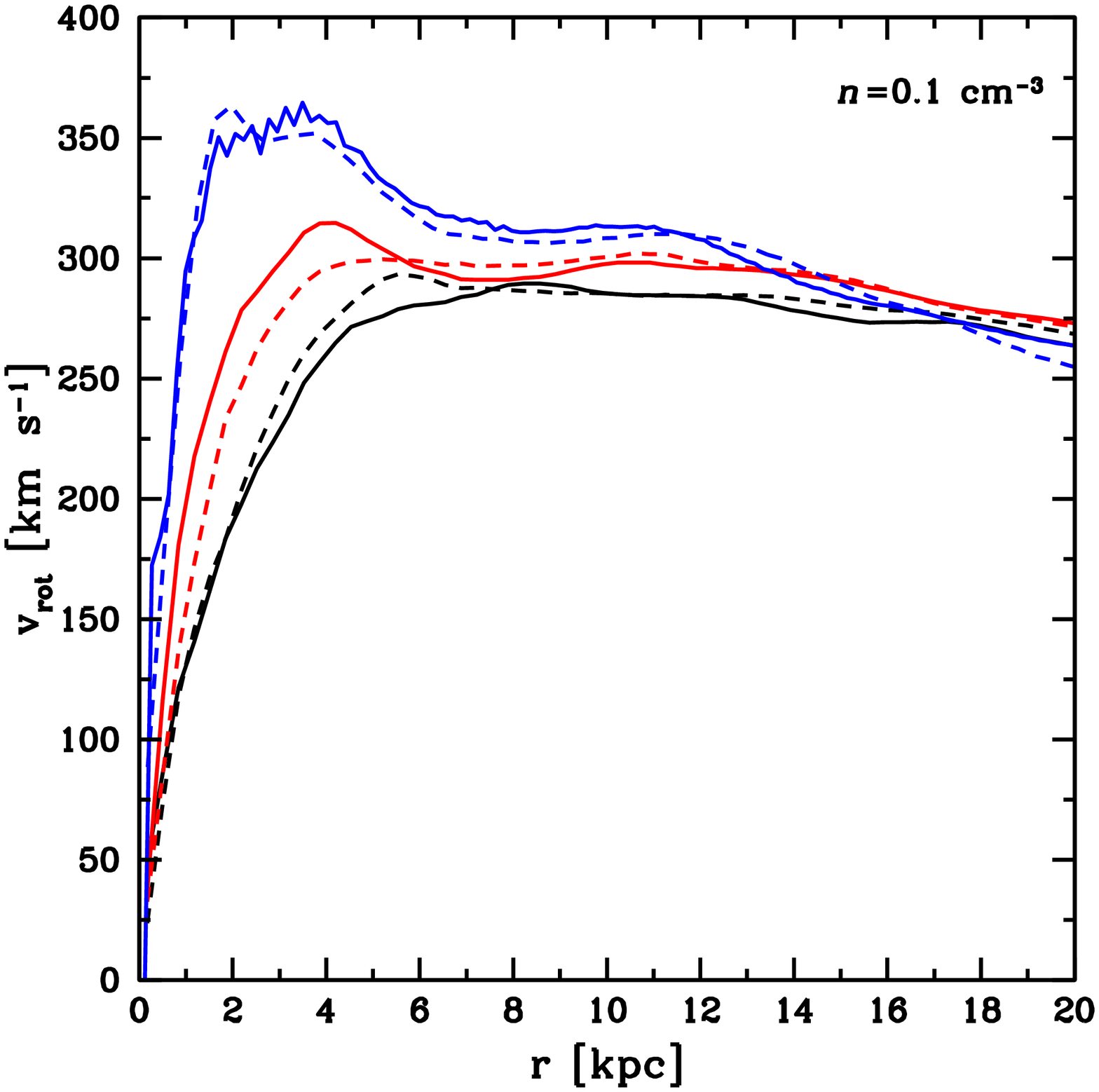,width=156pt} \\
\end{tabular}
\caption[]{The effect of the Schmidt-law star formation efficiency. The panels show the stellar surface density (left), gas surface density (middle) and rotational velocity measured from the gas (right). We consider all material within a height of $|z|<2.5\,\kpc$ for all components. The Schmidt-law density threshold is $n=0.1\,{\rm cm}^{-3}$ in all simulations. The different colours are described in the first row, and a dashed line indicates that we use the extended feedback model (see text).}
\label{fig:sigma2}
\end{figure*}
\section{Effect of star formation parameters}
\label{sect:effect}
In this section we study the influence of star formation parameters, i.e. in essence the small scale physics, on disc properties at $z=0$. The resulting stellar surface densities ($\Sigma_{\rm s}$), gas surface densities ($\Sigma_{\rm gas}$) and rotational velocities measured from the gas ($v_{\rm rot}$) are presented in Figs \ref{fig:sigma} and \ref{fig:sigma2} for the first 11 simulations in the SR6 suite at $z=0$ (see Table.\,\ref{table:simsummary1}). 

\subsection{The star formation density threshold, $n_0$}
\label{sect:thresh}
We start by focusing on the data presented in Fig.\,\ref{fig:sigma}. By keeping $\epsilon_{\rm ff}$ fixed to 1 per cent, while varying $n_0$, we observe a strong change in the ability to form stars at large radii. The galaxies adopting a large threshold have a more concentrated distribution of stars, smaller stellar disc scale-lengths as well as larger $\Sigma_{\rm gas}$ at all radii. The scalelengths are $r_{\rm d}>4,\kpc$ for $n_0=0.1\,{\rm cm}^{-3}$, but only $r_{\rm d}\sim2.5\kpc$ for $n_0=1\,{\rm cm}^{-3}$. The latter values are on the low side when compared to observations of late-type spirals at this mass range \citep{Courteau96,Courteau97,oleggnedin07}. The systematically lower $\Sigma_{\rm s}$ signals an under-resolved or missing' star formation throughout the disc: the average physical gas density does not efficiently cross the targeted $n_0$, even at intermediate radii. This is also reflected in the gas fractions of $\sim 25$ per cent, which is much larger than observed average values for galaxies of this size \citep{Garnett02,Zhang09}. The rotational velocities are large at smaller radii for $n_0=1\,{\rm cm}^{-3}$, regardless of choice of $\epsilon_{\rm ff}$. Naively one would expect this numerically induced star formation deficiency to alleviate the angular momentum loss at high redshift, hence forming a less concentrated galaxy. However, due to secular evolution in the disc, this is not the case: disc instabilities drive gaseous flows to the centre of the disc where, as the gas crosses the correct threshold, star formation can proceed. We conclude, that given our numerical resolution and simulated system, $n_0=0.1\,{\rm cm}^{-3}$ yields more realistic (average) disc galaxies when compared to observations. We discuss this numerical effect and its relationship to the adopted mesh resolution further in Appendix\,\ref{appendix:subgrid}. Note that this is not a fundamental result of galaxy formation, but serves only as tuning given our numerical resolution and is a basis for the subsequent tests. A discussed in Section\,\ref{sect:starform}, $n_0$ should be increased as the resolution is increased.

\subsection{The star formation efficiency, $\epsilon_{\rm ff}$}
\label{sect:stareff}
We now turn to the data presented in Fig.\,\ref{fig:sigma2}, where we keep the threshold fixed at $n_0=0.1\,{\rm cm}^{-3}$ and adopt $\epsilon_{\rm ff}=1, 2$ or 5 per cent. As $\epsilon_{\rm ff}$ is increased, $\Sigma_{\rm star}$ increases at small radii i.e. the bulge mass increases, singalling a lower disc angular momentum. The bulge to disc ratio increases from ${\rm B/D}=0.25$ to 1.25 as $\epsilon_{\rm ff}$ increases from 1 to 5 per cent. $\Sigma_{\rm gas}$ roughly follows a $1/r$-profile and the magnitude is lowered at all radii by approximately the relative change in efficiency. The stellar disc is less extended and the exponential scale-length increases for larger efficiencies (see Table.\,\ref{table:simsummary2}). For $\epsilon_{\rm ff}=1$ per cent, the disc scalelength is measured to be $r_{\rm d}\sim 4-5\kpc$, in good agreement with observed average values from the SDSS \citep{oleggnedin07}, while for $\epsilon_{\rm ff}=5$ per cent, $r_{\rm d}\approx 15\kpc$ is a $>2\sigma$ outlier. Large uncertainties exist for $r_{\rm d}>10\,\kpc$ as the stellar discs are small and feature almost flat stellar surface density profiles.

\begin{figure}
\center
\psfig{file=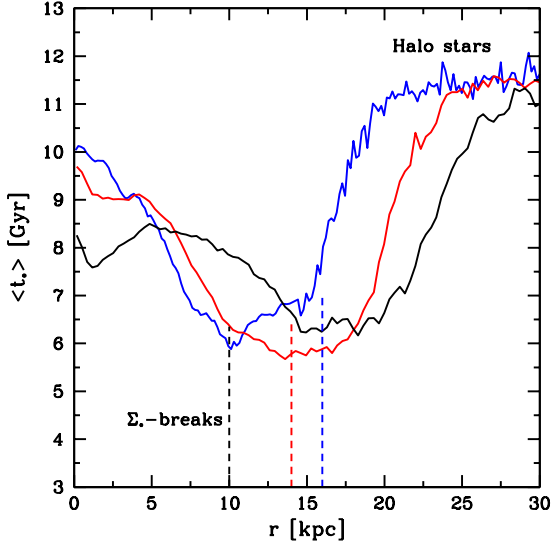,width=210pt}
\caption[]{Average stellar ages as a function of radius. The colours represent different star formation efficiencies, $\epsilon_{\rm ff}$= $1$ (black), $2$ (red) and $5$ per cent (blue). Dashed lines mark the measured stellar surface density breaks.}
\label{fig:break}
\end{figure}

At large radii in all simulations, $r_{\rm d}$ shifts to $\sim 2\kpc$. Disk breaking is a well observed phenomenon \citep{PohlenTrujillo06} correlated with a dip in the star formation rate, and has been studied numerically by \cite{Roskar08}. As larger star formation efficiencies lead to less extended discs, the disc breaks occurs at smaller radii: $r_{\rm break}\approx 16,14,10\,\kpc$ for $\epsilon_{\rm ff}=1,2$ and 5 per cent, respectively. The breaks can also be seen from the average stellar ages $\langle t_*\rangle$, shown in Fig.\,\ref{fig:break}. The central parts of the discs generally consist of older stars formed at $z>1$, and $\langle t_*\rangle$ decreases with larger radii, reaching $\langle t_*\rangle\approx\,6\,{\rm Gyr}$. Past the disc break, older stars appear which in part can be attributed to stellar migration as well as pollution by old halo stars with $\langle t_*\rangle\approx\,11-12\,{\rm Gyr}$. We note that $\langle t_*\rangle$ flattens or even declines towards the centre of the discs as $\epsilon_{\rm ff}$ is lowered. This is due to secular evolution: as spiral structure is more pronounced (as in n01e1), gas is transported towards the centre more efficiently and late-time star formation occurs; see Fig.\,\ref{fig:maps}.

The efficiency has a strong impact on the rotational velocity. The rotation curve in the n01e5 simulation features a strong peak in the inner parts of the disc. As $\epsilon_{\rm ff}$ is lowered, B/D decreases and the velocity profile flattens. Only when $\epsilon_{\rm ff}<2$ per cent can a flat rotational velocity profile be produced! In n01e1, the rotational velocity reaches $v_{\rm rot}\approx 275\,{\rm km\,s}^{-1}$ and stays roughly flat. For n01e2 and n01e5, $v_{\rm rot}$ peaks at $\approx 310$ and $360\,{\rm km\,s}^{-1}$, but converges at $275\,{\rm km\,s}^{-1}$ close to $r=20\,\kpc$. 

\begin{figure*}
\center
\begin{tabular}{cc}
\psfig{file=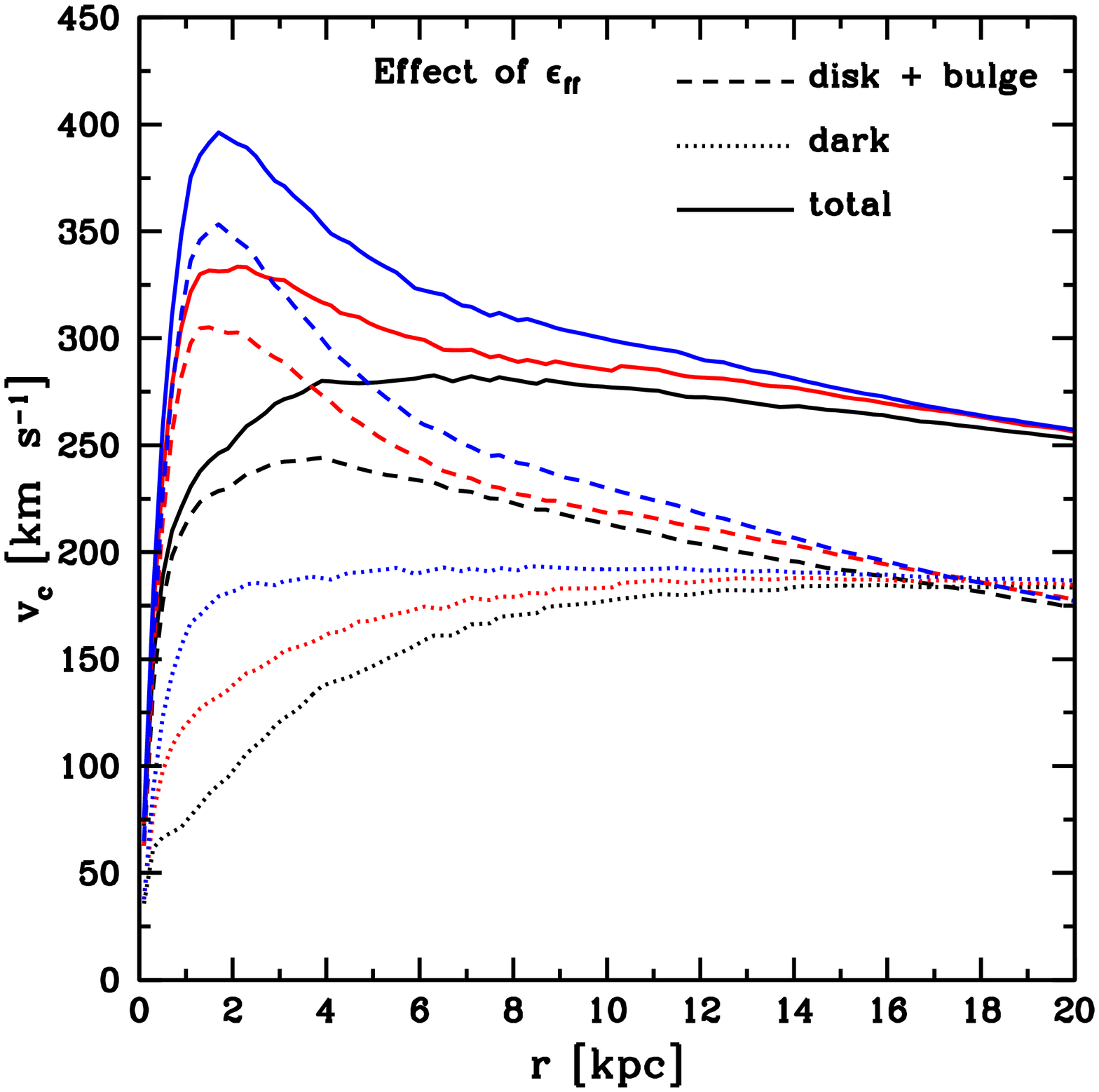,width=200pt} &
\psfig{file=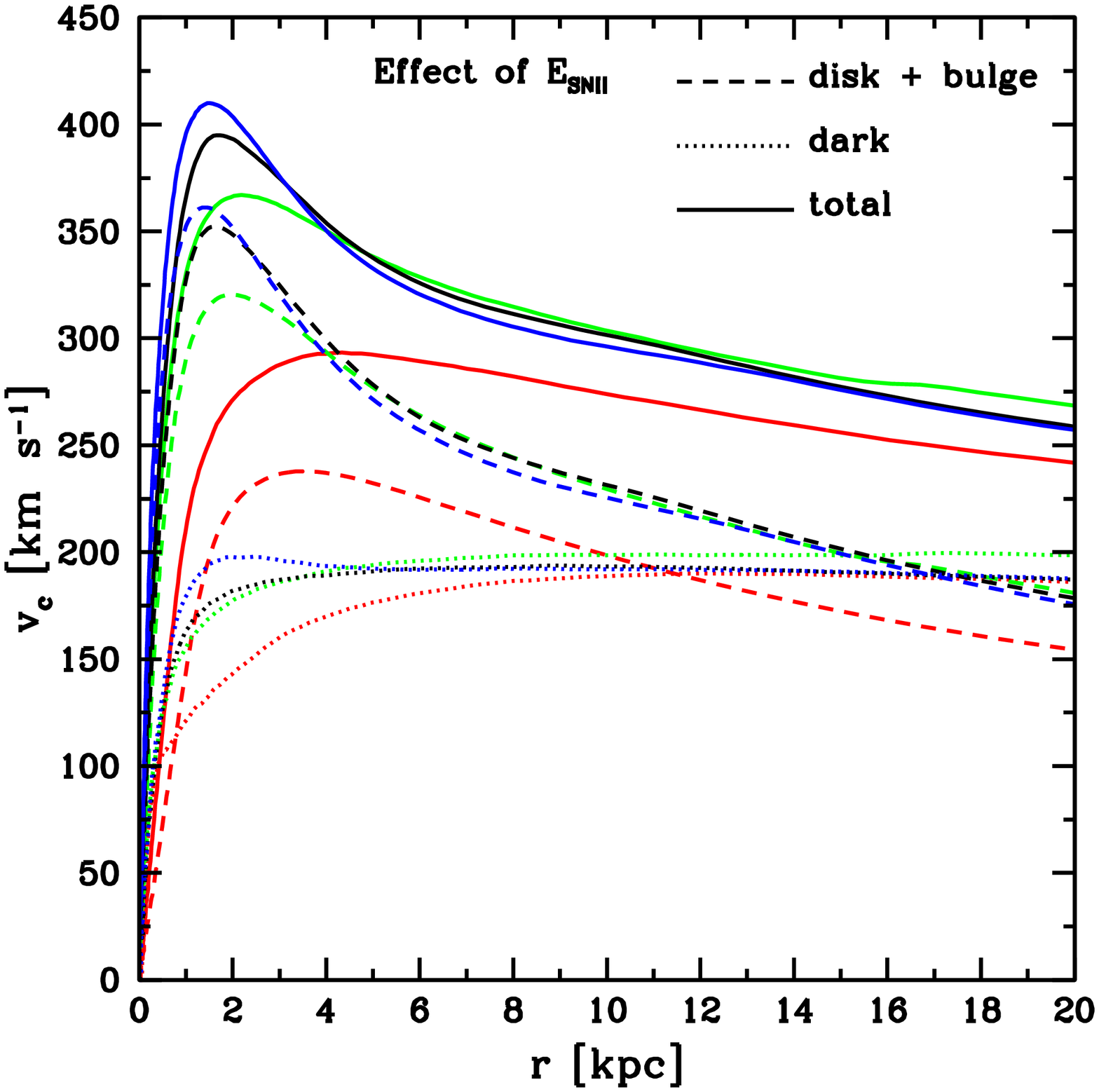,width=200pt} \\
\end{tabular}
\caption[]{Circular velocities, $v_{\rm c}=\sqrt{GM(<r)/r}$, of the stars (solid), dark matter (dotted) and total mass (dashed). (Left) The effect of the star formation efficiency: $\epsilon_{\rm ff}$= 1 (black), 2 (red) and 5 per cent (blue). (Right) The effect of supernovae feedback strengths (adopting $\epsilon_{\rm ff}$= 5 per cent): $E_{\rm SNII}=0$ (blue), $10^{51}$ erg (black), $2\times10^{51}$ erg (green) and $5\times10^{51}$ erg (red).}
\label{fig:vc}
\end{figure*}

In the left-hand panel of Fig.\,\ref{fig:vc}, we plot the \emph{circular} velocities, i.e. $v_{\rm c}(r)=\sqrt{GM(<r)/r}$, for the dark matter, disc + bulge baryons and total mass of the galaxies. We note that the $v_{\rm c}$-profiles are well traced by the cold gas rotation curve (the stellar rotational velocities are lower than what can be expected from $v_{\rm c}$ due to a larger velocity dispersion). While all simulations converge at large radii, and have an equal dark matter and baryon contribution within $r\sim 17\,\kpc$, the mass distribution (and angular momentum distribution) differs dramatically leading to large difference in circular velocities. As we will demonstrate in the next section, a majority of the mass within the bulge component originates from the intense star formation epoch at $z\sim2-3$ where the value of $\epsilon_{\rm ff}$ matters the most. We also observe a significantly enhanced dark matter contraction at large $\epsilon_{\rm ff}$. This effect in our simulation suite, and its relevance for direct dark matter detection, has recently been analysed by \cite{Pato2010}.

\subsection{Star formation histories}
\begin{figure*}
\center
\begin{tabular}{cc}
\psfig{file=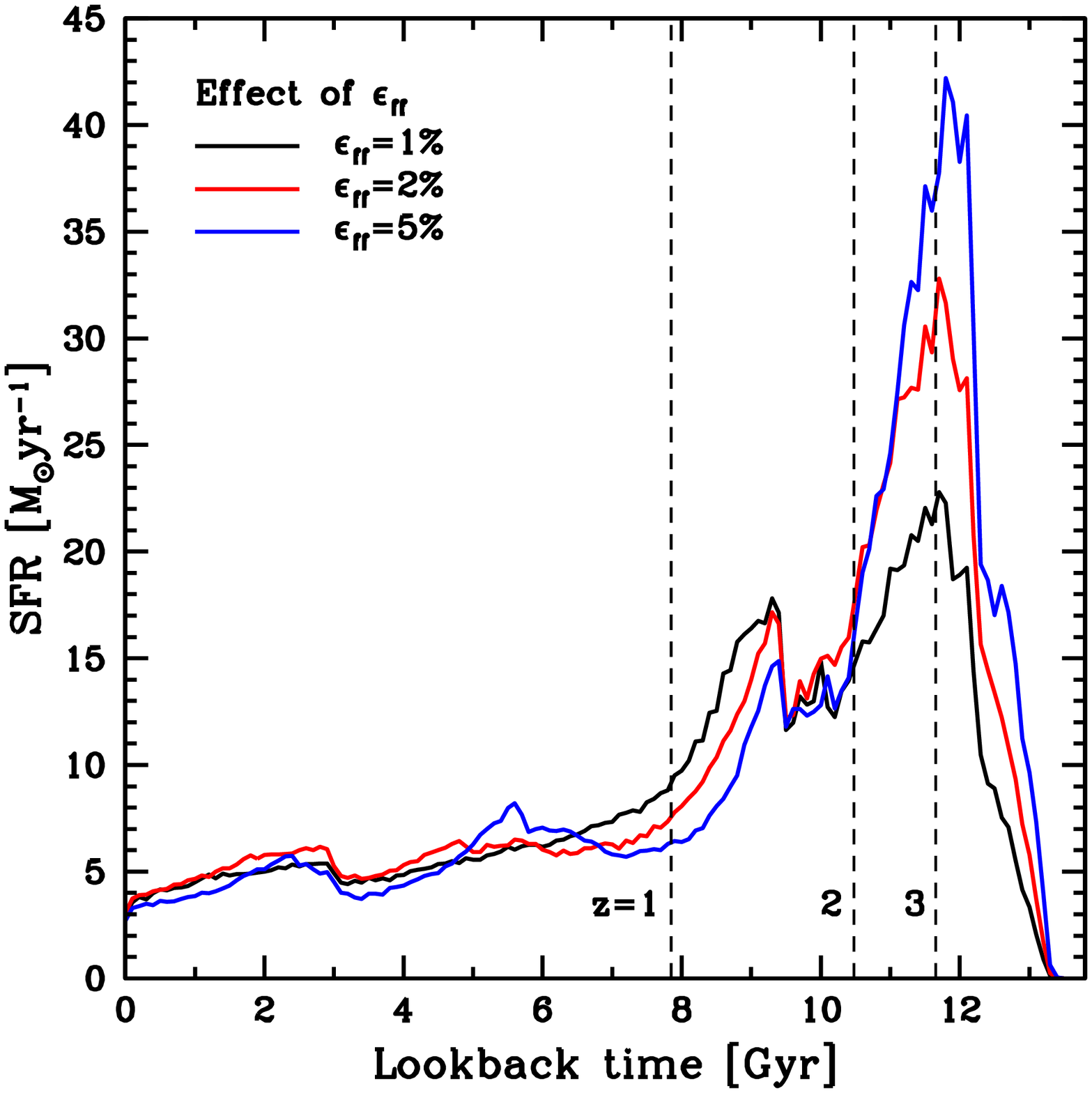,width=200pt} &
\psfig{file=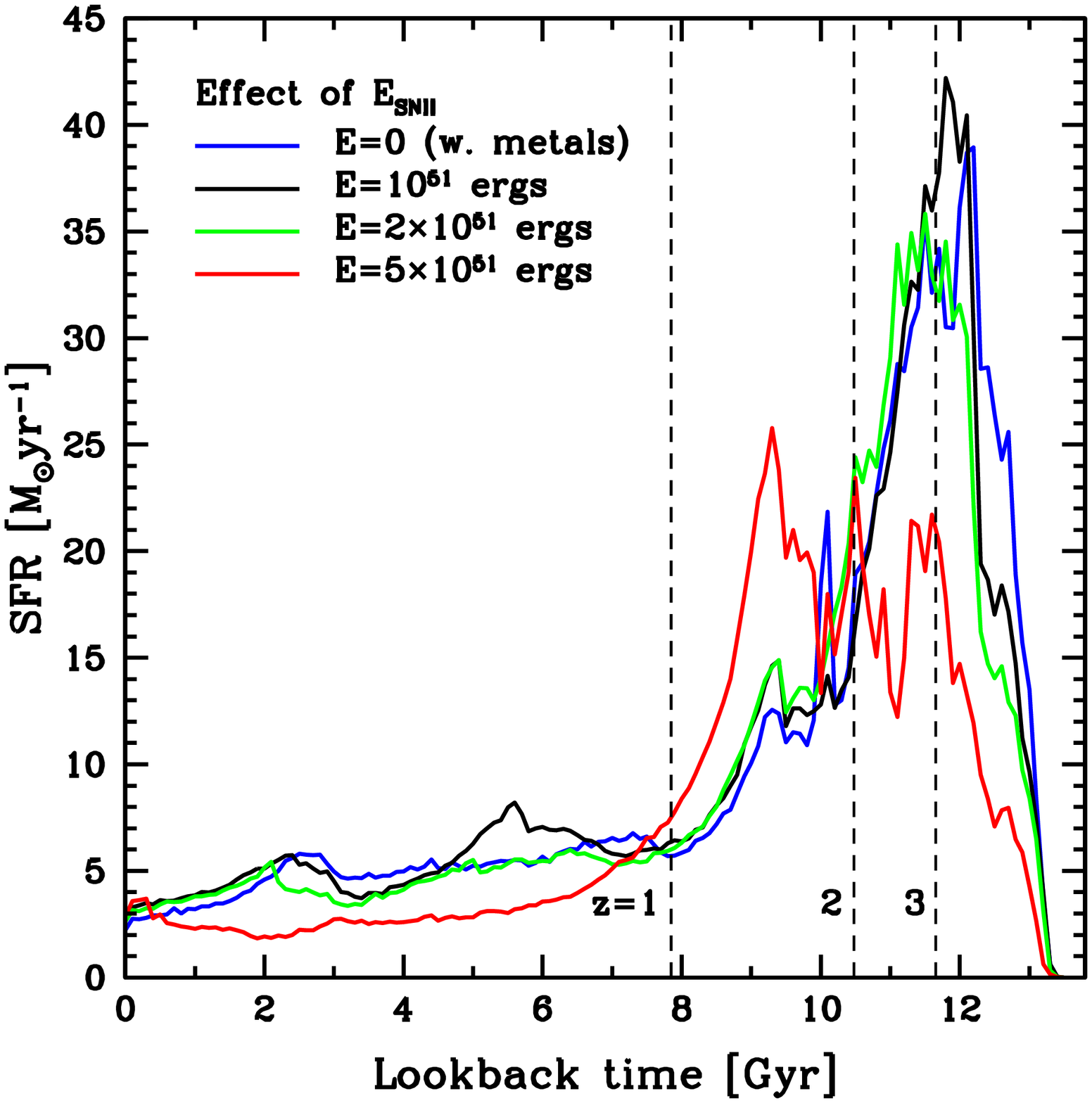,width=200pt} \\
\end{tabular}
\caption[]{(Left) Star formation histories using different values of $\epsilon_{\rm ff}$. At $z<1$, all simulations regulate to similar SFRs, ending up at $\sim3-4\Msol$ at $z=0$, regardless of star formation parameters. Using a high efficiency leads to central galaxies and dwarfs burning their fuel quickly at high redshift during galaxy assembly, resulting in excess angular momentum loss and a prominent central spheroid. A lower efficiency avoids this issue, leaving more gas left for star formation at lower redshifts in a more disc-like configuration. (Right) Star formation histories for a set of simulations of increasing supernovae feedback strength ($E_{\rm SNII}$). We note that the large SFR peak at $z=3$ is only lowered when a very large amount of energy is injected into the ISM.}
\label{fig:SFR}
\end{figure*}

\begin{figure*}
\center
\psfig{file=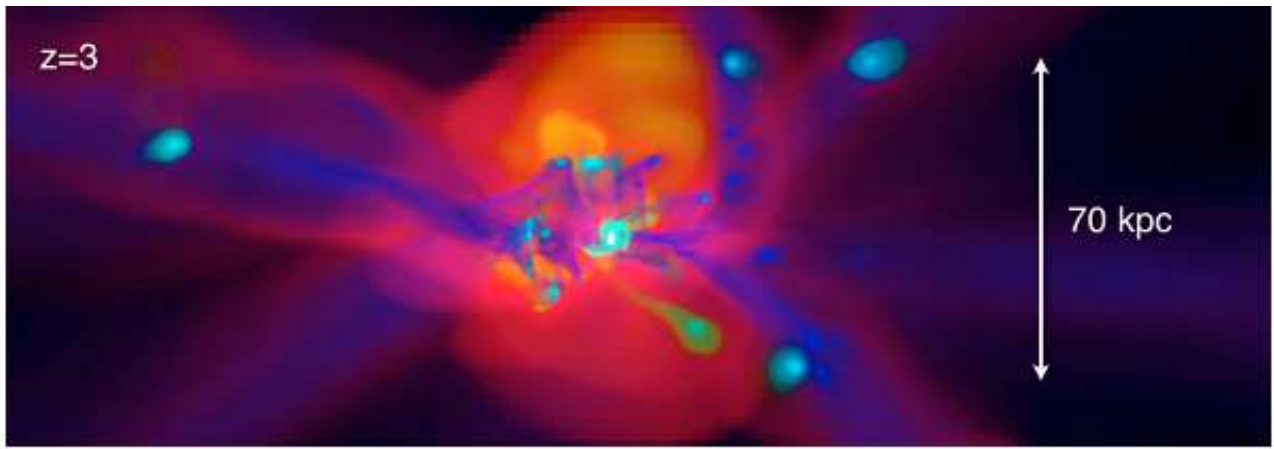,width=500pt} \\
\caption[]{A large scale view of the assembling spiral galaxy from the SR5 simulation at $z\sim3$; the most intense epoch of star formation for this system. The \emph{RGB}-image$^2$ shows the gas component using temperature (red), metals (green) and density (blue). We can clearly distinguish accretion via streams of cold pristine gas (in blue) penetrating the shock-heated gas (in red), reaching the heart of the halo. Dwarf galaxies outside of the large gaseous halo are surrounded by puffy distributions of enriched gas originating from stellar outflows. Gas is efficiently lost via tidal and ram-pressure stripping as the dwarfs interact with the main galaxy and its hot gaseous halo. The distance measure is in physical units.}
\label{fig:z3}
\end{figure*}

The left-hand panel of Fig.\,\ref{fig:SFR} shows the star formation histories for all stars belonging to the discs in n01e1, n01e2 and n01e5 at $z=0$. The average SFR at the current epoch is $\sim 3-4\,\Msol\,{\rm yr}^{-1}$ in all simulation, regardless of numerical setting. Moreover, the star formation history during the quiescent phase of disc evolution, i.e. after $z\sim1$, is relatively flat and roughly the same in all simulations. Significant differences occur at intense epochs of star formation, especially at $z=3$ where the proto-disc is assembled via cold streams, satellite mergers and gas accretion from the hot halo, as seen in the simulation snapshot in Fig.\,\ref{fig:z3}\footnote{The image is an $RGB$ composite image where \emph{red} is temperature, \emph{green} is metals and \emph{blue} is density. Each quantity is a mass weighted average along the line of sight. For each image pixel, we calculate the $RGB$ triplet as 
\begin{equation}
\nonumber
(R,G,B)=255\left(\frac{\log(T/T_{\rm m})}{\Delta T},\frac{\log(Z/Z_{\rm m})}{\Delta Z},\frac{\log(\rho/\rho_{\rm m})}{\Delta \rho}\right),
\end{equation}
where $\log\{T_{\rm m}\,{\rm [K]},Z_{\rm m}\,[Z_\odot],\rho_{\rm m}\,{\rm [cm^{-3}]}\}=\{4.1,-3,-4.8\}$ and $\{\Delta T,\Delta Z,\Delta \rho\}=\{2,2,6\}$.}. Here the SFR peak changes dramatically from $\sim 43\,\Msol\,{\rm yr}^{-1}$ in n01e5 to $\sim 23\,\Msol\,{\rm yr}^{-1}$ in n01e1. 

In n01e5, stars form efficiently everywhere, even in satellites. The gas is quickly consumed locally during the high-redshift assembly, and merging systems lose angular momentum to the dark halo, ending up in the central part of the galaxy. Accretion via cold streams and from the hot gaseous halo will still supply the galaxy with unprocessed gas \citep{Dekel09,Agertz09b}, but now in a more bulge-dominated environment. In n01e1, star formation is less efficient and a significant portion of the mass in the merging clumps is in gaseous form. This material is lost to the hot gaseous halo via ram-pressure and tidal stripping, or expelled during SNe events, which later cools down to join the disc. Hence, the material that is not consumed by star formation at $z=3$ is processed at a later epoch, closer to $z=1-2$ (see Fig.\,\ref{fig:SFR}), but now in a more disc-like, higher angular momentum configuration. In fact, the trend at $z=3$ is reversed at this later epoch, and the largest SFR is found for $\epsilon_{\rm ff}=1$ per cent. These two modes of star formation are related to the classical angular momentum problem \citep{NavarroWhite1994}, and they lead to fundamentally different modes of disc assembly and morphology.

\begin{figure*}
\center
\begin{tabular}{cc}
\psfig{file=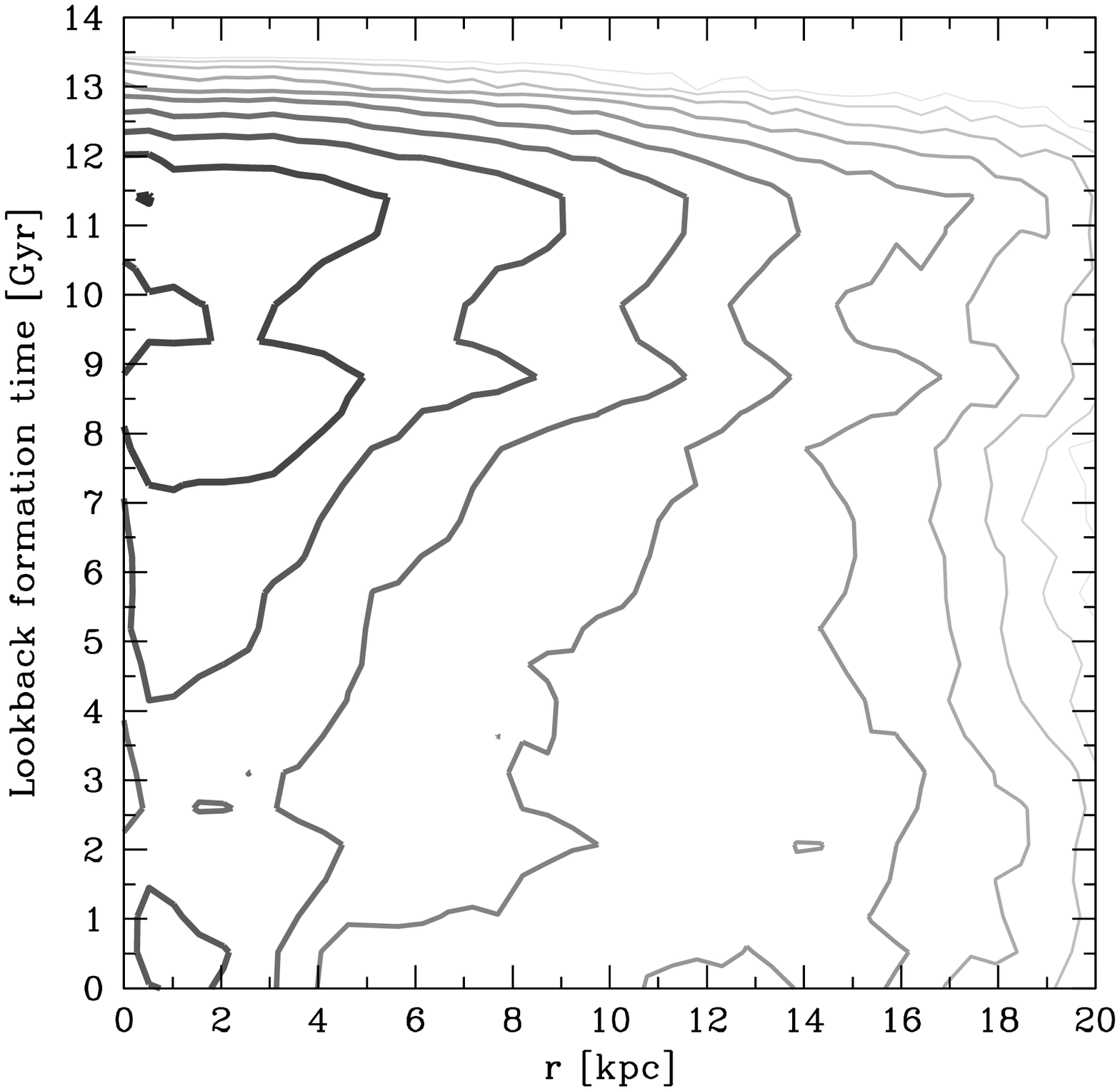,width=190pt} &
\psfig{file=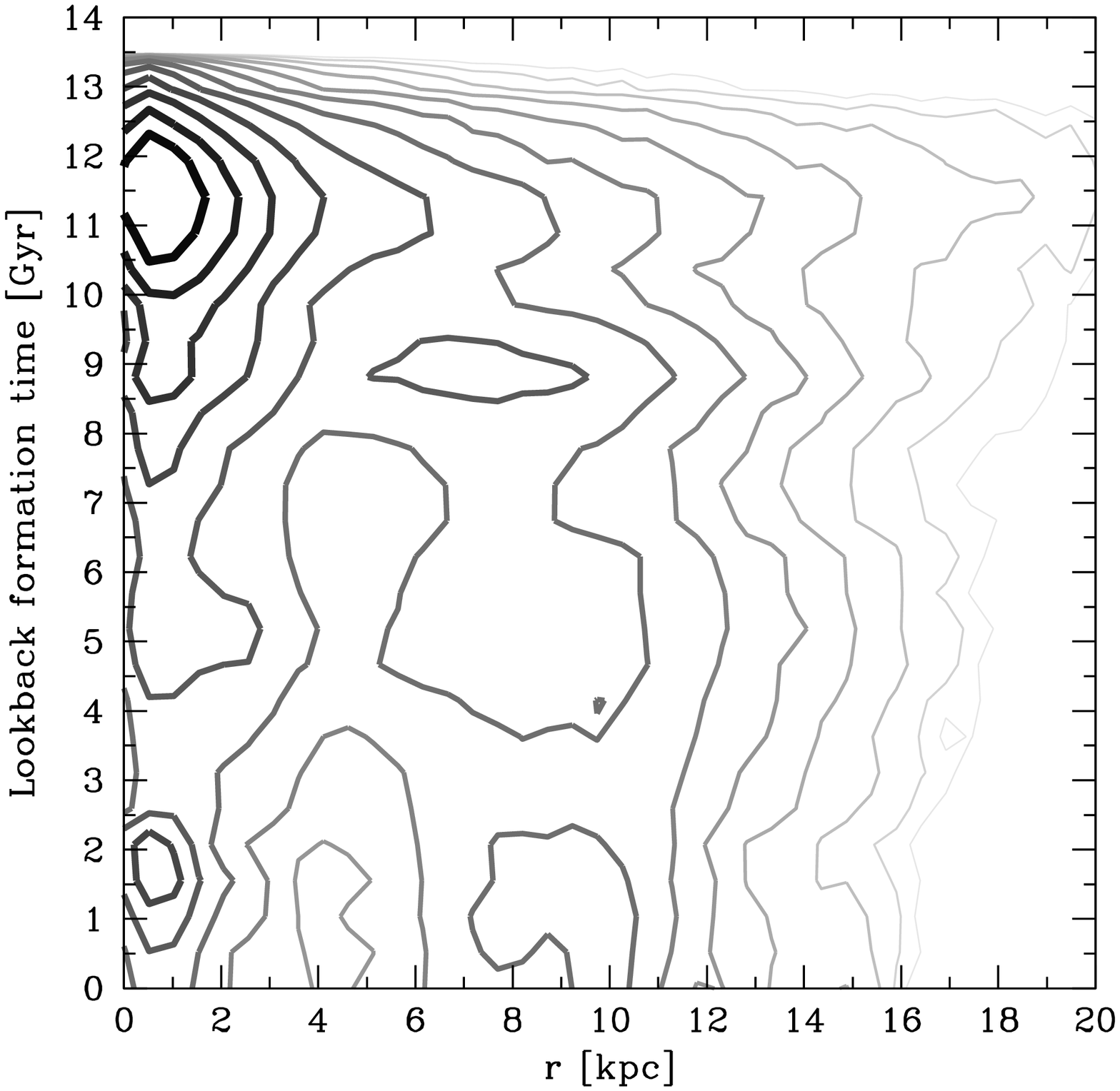,width=190pt} \\
\end{tabular}
\caption[]{Formation time of stars and their radial distribution for the discs in n01e1 (left) and n01e5 (right) at $z=0$. The contours trace regions of binned mass using bins of size $\Delta t=0.5\,{\rm Gyr}$ and $\Delta x=0.5\,{\rm kpc}$. The contour lines trace, from thin lines with light shades, to thick lines with dark shades, the formed stellar masses from $\log(M_*)=6.5$ to $ \log(M_*)=9.5 $ in steps of 0.25 dex. While the formed stellar mass in n01e1 is smoothly distributed across the disc at all times, the n01e5 simulation shows a strong central concentration of stars formed at $t=11.5\,{\rm Gyr}$ ($z\sim3$).}
\label{fig:agecont}
\end{figure*}

A confirmation of the above discussion is shown in Fig.\,\ref{fig:agecont}, where contours of formed stellar masses are outlined in the star formation time vs. disc radius plane. Note that this mass refers to all the stars at $z=0$ contained in the disc, and is hence the sum of the stars formed in merging satellites as well as in situ. While the formed stellar mass in n01e1 is smoothly distributed in a roughly  exponential profile across the disc at all times, without a clear sign of extreme star formation bursts, the n01e5 simulation shows a strong central concentration of stars formed at $t=11.5\,{\rm Gyr}$ ($z\sim3$). This analysis confirms the notion of efficient star-forming satellites loosing angular momentum and being dragged into the central parts of the galaxy.

\label{sect:SFH}


\subsection{Hubble types}
\label{sect:hubble}
Fig.\,\ref{fig:maps} shows mass weighted projections of the stellar and gas surface densities for n01e1ML, n01e1, n01e2ML, n01e2, n01e5ML and n01e5. The gaseous discs are thin and extended in all simulations, and are surrounded by a warped layer of cold/warm gas, probably associated with misaligned accretion events \citep{ShenSellwood2006}. A hot gaseous halo surrounds the discs, and a temperature projection (not shown) reveals an extended disc-halo interface of warm/hot gas. We will explore this in future work.

As discussed in Section\,\ref{sect:stareff}, we find a very strong trend in disc and bulge mass with increasing star formation efficiency. The n01e1 simulation feature a $8.6\times 10^{10}\Msol$ stellar disc with a $2\times 10^{10}\Msol$ bulge, hence B/D$\sim 1/4$. In n01e5 the disc is 35 per cent less massive and the bulge 3.5 times more massive with B/D$\sim 1.25$. Roughly the same scaling holds when including additional SNIa feedback and stellar mass-loss.
 
All discs show spiral pattern in the gas component with a larger amplitude in the more gas-rich discs, having lower $\epsilon_{\rm ff}$. We also observed spiral structure in the stellar component, which is the most pronounced in n01e1 and n01e1ML and n01e2ML simulations. As $\epsilon_{\rm ff}$ is increased, B/D increases and the spirals arms become more tightly wound as marginal gravitational instabilities can no longer excite pronounced open spiral arm structure. All discs feature a stellar bar, and viewed edge-on, we observe how the inner stellar distribution flattens as $\epsilon_{\rm ff}$ is decreased. The flattened central parts of n01e1 and n01e1ML, and the fact that the bulge is well fitted using an exponential, is indicative of a bulge formed via secular processes \citep{KormendyKennicutt04} e.g. bar buckling \citep{debattista06}. The gaseous bar strengthens at lower $\epsilon_{\rm ff}$, and in n01e1 and n01e1ML gas is transported towards the disc centre, triggering star formation. In these simulations, close to $50$ per cent of the stars associated with the bulge formed in situ of the disc at $z\lesssim1$, and only $\sim 25$ per cent formed at the intense star formation peak at $z\sim 3$. This indicates that a significant portion of the flattened bulge has formed via secular evolution, leading to a pseudo-bulge. This is in stark contrast to the bulge formation epoch seen in the central parts of the n01e5 simulations (right-hand panel of  Fig.\,\ref{fig:agecont}), where essentially all bulge stars form at $z\sim 3$. 

\cite{Weinzirl09} \citep[see also][]{Laurikainen10}, recently analysed 182 $H$-band images from the OSUBSGS survey \citep{Eskridge02} to obtain B/D and B/T values across the Hubble sequence. Comparing their sample averages (see e.g. their fig. 14) to our set of simulations (B/D in Table\,\ref{table:simsummary2}) suggests that the final disc in n01e5 is of S0/a type, in n01e2 it is of Sa/Sab type and in both n01e1 and n01e1ML it is of Sb/Sbc type. We consider this agreement only as indicative as each Hubble type spans a wide range of B/D and B/T values. \cite{GrahamWorley2008} presented B/D and B/T flux ratios using a sample of over 400 galaxies observed in the $K$ band. Their B/D estimates for different Hubble types confirm the classification of our simulated discs. There is no doubt that we are measuring a transformation along the Hubble sequence. 

\begin{figure}
\center
\psfig{file=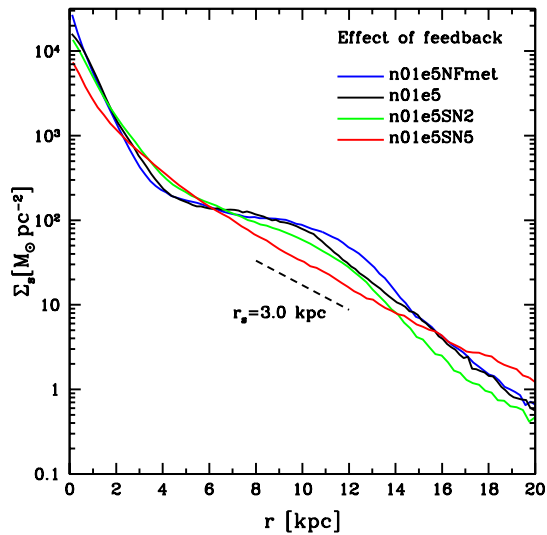,width=210pt} 
\caption[]{Stellar surface densities of the $z=0$ discs in the supernova feedback test suite. As the SNII feedback energy input is increased, the disc becomes more extended and the bulge component less massive.}
\label{fig:SigmaFB}
\end{figure}
\section{Effect of supernova feedback}
\label{sect:FB}
As we have demonstrated in the previous sections, a high SFE overproduces the central stellar mass of the galaxy. The inclusion of additional SNIa feedback and stellar mass-loss did not drastically change the galaxy properties, even though differences can be seen in Fig.\,\ref{fig:maps} (the discs in n01e1 and n01e2 feature much stronger spiral structure) and Fig.\,\ref{fig:sigma}, and more late time star formation is made possible (see Section\,\ref{sect:TF}). The effect of stellar mass-loss was studied by \cite{MartigBournaud2010} who found a stronger effect on the bulge mass in a similar setting, perhaps due to implementation differences.

In Figs \,\ref{fig:vc} and \ref{fig:SigmaFB} we present $v_{\rm c}$ and $\Sigma_{\rm s}$ for the n01e5 simulation, but with different amounts of injected SNII energies; see Table\,\ref{table:simsummary2}. Note that we still enrich the ISM with metals in the simulation with $E_{\rm SNII}=0$. Without this, the effect of metal cooling will not be present in all simulations. We find that the bulge mass is lowered as we increase $E_{\rm SNII}$, but only for a very large injected value of $5\times 10^{51}\,{\rm erg}$ can the disc rotation curve peak at a reasonable $v_{\rm c}<300\,\kms$, resembling that of the n01e1 simulation. As for the standard feedback runs in the previous section, the dark matter halo is more contracted as star formation is less regulated. The difference in $v_{\rm c}$ at $r=20\,\kpc$ between the n01e5SN5 and the other simulations, corresponding to a few $10^{10}\Msol$, is due to the expelled gas during galaxy assembly which can be accounted for in the more massive gas halo. This effect can also be seen in $\Sigma_{\rm s}$ as the central values are decreased, the discs scale radius decreases and the break radius is shifted to larger radii. As seen in Table\,\ref{table:simsummary2}, a massive disc still forms. The effect on the star formation histories are shown in the right-hand panel of Fig. \ref{fig:SFR}; we find no significant difference among the simulations, apart for the very energetic n01e5SN5 simulation. The SFH now resembles that of n01e1 where the $z=3$ amplitude is lowered to $\approx 20\Msol{\rm yr}^{-1}$ and more gas is left to form stars in a disc-like environment at $z\sim1-2$.

The projected gas density and stellar maps were shown in Fig.\,\ref{fig:mapsFB}. While the standard feedback simulations shown in Fig.\,\ref{fig:maps} showed a clear Hubble sequence of open to tightly wound spiral structure as B/D was lowered, this is not the case for the feedback test suite. In n01e5SN5, the gaseous disc is heavily distorted, warped and puffed up by the large SNII energy injections. Star formation is here very different compared to n01e1 as stars form in filaments and shells from SNe explosions rather than in gas-rich spiral arms. 

The effect of metal cooling is not always accounted for in cosmological simulations. \cite{Piontek09a} included this effect and reported on difficulties in suppressing the initial high-$z$ peak, even with sophisticated feedback models. In our simulations metal cooling is roughly counter-acted by the standard SNII feedback. If metal enrichment is turned off together with the feedback, we do not find a significant modification to our discs. For example, the n01e1NFB simulation shows a surprisingly successful set of characteristics when compared to n01e1 (see Table\,{\ref{table:simsummary2}). As a zero metallicity gas cools inefficiently below $10^{4}\,{\rm K}$, as well as in the range $10^{5}\,{\rm K}$ to $10^{7}\,{\rm K}$, the n01e1NFB disc essentially behaves as higher metallicity counterpart but with SNII heating balancing cooling. This is the philosophy behind subgrid multiphase models \citep[e.g.][]{springelhernquist03} in which feedback is implicitly treated as a stiff gas equation of state. Note that our polytropic EOS is slightly stiffer than what is usually adopted ($\gamma=2$ instead of $\gamma=5/3$). 

\section{Relationship to observations}
\label{sect:observations}
\subsection{Angular momentum of the baryons}
For each galaxy we calculate the cumulative specific angular momentum vector, defined as 
\begin{equation}
\label{eq:angmom}
\boldsymbol{j}_{\rm bar}(\leq r)=\frac{1}{M(\leq r)}\sum_{i=1}^{N(\leq r)} m_i\boldsymbol{x}_i\times\boldsymbol{v}_i,
\end{equation}
including all bulge and disc baryons. Here $\boldsymbol{x_i}$ and $\boldsymbol{v_i}$ are positions and velocities of the gas cells and star particles of the $N$ elements within a radius $r$ encapsulating the mass $M(\leq r)$. The resulting $j_{\rm bar}=|\boldsymbol{j}_{\rm bar}|$ are presented in Table \ref{table:simsummary2}.

Using the sample of \cite{Courteau97} and \cite{mathewson92}, \cite{navarrosteinmetz00} calculated the specific angular momenta vs. rotational velocity for late-type spiral galaxies and compared them to numerical simulations. The discs were assumed to follow an exponential profile for which the peak rotational velocity, $v_{\rm rot,2.2}$, occurs at $r=2.2 r_{\rm d}$, and it follows that $j_{\rm bar}=2r_{\rm d}v_{\rm rot,2.2}$. This assumption can be misleading when comparing to simulated galaxies as the true $v_{\rm rot}$-peak can be significantly underestimated in the case of bulge-dominated disc galaxies. The sample of \cite{Courteau97} concerned Sb-Sc galaxies for which B/D is low and a dominating exponential disc assumption is roughly valid. The difference in measured and estimated angular momentum content makes it difficult to compare simulated and observed galaxies, as discussed in \cite{Abadi03b} and \cite{Piontek09b}. A simulated galaxy can be considered as a successful realizations of a late-type (Sb-Sc/Sd) galaxy if the estimated and measured angular momenta are in agreement.

Focusing on the $n=0.1\,{\rm cm}^{-3}$ suite, we find that the n01e1 and n01e1ML simulations are in good agreement with the observed galaxies, both when analysed using Eq.\,\ref{eq:angmom} and the exponential disc approximation. Typical measured and estimated values are here $j_{\rm bar}\sim 2000\,{\rm km}\,{\rm s}^{-1}{\rm kpc}$ and $j_{\rm bar}\sim 2750\,{\rm km}\,{\rm s}^{-1}{\rm kpc}$ respectively. As $\epsilon_{\rm ff}$ is increased, the calculated angular momentum decreases. The $\epsilon_{\rm ff}=2$ per cent simulations are still a part of the observed scatter while higher values create more significant outliers in the observed distribution. When using the exponential disc approximation, all simulated galaxies are in good agreement with the observed data as the velocities are quite comparable at larger radii, and for the fact that the discs, although less extended, have larger $r_{\rm d}$ in the higher efficiency cases (see Fig.\,\ref{fig:sigma2}). We conclude that an angular momentum reservoir comparable to Sb/Sc galaxies have been reproduced for the baryons in the case of low SFE (i.e. $\epsilon_{\rm ff}=1$ per cent).

The lack of correlation between $E_{\rm SNII}$ and the baryonic angular momentum content might come as a surprise. However, while the B/D ratio decreases for large supernova energy injections, the actual disc mass changes little, and is $\sim 6\times 10^{10}\Msol$ for all $\epsilon_{\rm ff}=5$ per cent simulations. As the net contribution of the bulge to the angular momentum content is roughly zero, similar $j_{\rm bar}$ is to be expected. All $\epsilon_{\rm ff}=5$ per cent simulations have measured $j_{\rm bar}\sim 1300-1450\,{\rm km}\,{\rm s}^{-1}{\rm kpc}$ which is close to the estimated $j_{\rm bar}\sim 1600\,{\rm km}\,{\rm s}^{-1}{\rm kpc}$ in n01e5SN5.

In summary, the largest measured baryonic specific angular momentum reservoir can be found in simulations using $\epsilon_{\rm ff}=1$ per cent due to a massive disc component, regardless of including feedback or not. At higher efficiencies, $j_{\rm bar}$ decreases, again regardless of feedback. 

\subsection{The Tully-Fisher relationship}
\label{sect:TF}
\begin{figure}
\center
\psfig{file=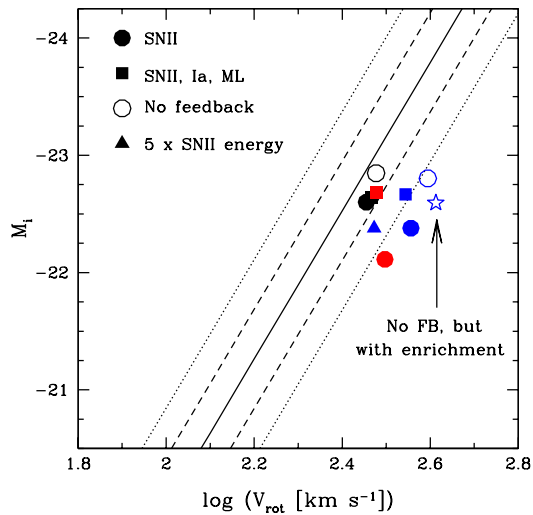,width=210pt} 
\caption[]{The $i$-band Tully-Fisher relationship from the SDSS \citep{Pizagno07}. We show the observed average (solid line), $1\sigma$ (dashed line) and $2\sigma$ (dotted line) relation. The symbols are results from our simulated galaxies, which use $n_0=0.1\,{\rm cm}^{-3}$ and $\epsilon_{\rm ff}=1$ (black symbols), 2 (red symbols) and 5 (blue symbols) per cent.}
\label{fig:TF}
\end{figure}

The photometric Tully-Fisher (TF) relation \citep{TF77} links the characteristic rotational velocity of a galaxy with its total absolute magnitude. This correlation holds in all typical photometric bands but with variation in functional form \citep[e.g.][]{Pizagno07}. Early attempts in forming realistic galaxies \citep[e.g.][]{Abadi03b} showed off-sets in the observed relation owing to the formation of very concentrated bulge-dominated galaxies with a low star formation activity at late times. Their velocity-magnitude relation had more in common with S0 galaxies \citep{Mathieu02}. Recent work seems to have improved on these results by SN feedback regulated star formation \citep{Governato07,Piontek09b}. These studies place galaxies closer to the observed relation, but this is in part achieved by circumventing the large measured $v_{\rm rot}$ (caused by the dominant bulge) by using the exponential disc assumption discussed above \citep[but see][]{Governato09}, i.e. using $v_{\rm rot,2.2}$. Observationally, the measured quantity is often half of the HI velocity width at $20\,(W_{20})$ or $50\,(W_{50})$ per cent of the peak intensity. 

In Fig.\,\ref{fig:TF} we present the measured $i$-band magnitudes of several of our simulated galaxies as a function of their peak rotational velocities measured from the gas component. These are compared to the observed TF relation from the SDSS \citep{Pizagno07}. Pizagno et al. measured the velocity at a radius containing 80 per cent of the $i$-band flux. This measure ($V_{80}$) is equivalent to measuring $v_{\rm rot}$ at $\sim 3r_{\rm d}$ for a pure exponential disc. By using the true peak of $v_{\rm rot}$, we provide an absolute lower limit to the agreement with observations, and can clearly separate disc and bulge-dominated galaxies. We note that the low efficiency models agree well with the average data, regardless of adopted feedback scheme, and even without. At higher efficiencies, the discs are off-set by more than $2\sigma$, mostly due to their peaked rotation curves. In these circumstances, the inclusion of additional recycling via SNIa and stellar mass-loss increases the magnitudes by $\sim 0.5$ dex in n01e2ML and n01e5ML simulations. The n01e5NFB simulation is brighter than the corresponding simulations including feedback due to exclusion of metal enrichment, leading to less efficient cooling and more gas left to form stars at later times. Allowing for enrichment without any energy deposition demonstrates this fact (see figure). From a photometric TF point of view, the $\epsilon_{\rm ff}=$ 5 per cent discs correspond to S0 systems or early type spirals \citep{Mathieu02}. 

As described in the previous sections, $v_{\rm rot}$ and $\Sigma_{\rm s}$ in the n01e5SN5 simulation agrees fairly well with the disc values found in n01e1. The strong feedback brings the galaxy closer to the observed values but the absolute magnitude is still lower than in the $\epsilon_{\rm ff}=1$ per cent simulations. The SFH in Fig.\,\ref{fig:SFR} tells us why: after $z=1$ the SFR is lower in n01e5SN5 compared to n01e1 by almost a factor of 2 (even though the $z=0$ values agree) due to strong gas expulsion, resulting in a less bright disc by $\sim 1/3$ dex in $i$-band magnitude. 

As for the specific angular momentum analysis, adopting the $v_{\rm rot,2.2}$ measure (or $V_{80}$), \emph{all} discs would agree statistically with the observed TF relation, especially when including SNIa feedback and stellar mass-loss.

Similar to the photometric TF is the `baryonic TF' relation \citep{mcgaugh00,mcgaugh10} which links characteristic rotation velocity with total galaxy baryonic mass. The baryonic TF relation therefor accounts for the fact that less massive galaxies are more gas rich, and their stars only account for a small fraction of the total disc mass. The same conclusion as above holds for the baryonic TF: the low-efficiency simulations agrees well with the observations. As the baryonic masses for the discs are not strongly affected, even in the case of extreme feedback (n01e5SN5) the data points shift only with the increase of the $v_{\rm rot}$ peak. As for the photometric TF, using $v_{\rm rot,2.2}$ puts all galaxies on the observed relation.

\subsection{The $\Sigma_{\rm SFR}$-$\Sigma_{\rm gas}$ relation}
\begin{figure*}
\center
\begin{tabular}{cc}
\psfig{file=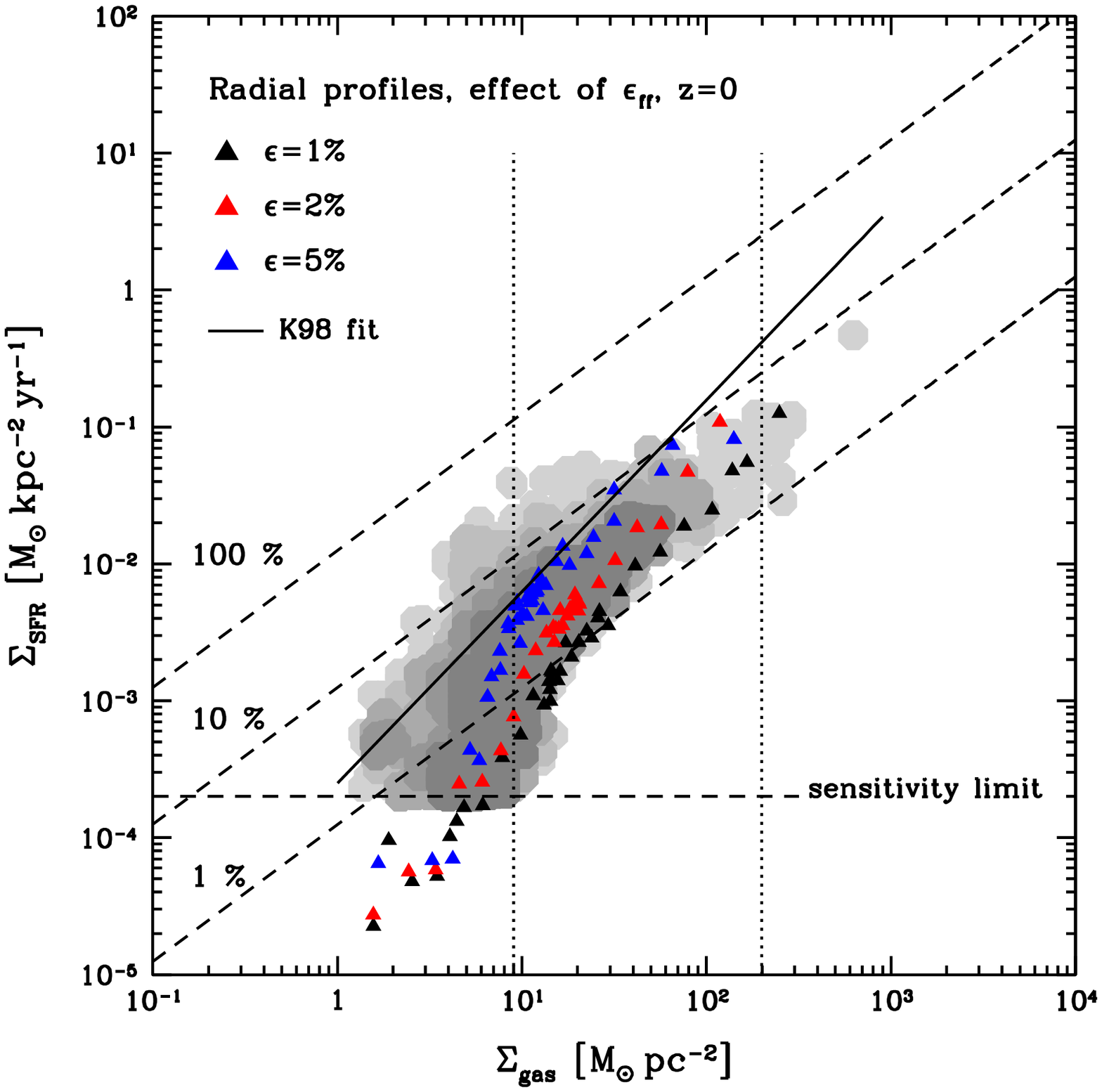,width=255pt} 
\psfig{file=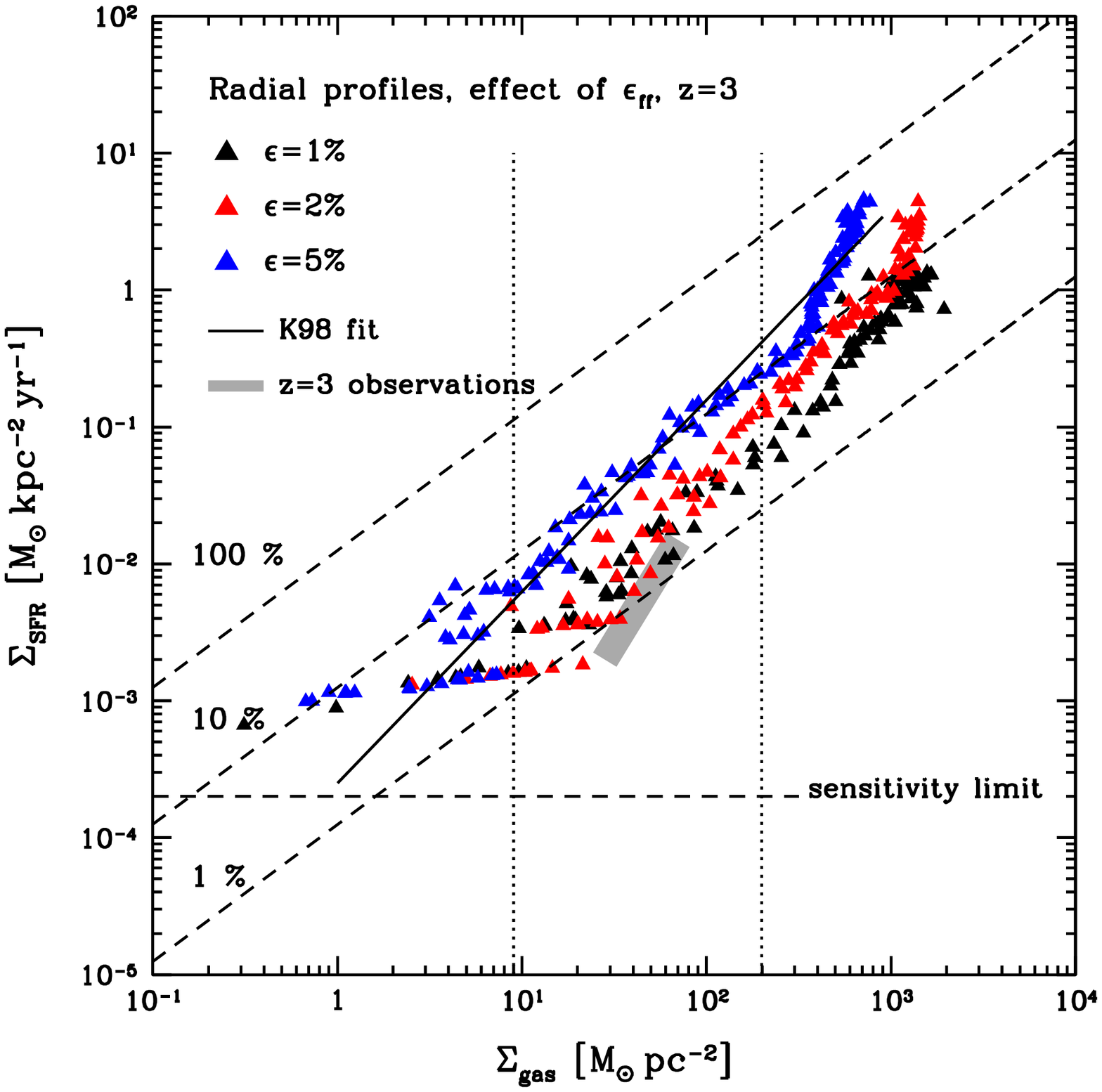,width=255pt} 
\end{tabular}
\caption[]{$\Sigma_{\rm SFR}$ vs. $\Sigma_{\rm g}$ for the resulting discs using $\epsilon_{\rm ff}=1$ (n01e1), 2 (n01e2) and 5 (n01e5) per cent at $z=0$ (left-hand panel) and at $z=3$ (right-hand panel). The filled circles are radial data for 7 spiral galaxies from the THINGS survey \citep{bigiel2008}, where $\Sigma_{\rm gas}$ includes the contribution from helium ($\Sigma_{\rm gas}=1.36\,\Sigma_{\rm HI+H_2}$). The data points represent, from lightest to darkest, $>1, >5, >10, >20$ and $>30$ detections. The vertical dotted lines separate regions where different star formation laws are conjectured to apply (see text). Diagonal dotted lines show lines of constant SFE=$\Sigma_{\rm SFR}/\Sigma_{\rm gas}$, indicating the level of $\Sigma_{\rm SFR}$ needed to consume 1, 10 and 100 per cent of the gas reservoir in $10^8$ years. The solid black line is the average relation from \cite{kennicutt98}. The $z=3$ observations approximately populate the region of the \cite{WolfeChen06} observations. THINGS data courtesy of F. Bigiel.} 
\label{fig:sf}
\end{figure*}
\begin{figure*}
\center
\begin{tabular}{cc}
\psfig{file=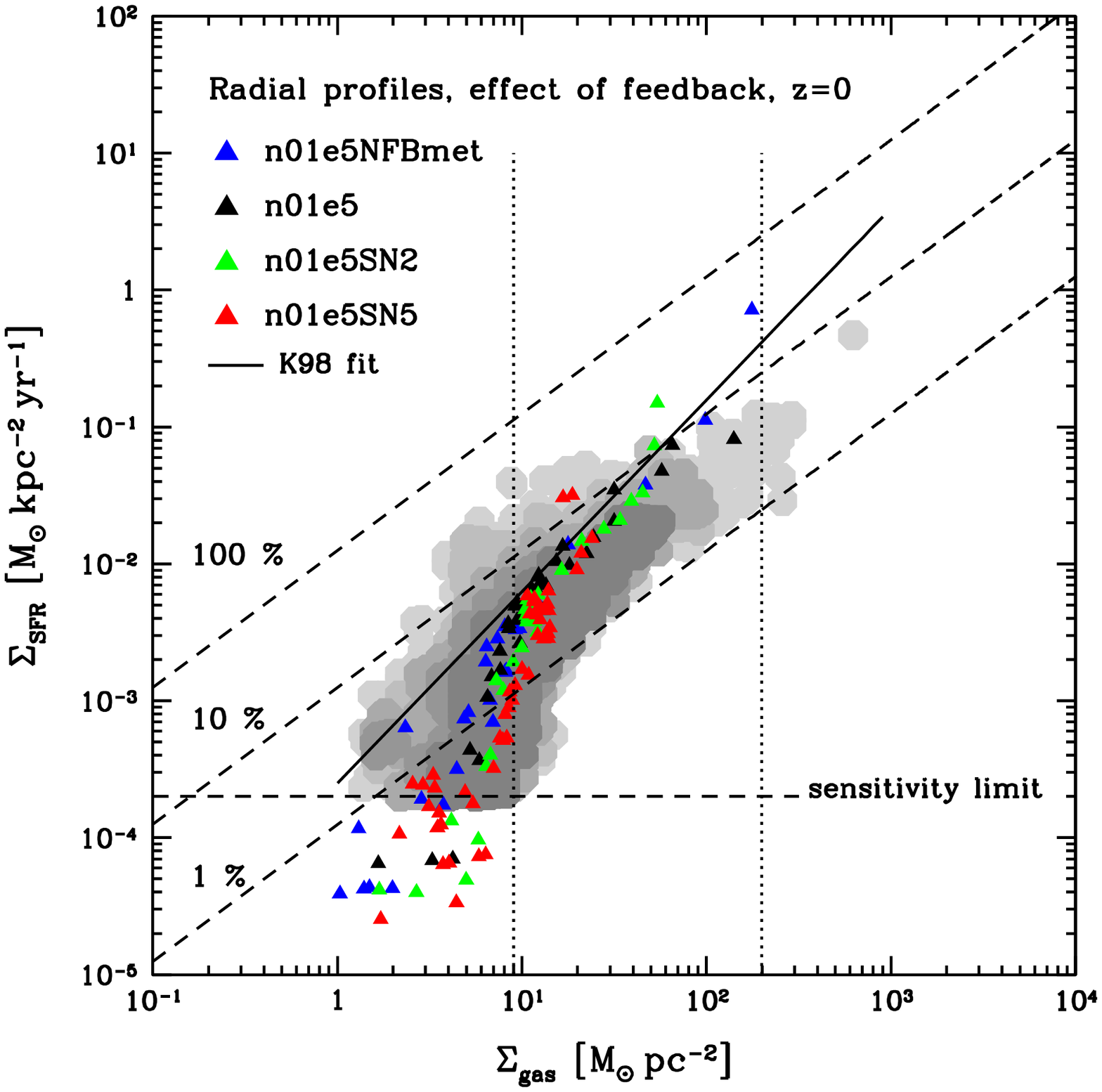,width=255pt} 
\psfig{file=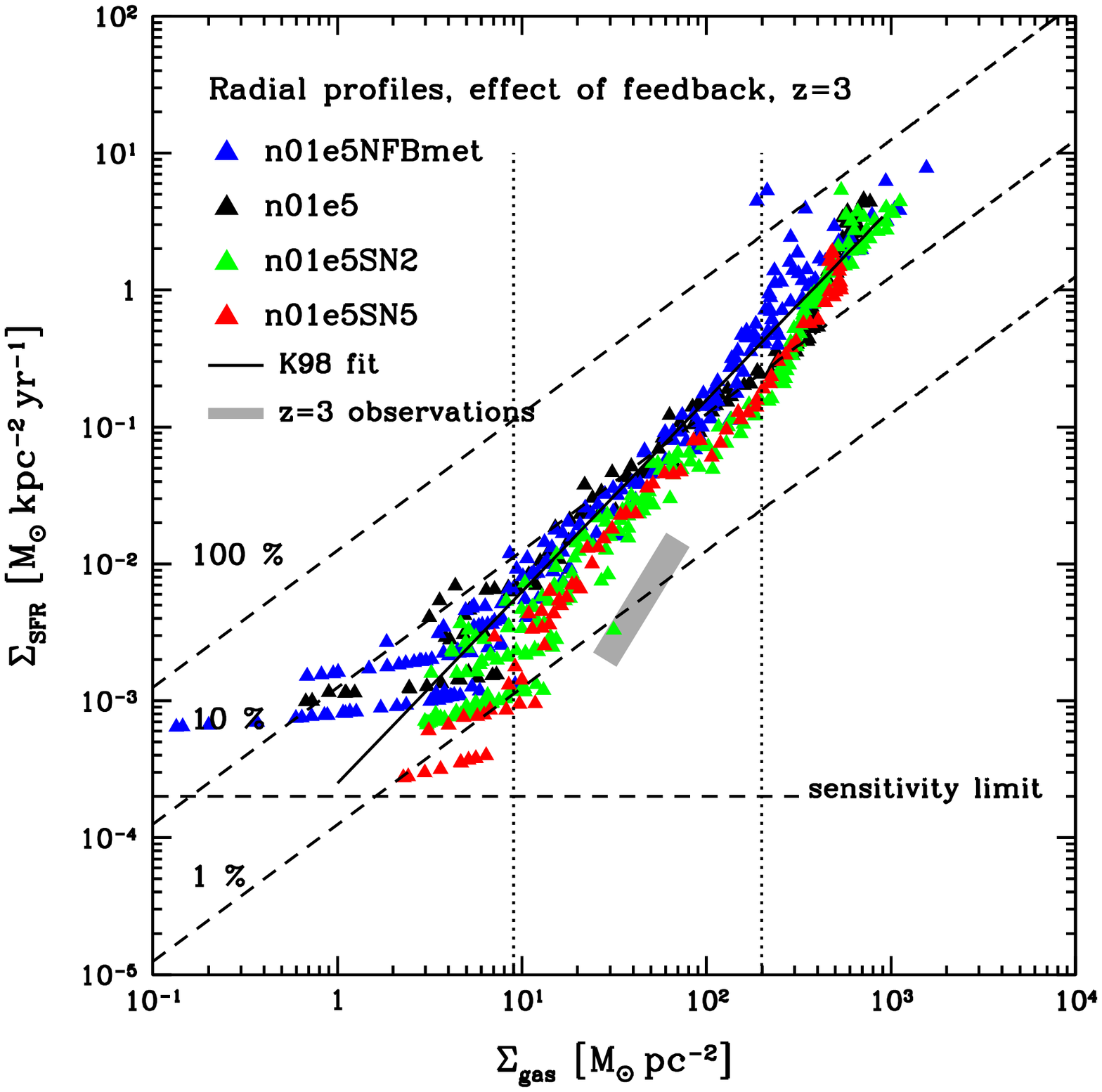,width=255pt} 
\end{tabular}
\caption[]{$\Sigma_{\rm SFR}$ vs. $\Sigma_{\rm g}$ for the resulting discs using a high star formation efficiency ($\epsilon_{\rm ff}=5$ per cent), but with different SN feedback strengths: $E_{\rm SNII}= 10^{51}\,{\rm erg}$ (n01e5), $2\times 10^{51}\,{\rm erg}$ (n01e5SN2), $5\times 10^{51}\,{\rm erg}$ (n01e5SN5) as well as with zero SNII feedback energy but metal enrichment (n01e5NFBmet). The panels show the results at $z=0$ (left) and at $z=3$ (right). The lines and symbols are described in the caption of Fig.\,\ref{fig:sf}. SN feedback has little effect on the $\Sigma_{\rm SFR}$-$\Sigma_{\rm g}$ relation at $z=0$ but does affect the high redshift relation, although only for very large energy injections ($E_{\rm SNII}\geq2\times10^{51}\,{\rm erg}$).}
\label{fig:sfFB}
\end{figure*}
The most famous study of the globally averaged relationship between the star formation rate and gas surface density is from \cite{kennicutt98} (from now on K98), where a sample of 61 nearby normal spiral galaxies and 36 infrared-selected starburst galaxies were considered. Assuming a Schmidt-law of the form 
\begin{equation}
\label{eq:schmidt}
\Sigma_{\rm SFR}=a\left(\frac{\Sigma_{\rm gas}}{1\Msol\pc^{-2}}  \right)^N,
\end{equation}
the full sample yielded $a=(2.5\,\pm\,0.7)\times 10^{-4}$ and $N = 1.40\,\pm\,0.15$. The $\Sigma_{\rm SFR}$-$\Sigma_{\rm gas}$ relation has been studied by many authors \citep[e.g.][]{WongBlitz02,Misiriotis06,Kennicutt07cut,Schuster07}, both locally and globally, using different star formation tracers and galaxy samples. A large range of power-law indices ($N\approx1-3$) have been found, suggesting that either different SF laws exist in different galaxies or that $N$ is very sensitive to systematic differences in methodology. \cite{bigiel2008} presented a comprehensive analysis of the $\Sigma_{\rm SFR}$-$\Sigma_{\rm gas}$ relationship using multifrequency data of 7 spiral galaxies and 11 late-type and dwarf galaxies. The analysis pointed to a great variation within the sample and a markedly different functional behaviour in atomic-and molecular-dominated gas. 

The THINGS data of Bigiel et al., relevant for spirals, as well as the K98 law (Eq.\,\ref{eq:schmidt}), is reproduced in the left-hand panel in Fig.\,\ref{fig:sf} together with the azimuthally averaged ($\Delta r=540\pc$) data from n01e1, n01e2 and n01e5. For the calculation of $\Sigma_{\rm SFR}$ we only consider stars younger than 50 Myr. At a given value of $\Sigma_{\rm gas}$ we find a clear trend of higher $\Sigma_{\rm SFR}$ values for higher $\epsilon_{\rm ff}$. All simulations are compatible with the range of observed values, having the same functional behaviour but with an off-set. We note that only the disc in the n01e5 simulations is compatible with the K98 relation. The n01e1 simulation is on the low side but can still statistically be associated with one of the THINGS spiral galaxies. However, at high redshift the argument can be reversed, as can be seen in the right-hand panel of Fig.\,\ref{fig:sf}. The observations of DLAs at $z\sim 3$ by \cite{WolfeChen06} are typically an order of magnitude lower than the K98 relation, agreeing only with measurement of the low density environment of the discs in our low efficiency simulations. This trend is also predicted by simulations including treatment of H$_2$ formation \citep{GnedinKravtsov2010,GnedinKravtsov10}. In essence, while n01e1 is on the low side at $z=0$, it is consistent with high redshift observations and the reverse argument is valid for n01e5. A higher efficiency is acceptable at lower redshift, and is predicted due to e.g. higher gas metallicity. As the bulge component is assembled at high redshift, the efficiency of star formation during this epoch is crucial in setting the morphology of the galaxy. 

The same analysis is performed for the feedback test suite in Section\,\ref{sect:FB}, and shown in Fig.\,\ref{fig:sfFB}. At $z=0$, all simulations show a similar functional behaviour, but with a weak trend of lower $\Sigma_{\rm SFR}$ as $E_{\rm SNII}$ is increased, while remaining comparable to the K98 law. At $z=3$, a slightly greater effect is found, but only for very large energy injections ($E_{\rm SNII}\geq2\times10^{51}\,{\rm erg}$). The extreme case of $E_{\rm SNII}=5\times10^{51}\,{\rm erg}$ (n01e5SN5) is comparable to a lowering star formation efficiency to $\epsilon_{\rm ff}=2$ per cent (n01e2). None of the strong feedback simulations regulate star formation enough to reproduce the low $\Sigma_{\rm SFR}$ values found for $\epsilon_{\rm ff}=1$ per cent (n01e1). This $z\sim 3$ insensitivity of the \emph{K-S} relation to feedback was also found by \cite{Kravtsov2003}.

\section{Discussion and conclusions}
\label{sect:discussion}
In this paper we have presented a set of AMR simulations studying the assembly of large Milky Way-like disc galaxies. The self-consistent formation of a late-type disc galaxy has remained elusive in the field of numerical galaxy formation, mainly due to the strong loss of angular momentum in the galaxy assembly process. A popular solution to this problem is to regulate star formation at high redshift via supernova explosions that drive galactic winds, transporting material out of star-forming regions hence lowering the local star formation rate. 

We have investigated the plausibility of this mechanism in comparison to a small scale ($\sim 100\,\pc$) physical approach where star formation is made inefficient by modifying the Schmidt-law star formation normalization. In a very crude way, this mimics unresolved physics such as H$_2$ formation, small scale turbulence and radiative effects. We find that the Schmidt-law efficiency of star formation is far more successful way of regulating star formation towards realistic galaxies than what can be achieved via supernova feedback. Our most successful models reproduce Milky Way galaxies with flat rotation curves, where the small bulge component is formed via secular processes. The main conclusions of this work can be summarized as follows.
\begin{enumerate}
\item Disk characteristics such as $\Sigma_*(r)$, $\Sigma_{\rm gas}(r)$, $v_{\rm rot}(r)$ and B/D strongly depend on the choice of star formation efficiency per free-fall time, $\epsilon_{\rm ff}$. The parameter will essentially set the mode of global star formation, hence governing the final spiral Hubble type, where low efficiencies of $\epsilon_{\rm ff}\sim1$ per cent render discs of Sb or Sbc type, while $\epsilon_{\rm ff}=5$ per cent moves the discs closer to Sa/S0 types. Simulations at low efficiencies agree well with observational constraints on disc characteristics \citep{Courteau97,oleggnedin07}, as well as the angular momentum content of disc galaxies \citep{navarrosteinmetz00}, the Tully-Fisher relationship \citep{Pizagno07} and the $\Sigma_{\rm SFR}$-$\Sigma_{\rm gas}$ relation \citep{kennicutt98,bigiel2008}. The origin of the successful Milky Way-like galaxy formation is a well motivated suppression of star formation at $z\sim 3$, the epoch at which the violent assembly process would form a slowly rotating bulge in case of efficient star formation.

\item Supernova feedback does not regulate star formation efficiently at low input energies. Only when the injected energy per supernova event is five times the canonical value, i.e. $5\times 10^{51}\,{\rm erg}$, do we find lower and more realistic B/D ratios in the simulations tuned to the standard \cite{kennicutt98} star formation law, leading to a flatter rotational velocity profile, hence resembling the galaxies formed without strong feedback but with a low Schmidt-law efficiency. This comes at the cost of a significantly distorted gas disc at $z=0$, as well as a less bright stellar disc as gas is expelled into the halo, leaving less fuel for star formation at late times. In essence, we find that changes in $\epsilon_{\rm ff}$ can play a much greater role in shaping a spiral galaxy than gas redistribution via supernovae-driven winds.

It is plausible that at very high resolution, or using a drastically different recipe of supernovae feedback, lower values of $E_{\rm SNII}$ may be successful in regulating the SFE. If so, it will still need to mimic the low efficiency on scales of a few $100\pc$ which, as argued in this work, can be absorbed by the $\epsilon_{\rm ff}$-term.

\item If the star formation efficiency parameter is tuned to match the standard $z=0$ \emph{K-S} data \citep{kennicutt98}, i.e. requiring on the order of $\epsilon_{\rm ff}\ge 5\,{\rm per cent}$ \citep[e.g.][]{Stinson06}, star formation is likely to be overestimated at high redshift ($z=3$) where the amplitudes of $\Sigma_{\rm SFR}$ are an order of magnitude lower \citep{WolfeChen06,GnedinKravtsov2010}. All efficiencies studied in this work ($\epsilon_{\rm ff}=1-5$ per cent) are compatible with modern data of the THINGS survey \citep{bigiel2008} but only when $\epsilon_{\rm ff}\sim 1$ per cent can the constraints from $z=3$ data be met and late-type, disc dominated systems form. As the true SFE varies in space and time, being dependent on small scale physics governing H$_2$ formation \citep[see e.g.][]{Gnedin09}, present day simulations based on single valued efficiency parameter have little predictive power.
\end{enumerate}

We argue \citep[see also][]{Gnedin09} that the results presented in this paper indicate that other processes in the ISM in addition to, or in conjunction with, supernova feedback are important in explaining the evolution of the galaxy population, as well as regulating observed disc sizes. Some form of outflow process must be responsible for enriching the IGM \citep{oppenheimerdave06}, which together with an inefficient star formation might explain the faint end of the stellar mass function \citep{SomervillePrimack99,Keres09}. The same argument can be used for the mass-metallicity relationship \citep{Brooks07}, although \cite{Tassis08} demonstrated that it could be reproduced without supernova-driven outflows. Galaxies of masses considered in this work are situated at the knee of the stellar mass function, where the observed and simulated functions \citep[even without feedback; see][]{Keres09} are in closest agreement. This circumstance might explain why even our simulations without feedback resulted in realistic discs. At this galaxy mass, supernova driven winds cannot escape the deep potential well, and are impeded by the hot halo. On the other hand, AGN feedback, which recently has been introduced into galaxy formation simulations \citep{DiMatteo05}, is probably not relevant for the Milky Way since the black hole might not be massive enough for efficient AGN radio-heating. At higher masses, and/or at high redshift, the inclusion of AGN is probably necessary to correctly reproduce the observed abundances and stellar masses. This is the greatest uncertainty of our work, which we leave for a future study.

The way in which galaxies populate dark matter haloes is an important topic, see e.g. \cite{Dutton2010} and references within for a compilation of recent observational data and theoretical work. Recently, \cite{Guo2010} [see also \cite{Moster2010} and \cite{Behroozi2010}] matched dark matter halo mass function from cosmological $N-$body simulations to the stellar mass function of the galaxies from the SDSS \citep{LiWhite2009}. This analysis yields the required galaxy formation efficiency, $\eta=(M_{*}/M_{\rm halo})(\Omega_{\rm m}/\Omega_{\rm b}$), i.e. what fraction of the universal baryons that must have condensed into stars at a given halo mass. In our "best-case" model (n01e1ML, see Table\,\ref{table:simsummary1}), the total stellar and dark matter halo virial mass is $\sim10^{11}\,\Msol$ and $\sim 10^{12}\,\Msol$ respectively. This results in a stellar fraction of 10 per cent, which corresponds to almost 60 per cent of the cosmic baryon fraction. The rest of the baryons reside in the stellar halo, gaseous disc and ionized gas halo. At this halo mass, abundance matching requires that the stellar disc accounts for only $\sim 20$ per cent of the cosmic baryon fraction, i.e. a factor of three lower. Similar discrepancies exist in all modern work of numerical galaxy formation \citep{Abadi03b,Okamoto05,Governato07,Scannapieco09,Piontek09b}, and its origin is not yet know, although AGN is a compelling mechanism at the high mass end, as discussed above. This issue is the topic of a follow-up paper in preparation.

\cite{Behroozi2010} performed a comprehensive analysis of abundance matching, accounting for systematic errors in e.g. the stellar mass estimates, the halo mass function, cosmology etc.  Our simulated galaxy formation efficiencies would be in  $\sim 2\sigma$ agreement with their result (see their Fig. 11). We note that our own Galaxy and M31 also might be strong outliers in this analysis, considering the inferred $\eta$ from mass modelling \citep{Klypin2002,Seigar2008} as well as via recent MW halo mass estimates \citep{Xue2008}. 

Abundance matching is insensitive to the actual Hubble types of the galaxies, forcing all galaxies of a specific mass to be linked to only one halo mass. At  the stellar mass scale of the Milky Way ($5-7\times 10^{10}\,\Msol$), only $\sim 25$ per cent of galaxies are of Sb/Sbc type \citep{NairAbraham2010}. It is plausible that the more active merger histories associated with ellipticals and early type disc galaxies have led to a stronger mass expulsion, via e.g. AGN, in comparison to the more disc dominated counterparts. In this scenario, late-type discs are expected to be outliers in the galaxy formation efficiency vs. stellar mass relation, considering the strong bias towards early type systems. A detailed sub-division into Hubble types has not yet been performed when matching galaxies to haloes, although color separations into red and blue systems have been made in studies using weak-lensing \citep{Mandelbaum2006} and satellite kinematics \citep{more2010}. These studies indicate a different galaxy formation efficiency for galaxies similar in mass to the Milky Way; a late-type galaxy is associated with a halo of $\sim 0.5$ dex lower halo mass compared to an equally massive early type \cite[see e.g. fig. 11 in][]{more2010}.

Understanding, from a numerical perspective, the spread of baryon fractions across dark matter haloes of different masses, accretion histories and environments is a complicated problem, and will require a large sample of high-resolution simulations, which we leave for a future investigation.

\section*{acknowledgments}
We are very grateful to D. Potter for generating the initial conditions used in this paper. We thank F. Bigiel for providing a copy of his data. We thank Andrey Kravtsov, Joseph Silk, Simon White, Joop Schaye, Francois Hammer and Alister Graham for valuable comments. This work was granted access to the HPC resources of CINES and CCRT under the allocation 2009-SAP2191 made by GENCI (Grand Equipement National de Calcul Intensif). We have also made use of the zBox3 and Schr\"odinger supercomputers at the University of Z\"urich.

\bibliographystyle{mn2e}
\bibliography{galform.bbl}

\appendix
\section{Effect of resolution}
\label{appendix:resolution}
The numerical setup in the SR5-n1e1ML simulation is identical to SR6-n1e1ML, apart from having twice as high spatial resolution and hence eight times the mass resolution (see Table \ref{table:simsummary1}). In Fig.\,\ref{fig:RES} we plot the resulting stellar surface densities and circular velocities from the simulations. We find very little difference apart from more gas being consumed in SR5-n1e1ML resulting in a more massive disc (by $\sim 35$ per cent) at higher resolution and the bulge mass is lowered by $\sim 20$ per cent. The dark halo is less contracted but as $M(<r)$ remains roughly the same, the circular velocities changes little.

Why do we not see any striking differences in the higher resolution simulation? At the adopted resolutions we do not uncover more physics with only a factor of two increase in spatial resolution: we are still under-resolving the scale height and are not resolving individual star-forming clouds. At much higher resolution this conclusion will change and the numerical parameters will have a different meaning.
\begin{figure}
\center
\begin{tabular}{c}
\psfig{file=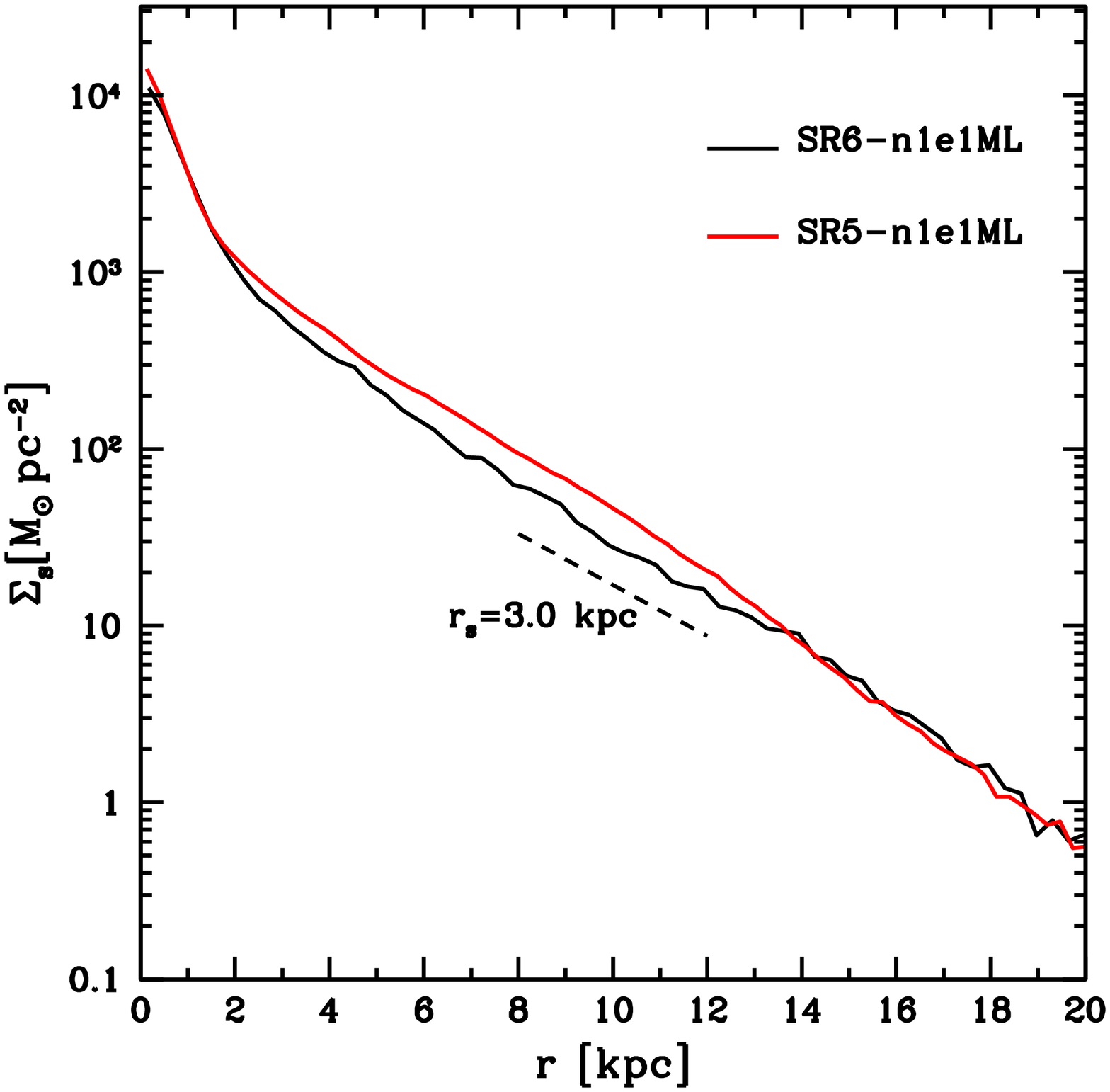,width=200pt} \\
\psfig{file=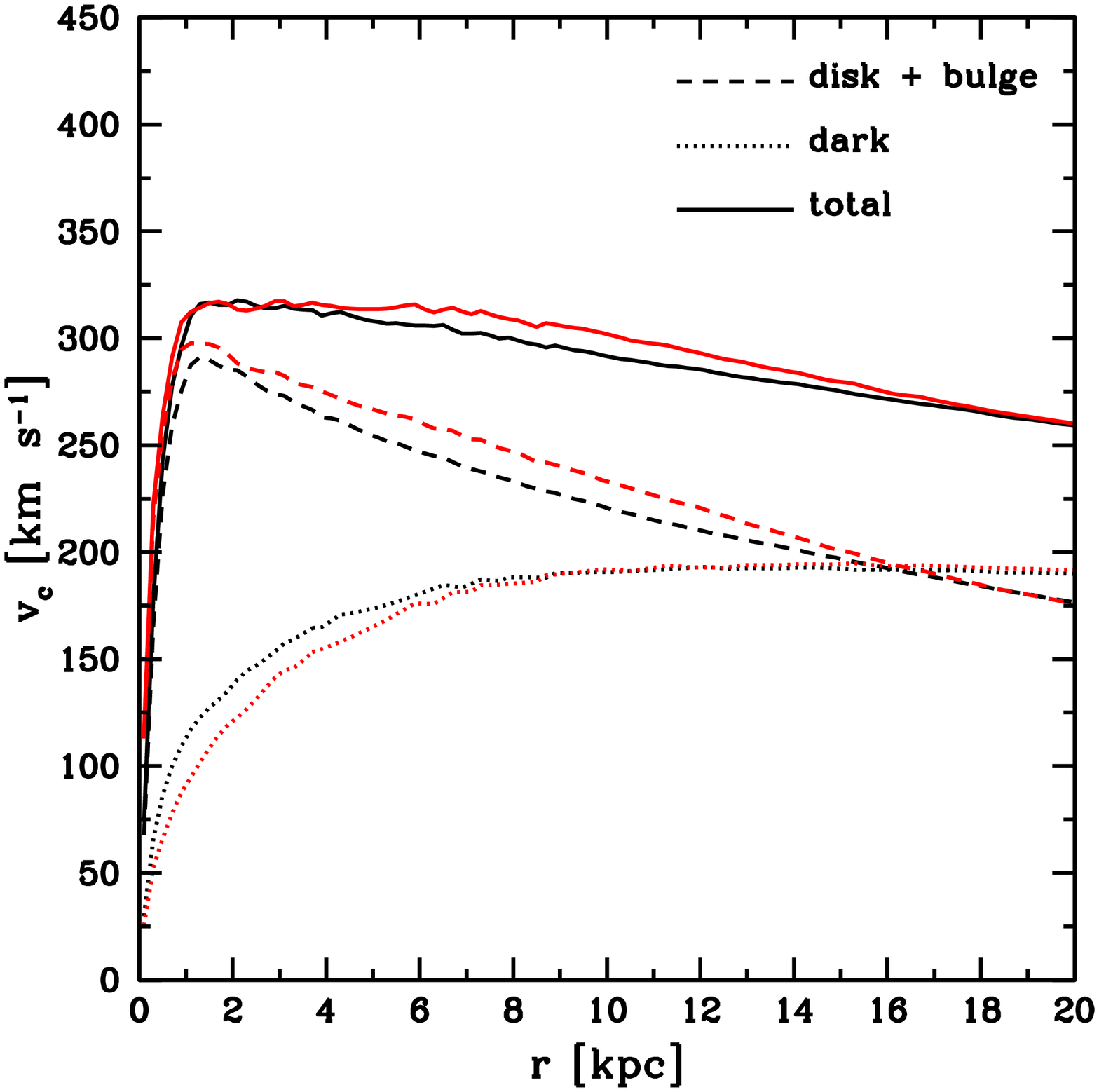,width=200pt} 
\end{tabular}
\caption[]{Stellar surface densities (top) and circular velocities (bottom) in SR6-n1e1ML (black line) and SR5-n1e1ML (red line).}
\label{fig:RES}
\end{figure}

\section{Resolving star formation}
\label{appendix:subgrid}
In this appendix we derive a simple relationship of the required resolution ($\Delta x$) and star formation density threshold ($n_0$) required to capture star formation in an extended disc at the current epoch. Let us assume that a gas disc at $z=0$ follows an exponential radial density profile with a ${\rm sech}^2$ vertical profile, i.e. 
\begin{equation}
\label{eq:rho}
\rho(r,z)=\rho_0\,{\rm sech}^2(-z/h)\exp(-r/r_d).
\end{equation}
The mass of this profile can be integrated from Eq.\,\ref{eq:rho} to give us the characteristic scale disc density
\begin{equation}
\rho_0=M_{\rm d}/4\pi h\,r_d^2.
\end{equation}
Assume that this density distribution is coarsened on a regular mesh with a cell size $\Delta x$, and that the disc is aligned to a mesh axis. The physical density in a central strip of cells can then be calculated as
\begin{align}
  \rho(r)_{\Delta x} &= \frac{M_{\rm d}}{4\pi h\,r_d^2 \Delta x }\exp(-r/r_d)\left(\int_{-\Delta x/2.0}^{\Delta x/2.0}{\rm sech}^2(-z/h){\rm d}z    \right) \\
   &=\frac{M_{\rm d}}{4\pi h\,r_d^2 \Delta x}\exp(-r/r_d)h\left[-{\rm tanh}(-z/h)\right]_{-\Delta x/2.0}^{\Delta x/2.0}
\end{align}
The true star-forming scale height of cold molecular gas is on the order of $\sim 10\pc$, and for the neutral atomic ISM it is $\sim 100\pc$.  Galaxy formation simulations under-resolve this structure, and it is safe to assume that the effective numerical scale height $h\sim\Delta x$, until $\Delta x$ goes below a few tens of parsecs. Under this assumption the integration limits in the ${\rm sech}^2$ term above will always be from $-z/h=-0.5$ to $z/h=0.5$. Eq.\,5 can now be simplified to
\begin{equation}
\label{eq:rhodx2}
  \rho(r)_{\Delta x}=0.924\frac{M_{\rm d}}{4\pi h\,r_d^2\Delta x}\exp(-r/r_d)
\end{equation}
which in units of particles per ${\rm cm}^{-3}$ this is approximately
\begin{equation}
\label{eq:rhodx3}
  n(r)_{\Delta x}\approx\left(\frac{M_{\rm d}}{10^9\Msol}\right) \frac{3\exp(-r/r_d)}{\Delta x \,r_d^2}\,\,\,\, {\rm cm^{-3}}
\end{equation}
We can rearrange this equation and write down the following useful relationship:
\begin{equation}
\label{eq:rad}
	r=r_{\rm d}\ln\left[\left(\frac{M_{\rm d}}{10^9\Msol}\right) \frac{3}{\Delta x \,r_{\rm d}^2 n_{\Delta x}}  \right].  
\end{equation}
This equation tells us at which radius, given a cell size, we cross the density $n_{\Delta x}$. In Section\,\ref{sect:thresh} we argued that using $n=1\,{\rm cm}^{-3}$ and $\Delta x=340\,\pc$ resulted in less realistic galaxies as star formation was "missing" at large radii. Inserting these values into Eq. \ref{eq:rad} gives us $r\sim 6-8\kpc$ for any reasonable choice of $r_{\rm d}$, confirming the arguments of Section\,\ref{sect:thresh}. Adopting $n=0.1\,{\rm cm}^{-3}$ increases the star-forming radius by a factor of 2.3. 

\end{document}